\documentclass[acmsmall, screen,nonacm]{acmart}
\usepackage{url}
\usepackage{amsmath,amsfonts}
\usepackage{algorithmic}
\usepackage{graphicx}
\usepackage{booktabs} 
\setcitestyle{numbers,sort&compress}

\usepackage{textcomp}
\usepackage{xcolor}
\usepackage{cuted}

\usepackage{threeparttable}
\usepackage{tikz}
\usepackage{paralist}
\usetikzlibrary{calc}
\usepackage{multirow}
\usepackage{amsthm}
\usepackage{pifont}
\usepackage{pgfplots}
\usepackage{subcaption}
\usepackage{comment}
\usepackage{circuitikz}
\usepackage{tcolorbox}
\usepackage{threeparttable}
\usepackage{makecell}
\usepackage{hyperref}
\usepackage[group-separator={,}, output-decimal-marker={.}, group-minimum-digits=4, group-digits = integer]{siunitx}
\usepackage{array}
\usepackage{subcaption}
\usepackage{xcolor}
\usepackage{enumitem}
\usepackage[utf8]{inputenc}
\usetikzlibrary{automata,positioning,fit,arrows.meta,backgrounds,shapes.geometric}
\usepackage{graphics}

\pgfplotsset{compat=1.16}

\newcommand{\customcitet}[1]{\textit{AuthorName (Year)}}

\xdefinecolor{mycolor}{RGB}{62,96,111} 
\colorlet{bancolor}{mycolor}

\tcbset{
    compactbox/.style={
        boxsep=1.8pt, 
        left=1pt, 
        right=1pt, 
        top=1pt, 
        bottom=1pt, 
    }
}

\usepackage{amsmath}
\usepackage{xspace}

\usepackage{amsmath,amsfonts}

\usepackage{algorithmic}
\usepackage{graphicx}
\usepackage{booktabs} 
\setcitestyle{numbers,sort&compress}

\usepackage{textcomp}

\def\BibTeX{{\rm B\kern-.05em{\sc i\kern-.025em b}\kern-.08em
    T\kern-.1667em\lower.7ex\hbox{E}\kern-.125emX}}

\usepackage{threeparttable}
\usepackage{tikz}
\usepackage{paralist}
\usetikzlibrary{calc}
\usepackage{multirow}
\usepackage{pifont}
\usepackage{pgfplots}
\usepackage{subcaption}
\usepackage{comment}
\usepackage{circuitikz}
\usepackage{tcolorbox}
\usepackage{threeparttable}
\usepackage{makecell}
\usepackage[group-separator={,}, output-decimal-marker={.}, group-minimum-digits=4, group-digits = integer]{siunitx}
\usepackage{array}
\usepackage{subcaption}

\usepackage{hanging}
\usepackage{enumitem}
\usepackage{ninecolors}
\usetikzlibrary{automata,positioning,fit,arrows.meta,backgrounds,shapes.geometric}
\usepackage{graphics}
\usepackage{enumitem}

\usepackage{currency}
\DefineCurrency{USD}{name={dollar},plural={dollars},iso={USD},kind=iso, symbol={\$}}
\usepackage{amsmath}

\usepackage{url}
\usepackage{siunitx}
\usepackage{graphicx} 
\usepackage{amsmath}
\usepackage{amssymb}
\usepackage{comment}
\usepackage{subcaption}
\DeclareSIUnit{\pp}{\textup{pp}}

\pgfplotsset{compat=1.16}

\xdefinecolor{mycolor}{RGB}{62,96,111} 
\colorlet{bancolor}{mycolor}

\tcbset{
    compactbox/.style={
        boxsep=1.8pt, 
        left=1pt, 
        right=1pt, 
        top=1pt, 
        bottom=1pt, 
    }
}

\usepackage{amsmath}
\usepackage{xspace}
\usepackage[normalem]{ulem}
\usepackage{pgfplotstable}
\usepackage{xfp}
\usepackage[]{xcolor}

\newcommand{\method}{\textsc{FlexLog}\xspace}
\newcommand{\task}{ULAD\xspace}
\newcommand{\slad}{SLAD\xspace}

\newcommand{\logevol}{LOGEVOL\xspace}
\newcommand{\hdfs}{HDFS\xspace}
\newcommand{\unstableadfa}{ADFA-U\xspace}
\newcommand{\unstablelogevol}{LOGEVOL-U\xspace}
\newcommand{\synlogevol}{SYNEVOL-U\xspace}
\newcommand{\synhdfs}{SynHDFS-U\xspace}
\newcommand{\datasize}{$\mathcal{D}_\#$\xspace}
\newcommand{\uniquesize}{$\mathcal{U}_\#$\xspace}
\newcommand{\uniquefullsize}{$\mathcal{U}_\#^{full}$\xspace}
\newcommand{\uniquepercent}{$\mathcal{U}_\%$\xspace}
\newcommand{\uniqueabnormal}{$\mathcal{U}^+_\%$\xspace}
\newcommand{\uniquenormal}{$\mathcal{U}^-_\%$\xspace}
\newcommand{\lcreduction}{$\Delta \mathcal{U}_\%$\xspace}
\newcommand{\fulltest}{Testing Dataset \xspace}
\newcommand{\woMistralError}{$\mathtt{MIS}_{\mathit{w/o\; Mistral}}$\xspace}
\newcommand{\woMLError}{$\mathtt{MIS}_{\mathit{w/o\; ML }} $\xspace}

\theoremstyle{definition}
\newtheorem{definition}{Definition}[section]

\title{LLM meets ML: Data-efficient Anomaly Detection on Unstable Logs}

\author{Fatemeh Hadadi}
\affiliation{
  \institution{University of Ottawa}
  \city{Ottawa}
  \country{Canada}
}
\email{fhada072@uottawa.ca}

\author{Qinghua Xu}
\affiliation{
  \institution{Lero, University of Limerick}
  \city{Limerick}
  \country{Ireland}
}
\email{et.qinghua@gmail.com}

\author{Domenico Bianculli}
\affiliation{
  \institution{University of Luxembourg}
  \city{Luxembourg}
  \country{Luxembourg}
}
\email{domenico.bianculli@uni.lu}

\author{Lionel Briand}
\affiliation{
  \institution{University of Ottawa}
  \city{Ottawa}
  \country{Canada}
}
\affiliation{
  \institution{Lero, University of Limerick}
  \city{Limerick}
  \country{Ireland}
}
\email{lbriand@uottawa.ca}

\setcopyright{cc}
\setcctype{by}
\acmJournal{TOSEM}
\acmYear{2025}
\acmVolume{1}
\acmNumber{1}
\acmArticle{1}
\acmMonth{1}
\acmPrice{}
\acmDOI{10.1145/3771283}

\begin{document}

\begin{abstract}

Most log-based anomaly detectors assume logs are stable, though in reality they are often unstable due to software or environmental changes. Anomaly detection on unstable logs (\task) is therefore a more realistic, yet under-investigated challenge. 
Current approaches predominantly employ machine learning (ML) models, which often require extensive labeled data for training. To mitigate data insufficiency, we propose \method, a novel hybrid approach for \task that combines ML models --- decision tree, k-nearest neighbors, and a feedforward neural network --- with a Large Language Model (Mistral) through ensemble learning. \method also incorporates a cache and retrieval-augmented generation (RAG) to further enhance  efficiency and effectiveness.
To evaluate \method, we configured four datasets for \task, namely \unstableadfa, \unstablelogevol, \synhdfs, and \synlogevol. \method outperforms all baselines by at least \num{1.2} percentage points (pp) in F1 score while using much less labeled data (\num{62.87} pp reduction). When trained on the same amount of data as the baselines, \method achieves up to a \num{13} pp increase in F1 score on \unstableadfa across varying training dataset sizes. Additionally, \method maintains inference time under one second per log sequence, making it suitable for most applications, except latency-sensitive systems. Further analysis reveals the positive impact of \method's key components: cache, RAG and ensemble learning. 

\end{abstract}


\maketitle
\begin{tcolorbox}[colback=gray!10, colframe=white, boxrule=0pt, left=2pt, right=2pt, top=2pt, bottom=2pt]
\small
\textit{This is the author's version of the work.  The definitive Version of Record was published in 
\textbf{ACM Transactions on Software Engineering and Methodology (TOSEM)}, 
\href{https://doi.org/10.1145/3771283}{http://dx.doi.org/10.1145/3771283}.}
\end{tcolorbox}

\section{Introduction}\label{sec:intro}
Various software-intensive systems, such as online service systems and Big Data systems, have permeated every aspect of people's daily lives. As the prevalence of such systems continues to grow, the potential impact of software failures has become increasingly significant. A critical software failure can result in service interruptions, financial losses and, in severe cases, poses threats to human safety~\cite{krasner2021cost}. 

Log-based anomaly detection has emerged as a promising approach to enhancing the dependability of software-intensive systems. An anomaly detector aims to discern anomalous patterns within system logs, which serve as vital indicators of the system's operational state. Early research predominantly employed classical machine learning techniques, such as Principal Component Analysis~\cite{detecting2009}, Isolation Forest~\cite{isolation2008}, and one-class SVM~\cite{hejazi2013one} for automated anomaly detection. However, these methods overlook the contextual information of the logs and, as a result, exhibit less effectiveness on more challenging cases~\cite{LogRobust, evlog, Neurallog}. 

In recent years, deep learning-based (DL) methods have gained significant traction in anomaly detection. Unlike classical machine learning methods, DL methods typically consist of a large number of trainable parameters, enabling them to model long contextual dependencies and complex semantic patterns in logs. In particular, log-based anomaly detection has significantly benefited from sequential DL models such as LSTM and transformers, achieving high predictive performance on multiple benchmark datasets~\cite{LogAnomaly,maddc,logbert,DeepLog}. Despite the success of DL-based methods, we highlight three key challenges prevalent in current practices of log-based anomaly detection:

\begin{compactenum}[\bfseries {C}1]
    \item \textbf{Existing approaches often assume a stable data distribution, which is unrealistic in real-world scenarios.} In practice, unlike current benchmark datasets, where log structures and contents remain stable, logs can be \emph{unstable} due to software evolution or environment changes. The majority of anomaly detection methods~\cite{DeepLog, LogAnomaly, lin2016log, Xu2010, PLELog, logbert, CNN, Neurallog} have been proposed for and evaluated on stable log datasets. In contrast, only a few studies~\cite{LogRobust, swisslog, evlog} have investigated anomaly detection on unstable logs, mainly due to a lack of public benchmarks. Earlier works leverage private or synthetic data. However, more recently, \citet{evlog} have proposed two public datasets for unstable logs.
    
    \item \textbf{Machine learning-based (ML) anomaly detection, especially when based on DL methods, often relies on substantial labeled data, which is costly to obtain. } The most effective methods in anomaly detection---particularly those based on DL---often rely on an extensive amount of labeled data for their training, due to their substantial number of trainable parameters. Collecting such data requires intensive labor and significant domain knowledge in practice. More recently, the study of~\citet{lightad} has demonstrated that simpler methods, such as Decision Trees (DT), exhibit greater effectiveness. However, they only evaluated their methods on stable logs.
    
    \item \textbf{The reported effectiveness of ML-based anomaly detection might be inflated due to data leakage issues.} \citet{boxi} found such issues in several benchmark datasets such as HDFS~\cite{HDFS} and BGL~\cite{bgl-spirit}, where the testing data contains training instances. This leakage can potentially boost the effectiveness of ML methods, especially DL methods, as their large parameter set allows them to memorize the training data. To address this issue, \citet{lightad} removed instances in testing data that were already seen in the training data, which led to a significant drop in anomaly detection effectiveness.

\end{compactenum}

In light of these challenges, we identify the next frontier of log-based anomaly detection as addressing a more realistic and demanding task: \emph{anomaly detection on unstable logs with limited labeled data (\task)}. Unlike \emph{anomaly detection on stable logs (\slad)}, \task reflects the practical reality where logs evolve due to software updates or environmental factors, resulting in instability (C1). This evolution results in changes such as the addition, removal, or modification of log messages, as well as shifts in their order. Furthermore, real-world constraints often limit the availability of labeled data, which is costly to collect and requires domain knowledge (C2). To ensure a realistic assessment of \task solutions, test instances already seen in the training data should be removed from testing data, addressing issues of data leakage (C3).

The literature on anomaly detection in the presence of unseen log templates~\cite{LogAnomaly, PLELog, logbert} is related to the \task challenge. However, \task is more challenging since, although unstable logs are test instances not present in training data, they further follow a different log distribution than historical data. Moreover, these works do not address the C2 and C3 challenges.

A promising approach to tackling  these challenges of \task lies in leveraging large language models (LLMs). Recently, LLMs have gained significant attention for their ability to mitigate the data insufficiency problem faced by ML-based methods. By pretraining on vast, diverse datasets, LLMs can excel in tasks with limited labeled data. Several researchers have investigated various prompts to instruct pretrained LLMs such as GPT
to perform anomaly detection directly (i.e., in-context learning)~\cite{LLMeLog, logprompt, LogRAG}. An alternative to in-context learning is fine-tuning, where extra training on domain-specific data is required. While in-context learning has been more widely studied because it does not require additional training and can be applied directly with prompts, \citet{mosbach-etal-2023-shot} highlighted its poor generalizability on challenging tasks. Drawing from this observation, fine-tuning may be a more suitable strategy for \task when using LLMs. 

Most recent works~\cite{LLMeLog, LogRAG, lilac, logbatcher, logparsingchatgpt, YALP} in log analysis focused on using closed-source LLMs from OpenAI, due to their user-friendly environment and effective performance. However, these LLMs induce a significant financial cost and show unpredictable latency during training and inference~\cite{10590016}. On the other hand, open-source LLMs are free to use, and we have some degree of control in terms of fine-tuning algorithms and inference time.

Though a fine-tuned LLM can address challenges C1, C2, and C3 faced by ML methods, they are inherently designed for textual understanding and generation rather than the detection of anomalous patterns in logs. Conversely, existing anomaly detectors using ML models such as Decision Tree (DT) and Single-layer Feedforward Network (SLFN) have proven to be effective in the SLAD task when abundant data is available for training~\cite{boxi}, demonstrating their capacity to detect anomalous patterns. This indicates that combining ML methods with an LLM would leverage the strengths of both approaches, ML models' ability to detect anomalous patterns and fine-tuned LLM's capacity to handle scarce labeled data effectively.

To this end, we propose \emph{\method}, a novel approach that requires significantly less labeled data for \task compared to ML methods.   \method integrates ML-based anomaly detectors and an LLM, combining their strengths to enhance effectiveness and data efficiency. We summarize our contributions as follows. 

\begin{itemize}
    \item \textbf{Dataset configuration for  \task.} Most existing benchmark datasets contain stable logs, used for the \slad task. In this paper, we selected three of these datasets---HDFS, LOGEVOL, and ADFA-LD (referred to as ADFA for brevity hereafter)---and configured four unstable datasets for \task, namely  \synhdfs, \unstablelogevol, \synlogevol, and \unstableadfa,  by deliberately introducing disparities between the training and testing datasets. To eliminate the influence of data leakage (C3)~\cite{boxi} and further increase instability, we performed de-duplication on each dataset, ensuring that any data samples included in the testing datasets were excluded from the training datasets. 
\item \textbf{\method, a novel approach for \task equipped with practical strategies to boost effectiveness and efficiency.}  \method uses average-based ensemble learning to combine the predictive strengths of a fine-tuned LLM and ML methods, capturing intricate anomalous patterns with only limited labeled data for training. Specifically, to address C2, \method employs parameter-efficient fine-tuning (PEFT) of a pretrained LLM, leveraging its vast embedded knowledge to mitigate the constraints of scarce labeled data. To tackle C1, \method integrates retrieval-augmented generation (RAG) to dynamically incorporate external knowledge and enhance the model's adaptability to unstable log distributions.  Additionally, a cache mechanism improves computational efficiency by eliminating redundant operations.
 
    \item \textbf{State-of-the-art effectiveness and data efficiency for \task.} To evaluate \method, we first compare it against baselines trained on full datasets, even though \method itself is trained on significantly smaller datasets. This comparison is conducted on two real-world datasets (\unstableadfa and \unstablelogevol) and two synthesized datasets (\synhdfs and \synlogevol). Experimental results show that \method achieves state-of-the-art effectiveness, outperforming the top baseline by at least \num{1.2} percentage points (pp) in terms of F1 score, while reducing the usage of labeled data by more than \SI{62.87}{\pp}. Further, we assess the data efficiency of \method by comparing it with baselines when trained on the same datasets. Experiment results on \unstableadfa show that \method consistently outperforms all baselines across varying training dataset sizes, except the extreme data scarcity scenario (the training dataset size is 50), where all methods exhibit poor performance due to insufficient labeled data. \method achieves a maximum gain of \SI{13}{\pp} in F1 score when the training dataset size is 500. This confirms \method is the most effective choice when only limited labeled data is available.  
\end{itemize}

The rest of the paper is organized as follows. Section~\ref{sec:background} presents the basic definitions and concepts that will be used throughout the paper. Section~\ref{sec:GPT-basedAD} describes our data-efficient anomaly detection approach, \method. Section~\ref{sec:studydesign} presents our experimental design. Section~\ref{sec:results} outlines our results, discusses the implications, and describes the threats to the validity of our study. Section~\ref{sec:related-works} presents related works and finally, Section~\ref{sec:conclusion} concludes and suggests future directions for research
and improvement.

\section{Background}\label{sec:background}
\subsection{Logs}\label{sec:background-logs}
\begin{figure}[tb]
    \centering
    \includegraphics[width=0.8\columnwidth]{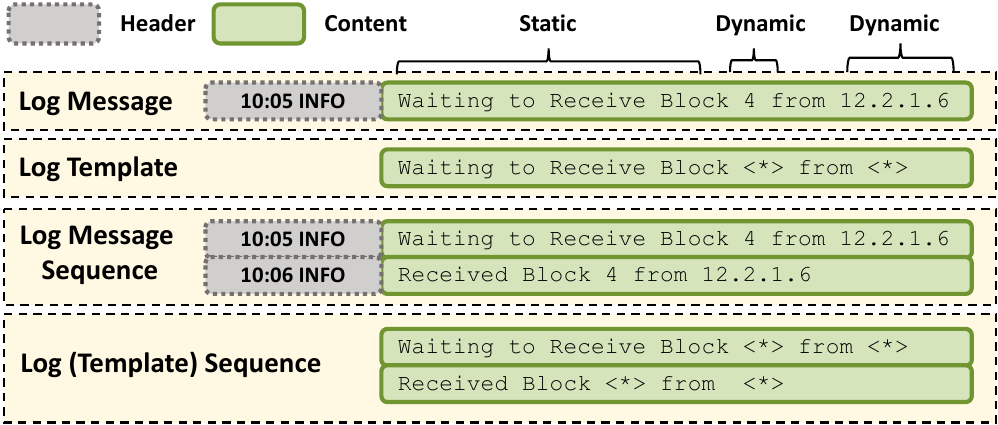}
     \captionsetup[figure]{skip=0pt}
     \captionof{figure}{Examples of log Message, Log Message Sequence, Log Template, and Log Template Sequence. }
    \label{fig:log_terms}
\end{figure}

Logs are semi-structured or unstructured text generated by logging statements (e.g., \textit{"logging.info\\('Received Block \%d from \%s',id, ip)"}) in source code~\cite{HeSurvey}. The main concepts related to logs are \textit{log message}, \textit{log message sequence}, \textit{log template}, and \textit{log template sequence}, which we explain below and exemplify in Figure~\ref{fig:log_terms}.

A \textit{log message} contains two main components: the \textit{header} (e.g., timestamp or log level), and the \textit{content}, depicted as grey-dotted and green boxes in Figure~\ref{fig:log_terms}, respectively. The content of a log message can be further divided into static and dynamic parts. The static parts refer to the fixed text written by developers in the logging statement, e.g.,``\texttt{Waiting to Receive Block from}'' and ``\texttt{Received Block from}''; the dynamic parts are expressions evaluated at runtime, such as the actual block id \texttt{``4''} and  IP address \texttt{``12.2.1.6''}. 

A \textit{log template}, also called \textit{event template} or \textit{log key}, is usually obtained through log parsing~\cite{logparsing-survey}, which masks the dynamic parts of the log message content with a special symbol, such as \texttt{``<*>''}. Compared to log messages, log templates eliminate the influence of specific values in the dynamic parts, enabling downstream tasks (e.g., anomaly detection~\cite{LANDAUER2023100470}, log summarization~\cite{logsummerization-survey}) to focus on analyzing patterns within logs,  without being confused by variations in concrete values.

A \textit{log message sequence} is a log fragment consisting of multiple log messages, which typically records the execution flow of a specific job or process. A log message sequence consists of log messages that either pertain to a specific task (i.e., session-based partitioning) or are grouped within a fixed-size window (i.e., sliding/fixed window partitioning). Session-based partitioning groups log messages based on their session IDs, thereby creating log sequences that encapsulate activities within sessions. On the other hand, sliding/fixed window partitioning employs a fixed-size window to group log messages, generating log sequences that capture a snapshot of system activity over time.

Parsing a log message sequence yields a \textit{log template sequence} (as shown at the bottom of Figure~\ref{fig:log_terms}).
We remark that ``log template sequence'' is often referred to simply as ``log sequence'' in the literature; therefore, for brevity, we adopt the term \emph{log sequence} throughout this paper. 

\subsection{Anomaly Detection on Logs}\label{sec:background-anomalydetection}
\textit{Anomalies} in logs refer to logs that do not conform to the normal behavior of a system~\cite{HeSurvey}. 
Log-based anomaly detection represents a binary classification task to identify anomalies from logs. 
Depending on their distributions, logs can be divided into two categories: \emph{stable logs} (Definition ~\ref{def:stablelogs}) and \emph{unstable logs} (Definition ~\ref{def:unstablelogs}).  
\begin{definition}[Stable Logs]
\label{def:stablelogs}    
Logs drawn from a single underlying distribution, i.e., their structure and semantics remain consistent in all the logs.
\end{definition}
\begin{definition}[Unstable Logs]
    \label{def:unstablelogs}
    Logs drawn from more than one underlying distribution.
\end{definition}

\emph{Stable logs} are typically generated from systems whose logging behaviors and operating environment remain unchanged over time, resulting in consistent structure and semantics of logs. In contrast, unstable logs originate from multiple distributions caused by system or environmental changes. \emph{System evolution} refers to internal changes within a software system, such as version upgrades. Developers often modify source code, including logging statements, resulting in changes to logs. As \citet{logevolution} reported, around \SI{24}{\percent}--\SI{40}{\percent} of log statements change during their lifetime. Taking the public dataset \logevol as an example, \SI{24}{\percent} of logging statements were modified during the system upgrade from Spark version 2 to 3~\cite{evlog}. As a result, \SI{14}{\percent} of logs collected in Spark~3 contain new log templates induced by system evolution. This figure represents a conservative estimate of the percentage of unstable logs, as other causes of instability (e.g., reordering of logging statements during execution) are not accounted for due to the lack of a mapping between execution paths and their log distributions. Nonetheless, these results clearly highlight that unstable logs are common in practice, underscoring the importance of handling instability caused by system evolution. \emph{Environmental evolution}, on the other hand, represents the changes of external factors, such as a shift of user distribution and the emergence of unseen attack types. These changes affect both normal and abnormal patterns in the logs by altering the structure, content, or frequency of log messages. For example, shifts in user distributions---such as changes in user geographic regions---may introduce new log sequences or alter the frequency of existing ones due to differences in usage patterns, device configurations, or regional preferences. Similarly, novel or previously unseen attacks may generate anomalous logs with sequences or templates that have not been observed in the earlier log distribution.

Built on the definitions of stable and unstable logs, we define two corresponding anomaly detection tasks, namely \textit{Anomaly Detection on Stable Logs (\slad)} and \textit{Anomaly Detection on Unstable Logs (\task)} as defined in in Definition \ref{def:slad} and ~\ref{def:ulad}, respectively. 
\begin{definition}[Anomaly Detection on Stable Logs]
\label{def:slad}  \slad is a binary classification task that aims to predict anomalies in stable logs, i.e., the training data and testing data follow the same distribution.  \end{definition}
\begin{definition}[Anomaly Detection on Unstable Logs]
\label{def:ulad} \task is a binary classification task that aims to predict anomalies in unstable logs, i.e., the training data and testing data are drawn from different distributions. \end{definition}
\slad is the predominant configuration in the literature ~\cite{LogRobust, le2022}, largely because existing benchmark datasets often assume a stable software system. In this configuration, the training dataset $D^{\mathit{train}}=\{ls_1, ls_2, ..., ls_n\}$ and the testing dataset  $D_S^{\mathit{test}}=\{ls_{n+1}, ls_{n+2}, ..., ls_{n+m}\}$ are sampled from the same distribution. Despite its wide adoption, \slad does not reflect challenges faced by real-world applications, e.g., evolving systems or operating environments. In contrast, \task (Definition \ref{def:ulad}) considers a more challenging yet realistic scenario. Specifically, the training dataset $D^{\mathit{train}}$ consists of stable logs collected under consistent conditions, while the test dataset $D_U^{\mathit{test}}$ contains unstable logs resulting from system or environmental evolution. \\

\noindent\textbf{Evaluation on De-duplicated Datasets.} As demonstrated in previous work~\cite{boxi}, data leakage is prevalent in existing benchmark datasets, artificially inflating the effectiveness of anomaly detectors. Data leakage entails an overlap between testing and training data, i.e., some log sequences in the testing dataset have already been seen in the training dataset. This phenomenon affects both the \slad and \task tasks. To eliminate the risk of data leakage, we remove \textit{seen} testing log sequences that are already present in the training dataset $D^{\mathit{train}}$, yielding a new testing dataset for \task and \slad, denoted as $D_{U\dagger}^{\mathit{test}}$ and $D_{S\dagger}^{\mathit{test}}$ respectively, defined in Equations~\ref{eq:dedup} and~\ref{eq:dedup_stable}.

\begin{equation}
    \label{eq:dedup}
    D_{U\dagger}^{\mathit{test}}=D_U^{\mathit{test}}\setminus D^{\mathit{train}}
\end{equation}

\begin{equation}
    \label{eq:dedup_stable}
    D_{S\dagger}^{\mathit{test}}=D_S^{\mathit{test}}\setminus D^{\mathit{train}}
\end{equation}

$D_{U\dagger}^{\mathit{test}}$ and $D_{S\dagger}^{\mathit{test}}$ consist of only \textit{unseen log sequences}. These sequences differ from the ones in the training dataset at two possible levels: 1) template level (when there is an \textit{unseen log template} in the sequence), 2) sequence level (when all the log templates are already mentioned in the training data but their order is new). Unseen log sequences can be either stable or unstable, depending on their underlying log distributions. As mentioned by \citet{lightad}, after de-duplication, the effectiveness of anomaly detection models on testing data drops, making it a more challenging task in this realistic scenario.\\
\begin{figure*}[tb]
    \centering
    \includegraphics[width=\columnwidth]{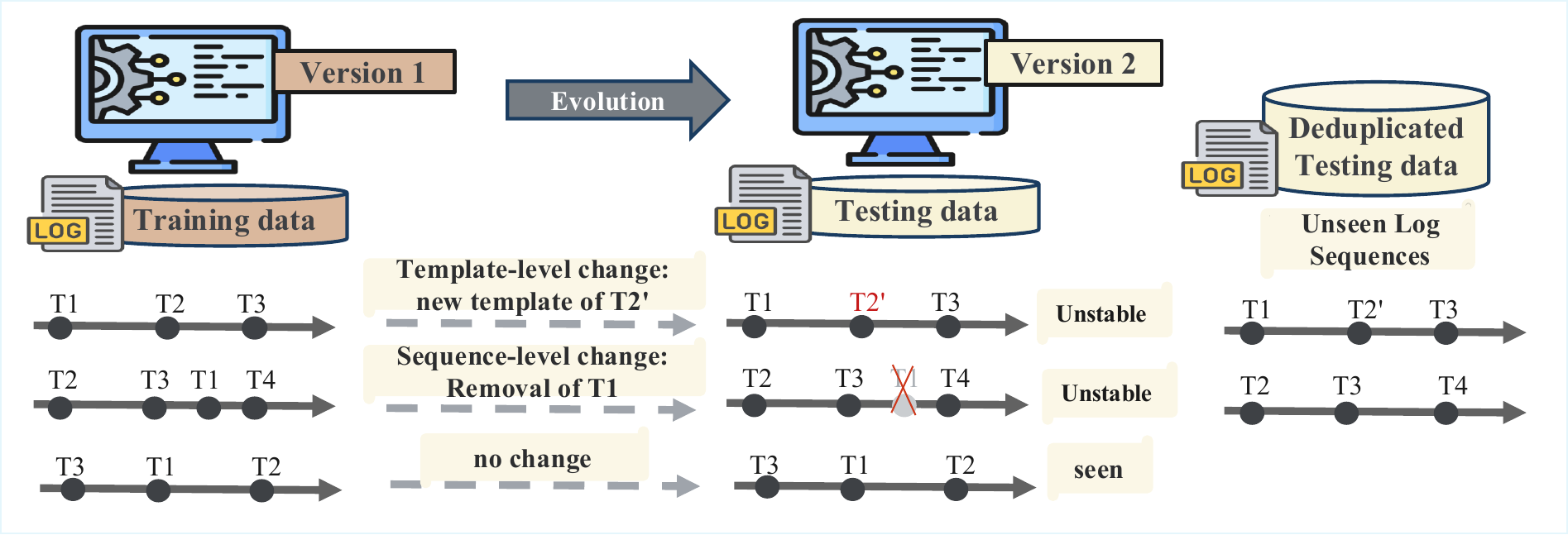}
     \captionsetup[figure]{skip=0pt}
     \captionof{figure}{Examples of Unstable Logs Resulting from Log Evolution.} 
     \label{fig:unseen_unstable}
\end{figure*}

\noindent\textbf{Illustrative Examples.} Figure~\ref{fig:unseen_unstable} provides an overview of the de-duplicated \task with example log sequences. After system evolution from version 1 to version 2, the first two sequences at the top undergo changes at different levels. In the first sequence, $T1 \rightarrow T2 \rightarrow T3$, template $T2$ is updated to a new template $T2^\prime$, representing a change at the template level. The second sequence, $T2 \rightarrow T3 \rightarrow T1 \rightarrow T4$, experiences a change at the sequence level, where template $T1$ is no longer present. The last sequence shown, $T3 \rightarrow T1 \rightarrow T2$, is regenerated in the later version without any change, and is therefore marked as a ``seen'' sequence relative to the training data. During de-duplication, the seen sequence is removed from the test set. The remaining sequences are referred to as unseen log sequences, as they contain no overlapping sequences.

\subsection{Task Adaptation Strategies for Large Language Models }
\label{subsec:task_adaption_llm}
LLMs typically consist of substantial parameters pretrained on vast and diverse datasets, possessing knowledge across various domains. However, how to effectively adapt pretrained LLMs to domain-specific tasks remains an open problem. Two predominant strategies for task adaptation are in-context learning (ICL) and fine-tuning (FT). 
 
ICL operates without altering the weights of the LLMs~\cite{brown2020language}. Instead, it leverages prompts—structured textual inputs—to guide the model’s behavior. These prompts typically include task instructions and, in some cases, a series of demonstrations in a conversation between the user and the assistant. In a classification task, e.g., \task, each demonstration consists of an input  $x$  paired with its corresponding ground-truth label  $y$. When no demonstrations are provided, the approach is referred to as zero-shot ICL, whereas the inclusion of a few demonstrations constitutes a few-shot ICL.

Although ICL is relatively easy to implement, it faces several challenges and limitations, including issues with efficiency, scalability, generalizability, and high financial cost when using closed-source LLMs~\cite{dong2022survey}. As an alternative, FT alleviates these issues by training pre-trained LLMs with domain-specific data. In practice, there are two main types of fine-tuning, namely API -based FT and Custom FT.

API-based FT refers to fine-tuning performed through dedicated APIs made available by the LLM provider, e.g., OpenAI~\cite{fine-tuning-format}. This is typically the case for closed-source LLMs,

such as
GPT-3.5~\cite{Ouyang2022TrainingLM} and GPT-4~\cite{Achiam2023GPT4TR}, for which neither full nor selective fine-tuning is allowed without accessing their APIs. 
These APIs support fine-tuning a set of prompt-completion pairs or conversations depending on whether LLMs are used in purely generative or conversational settings. 

Custom FT, on the other hand, is applicable to open-source LLMs, such as LLama~\cite{touvron2023llama} and Mistral~\cite{jiang2023mistral7b}. Common custom FT techniques include full fine-tuning and parameter-efficient fine-tuning (PEFT). Let the trainable parameter set of an LLM be denoted as $W$, the task-specific dataset as $D$, and its associated label set as $L$. 
\begin{itemize}
    \item \emph{Full Fine-Tuning}: This approach utilizes gradient descent-based optimizer to update $W$ to $W^f$, thereby adapting the LLM to a specific task. Specifically, prompts are constructed using $D$ and fed into an LLM. The model's output distribution $\hat{y}$ is then compared with the corresponding label distribution $y$, using a distribution-level loss, e.g., cross-entropy loss. This loss guides weight updates from $W$ to $W^f$  through backpropagation. 
    \item \emph{Parameter-Efficient Fine-Tuning:} PEFT preserves original LLM weights while training only a small number of task-specific adapter layers and parameters. There are several types of PEFT methods, including additive, selective, reparameterized, and hybrid PEFT~\cite{han2024parameterefficient}. The predominant PEFT techniques are LoRa~\cite{hu2021lora} and its derivative techniques such as QLoRa~\cite{qlora}. Essentially, LoRa uses a low-rank decomposition to reduce computational cost while maintaining performance similar to full fine-tuning as in Equation \ref{eq:lora}.
    \begin{equation}
    \label{eq:lora}
    \begin{aligned}
        W^f &=W+\Delta W \\
        &=W+AB
    \end{aligned}
\end{equation}
where $A\in  \mathbb{R}$ and $B\in  \mathbb{R}$ are lower rank matrices compared to $W$, with dramatically fewer trainable parameters. Such techniques significantly reduce the computational cost while maintaining comparable performance to fully fine-tuned LLMs.
\end{itemize}

\section{Methodology}\label{sec:GPT-basedAD}

\begin{figure*}[hbt]
\centering
\includegraphics[width=\columnwidth]{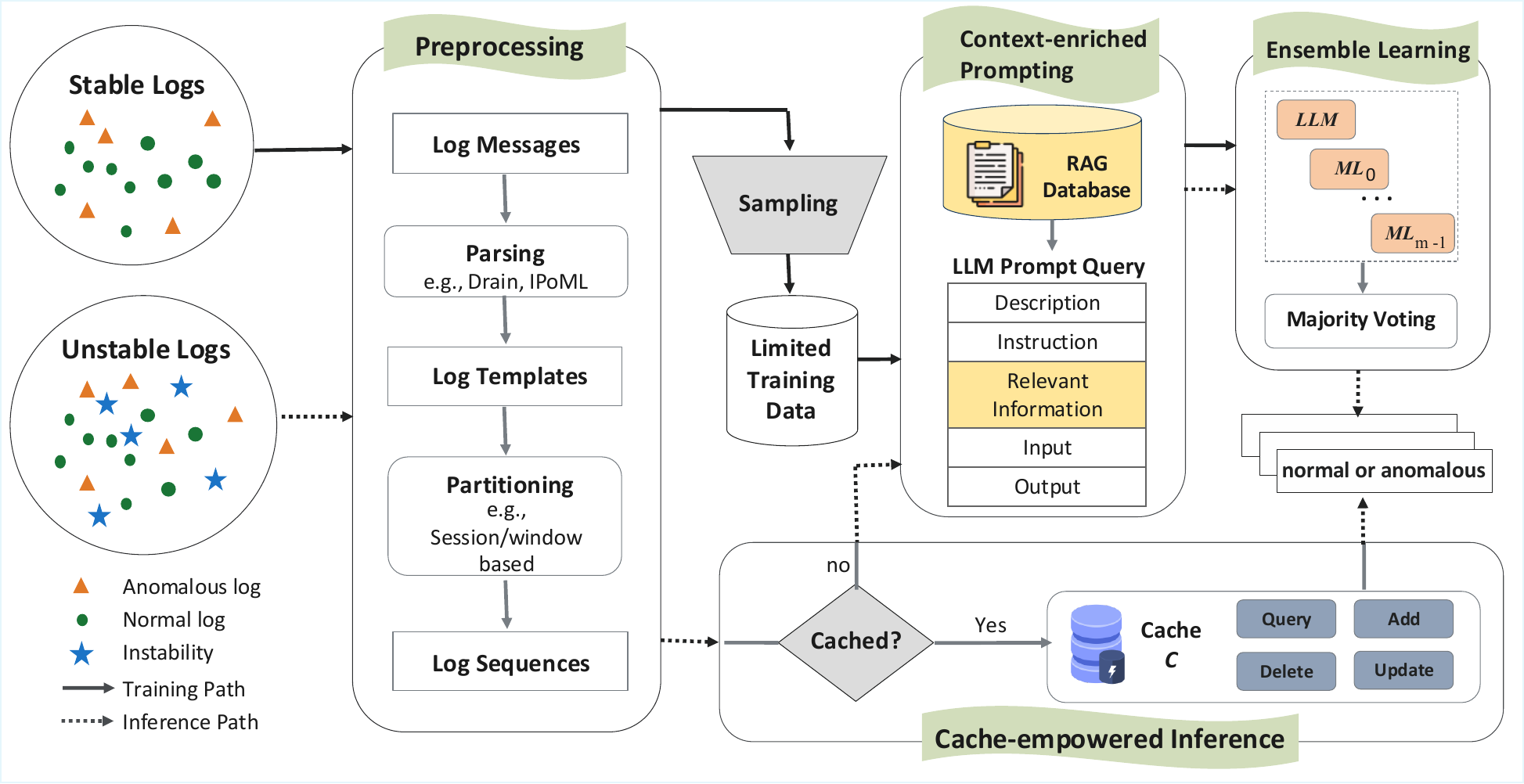}
 \captionsetup[figure]{skip=0pt}
     \captionof{figure}{Architecture of \method. }\label{fig:overviewmodel}
\end{figure*}

In this paper, we propose a novel approach, namely \method, to tackle \task by synergizing the capabilities of ML methods and LLMs via ensemble learning. Specifically, \method combines the predictions from trained ML methods and a fine-tuned LLM to make a final decision. This approach leverages the strengths of both paradigms: ML methods excel at capturing anomalous patterns within logs, while LLM brings broad prior knowledge from pretraining, allowing them to adapt to novel log patterns even with limited labeled data. Also, we tackle the three key challenges in \task --- unstable log distribution (C1), data insufficiency (C2), and data leakage (C3) --- by employing ensemble learning of ML and LLM, PEFT, and de-duplication in the testing data, respectively. Additionally, we further tackle unstable log distributions (C1) by using RAG in prompting when relevant external information is available regarding log sequences.

Figure~\ref{fig:overviewmodel} illustrates the architecture of \method, which comprises four main components: preprocessing (\S~\ref{subsec:preprocessing}), cache-empowered inference (\S~\ref{subsec:cache}), context-enriched prompting (\S~\ref{subsec:rag}), and ensemble learning (\S~\ref{subsec:fsel}). \method is designed to predict whether unstable logs—generated by software systems that have undergone software or environmental evolution—are anomalous. Specifically, the \emph{preprocessing component} converts raw stable and unstable logs
 into log sequences by extracting log templates and grouping them into log sequences using either window-based or session-based partitioning. For illustrative purposes, consider a log sequence \( \mathit{ls}_i \) as an example. The \emph{cache-empowered inference component} first checks if \( \mathit{ls}_i \) matches an existing entry in the cache. If a match is found, \method retrieves the stored prediction as the output label directly. Otherwise, the \emph{context-enriched prompting component} uses \( \mathit{ls}_i \) in a structured prompt enriched with contextual information, such as log event descriptions and Linux system call names. This prompt includes key fields such as description, instructions, relevant information (optional), input, and output. It serves as input both for fine-tuning of and for inference with an LLM, which acts as one of \method's base models. 
 To construct the ensemble, the \emph{ensemble learning component} fine-tunes the LLM-based model (e.g., Mistral or Llama) and trains the ML-based models (e.g., DT and KNN~\cite{lightad}) on a limited dataset sampled from the preprocessed stable logs to maintain data efficiency and reduce training overhead. During inference, if no matching log sequence is found in the cache, binary anomaly predictions from these base models are aggregated using majority voting to produce the final decision. This prediction is then stored in the cache to optimize future queries. In the following sections, we provide a detailed explanation of each component.

\subsection{Preprocessing}
\label{subsec:preprocessing}
 As illustrated in the first box of Figure~\ref{fig:overviewmodel}, the \emph{preprocessing component} transforms raw log messages to log sequences via two primary processes, namely parsing and partitioning. Given a raw log message (e.g., ``\texttt{12:03 INFO Sent Block 12}''), we leverage a log parser (e.g., Drain~\cite{drain}) to identify the static parts (e.g., ``\texttt{Sent Block}'') and dynamic parts (e.g., ``\texttt{12}'') and replace the latter with the symbol \texttt{"<*>"}. 
 
The parsing process identifies unique log templates and assigns the same identifier to all their occurrences, enabling the \textit{cache-empowered inference} component to track processed log sequences and the \textit{context-enriched prompting} component to incorporate relevant template information during prompt construction. 
 The partitioning process aggregates log templates into log sequences based on their session IDs or a fixed-size window (as described in \S~\ref{sec:background-logs}).
\subsection{Cache-empowered Inference}
\label{subsec:cache}

\method maintains a cache $C$ as illustrated at the bottom of Figure~\ref{fig:overviewmodel}. $C$ stores previously seen log sequences along with their predicted labels. Given a log sequence $ls_i$ under detection, \method first queries cache $C$ for a matching entry. If an identical log sequence is found, the corresponding label $l_i$ is retrieved and used directly, bypassing the need for additional computation by RAG and ensemble learning. Conversely, if no match is found in $C$, \method performs context-enriched prompting (\S~\ref{subsec:rag}) and leverages ensemble learning (\S~\ref{subsec:fsel}) to predict the label for $ls_i$ and subsequently adds the new log sequence and its label to the cache.  To reduce storage overhead and remain independent of template length, each log sequence is represented as an ordered sequence of log template IDs rather than raw log messages.

The maintenance of $C$ involves four core functions, namely \emph{query}, \emph{add}, \emph{update}, and \emph{delete}. The \emph{query} function compares the input log sequence against the entries in $C$ and returns the identical entry along with its associated label if a match is found. The \emph{add} function inserts a new log sequence and its predicted label into $C$ if it is not present.  The \emph{delete} function removes a specific entry from $C$, allowing \method to manage cache size or discard outdated information. The \emph{update} function modifies an existing entry's label, incorporating human corrections to improve future predictions. By leveraging the delete and update functions, the cache mechanism enables flexibility to adjust to memory constraints and future updates of the sequences' labels, respectively.

\subsection{Context-enriched Prompting}
\label{subsec:rag}
\begin{figure*}[hbt]
\centering
\includegraphics[width=\columnwidth]{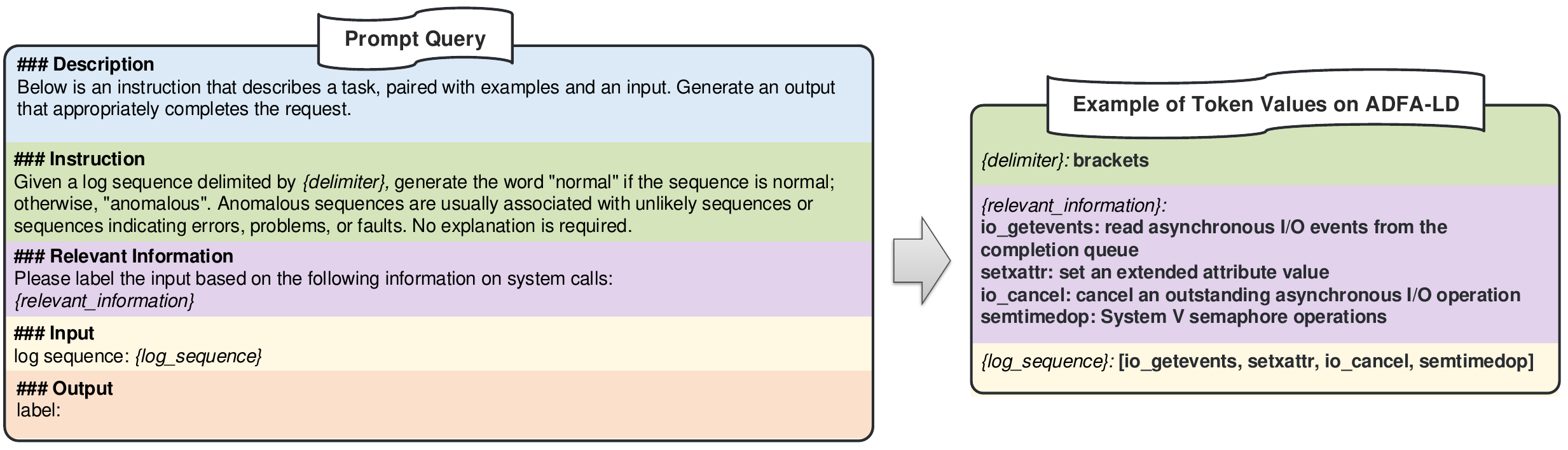}
 \captionsetup[figure]{skip=0pt}
     \captionof{figure}{\method's Prompt Design for LLM Fine-tuning and Inference. }\label{fig:template_design}
\end{figure*}
This component processes the log sequences that are not stored in the cache, preparing them for the LLMs used in \method. Log sequences are inherently challenging for LLMs to understand because these models are primarily designed for natural language processing (NLP) tasks and are better suited to processing and reasoning over textual data. To bridge this gap, we place log sequences with their log templates into semi-structured prompts that resemble natural language, enabling LLMs to leverage their NLP capabilities effectively. The RAG component inside the prompt aims to integrate relevant external information with log sequences acquired from preprocessing, formulating semi-structured prompts. 

To devise an effective prompt structure for \task, we adopt the tactics reported by~\citet{Winteringham2024AIAssisted}. Concretely, as illustrated in Figure~\ref{fig:template_design}, each prompt comprises five parts, namely, description, instruction, relevant information, input, and output. Notably, the retrieved context is included only when pertinent information is available. We provide the details of each part next.\\
\noindent \textbf{Description.} This part sets the overall context of the task for the LLM. It provides a high-level overview of what the model is expected to accomplish. For instance, as shown in Figure \ref{fig:template_design}, the description explains that the LLM should generate an output that appropriately completes the task request. While this part does not specify the input or output format explicitly, it prepares the LLM by providing a concise summary of the task objective.\\
\noindent \textbf{Instruction.} The instruction part formally introduces the task by describing its goal, input format, and output expectations. It specifies how the LLM should process the input and produce the desired label. After experimenting with multiple instruction formats, we present the most effective formulation in Figure \ref{fig:template_design}. In this part, we describe the delimiter for log sequences (i.e., ``brackets'') and add ``No explanation is required'', instructing the LLM to output only the label.\\
\noindent \textbf{Relevant Information (Optional).} This part includes additional information related to the log sequence, which enhances the LLM's ability to interpret the input. When available, contextual data is retrieved and presented in this part to provide background information. This information can include descriptions of specific log templates, system calls, or other contextual elements relevant to the log sequence under analysis. For example, the right-hand side of Figure~\ref{fig:template_design} includes system call descriptions, such as \texttt{setxattr} or \texttt{semtimedop}, which help clarify the function and purpose of the operations within the log sequence. Including such information allows the LLM to better understand relationships between the components of the log sequence, improving its ability to generate accurate predictions. However, if no relevant external context is available, this part of the prompt is omitted. \\
\noindent \textbf{Input.} 
This part presents the log sequence to be analyzed in the format \textit{``log sequence: \{log\_sequence\}''}, where \textit{\{log\_sequence\}} is a placeholder dynamically replaced with different log sequences during fine-tuning and inference. For example, the right-hand side of Figure~\ref{fig:template_design}  shows the replacement of the placeholder with the log sequence \textit{``[io\_getevents, setxattr, io\_cancel, semtimedop]''} from the \unstableadfa dataset. \\
\noindent \textbf{Output.} The output section guides an LLM in predicting the label of the input log sequence. It provides a formatted prompt that ends with ``label:'', prompting the LLM to generate the next token as either ``normal'' or ``anomalous'', based on its analysis of the log sequence.

\subsection{Ensemble Learning}
\label{subsec:fsel}
The goal of this component is to maximize the utility of limited labeled data by integrating different base models, each offering a unique perspective on performing \task effectively. To train the base models, we leverage the stable logs collected from software systems before undergoing the software or environment evolution.  A salient feature of \method is employing both ML- and LLM-based models as base models as introduced in \S~\ref{sec:intro}. Due to LLMs' pretraining on diverse corpora, they can be effectively fine-tuned with only limited data. Consequently, we sample only a subset from the stable logs to create the training dataset. Using this training dataset, \method fine-tunes an $\mathit{LLM}$ and fits $m$ ML models $\{\mathit{ML}_0, \dots, \mathit{ML}_{m-1}\}$.

For a given LLM $\mathit{LLM}$, \method adopts (see Equation~\ref{eq:llmft}) API-based fine-tuning for closed-source LLMs (e.g., GPT 4o)  and LoRa for open-source LLMs (e.g., Llama and Mistral); details about API-based fine-tuning and LoRa are provided in Section~\ref{sec:background}.  We denote the fine-tuned LLM as $\mathit{LLM}^f$. 
\begin{equation}
    \label{eq:llmft}
    \mathit{LLM}^f =
      \begin{cases}
      \mathit{API}\text{-}\mathit{based}\_FT(\mathit{LLM}, \mathit{data}=S_{\mathit{train}}, \mathit{label}=L) & \text{$\mathit{LLM}\in$ Closed-source LLMs} \\
      \mathit{LoRa}(\mathit{LLM}, \mathit{data}=S_{\mathit{train}}, \mathit{label}=L) & \mathit{LLM}\in\text{Open Source LLMs} 
      \end{cases}
\end{equation}

For a given ML model $\mathit{ML}$, gradient descent optimization is applied for neural network-based models, while model-specific fitting methods are used for non-parametric models such as DT (Equation \ref{eq:mlft}). Note that K-Nearest Neighbors (KNN) does not involve a traditional training or fitting process but instead relies on distance-based comparison during inference.  The resulting learned ML model is denoted as $\mathit{ML}^f$.
\begin{equation}
    \label{eq:mlft}
    \mathit{ML}^f =
      \begin{cases}
      \mathit{gradient\_descent}(\mathit{ML}, \mathit{data}=S_{\mathit{train}}, \mathit{label}=L) & \text{$\mathit{ML}\in$ Neural Networks} \\
      \mathit{fit\_dt}(\mathit{ML}, \mathit{data}=S_{\mathit{train}}, \mathit{label}=L) & \mathit{ML}=\text{ DT}\\
      \mathit{distance\_based\_comparison} & \mathit{ML}=\text{ KNN}
      \end{cases}
\end{equation}

After training individual models, the next step is to combine their outputs using \emph{majority voting}, a commonly used, simple yet effective ensemble learning technique in the literature ~\cite{ensemble1,ensemble2,ensemble3}. Formally, let $\mathcal{M}=\{M_0, M_2,\dots, M_{N-1}\}$ be the set of $N$ learned base models, with $\mathcal{M}_i\in \mathcal{M}$ representing either a learned LLM $\mathit{LLM}^f$ or an ML-based method $\mathit{ML}^f$. For a given log sequence $\mathit{ls}_i$, each base model  $\mathcal{M}_i$ predicts the label $y_i$ of $x$, equals $1$ if anomalous, and $0$ if normal. The final label is determined by 
a majority voting function $\mathit{MV}(\cdot)$  among all base models, as shown in Equation~\ref{eq:majority_vote}.
\begin{equation}
    \label{eq:majority_vote}
    \phi_i = \mathit{MV}(\mathit{ls}_i) =
    \begin{cases}
        1, & \text{if } \sum_{i=0}^{N-1} y_i > \frac{N}{2} \\
        0, & \text{otherwise}
    \end{cases}
\end{equation}
Here, $\phi_i$ represents the final prediction for $\mathit{ls}_i$. In case of a tie (when exactly half of the votes are normal), the sequence is classified as \textit{normal}, following our assumption that anomalies are rare.

\section{Experimental Design}\label{sec:studydesign}

\subsection{Research Questions}
\label{sec:rqs}

We investigate the following research questions:
\begin{compactenum}[\bfseries {RQ}1]
    \item (\textbf{effectiveness}) How effective is \method for \task compared to the baselines? 
    \begin{compactenum}[\bfseries{RQ1.}1]
        \item Can \method trained on limited labeled data achieve comparable effectiveness to baselines trained on full datasets?
        \item
        What impact does the level of log instability have on \method and the baselines?
    \end{compactenum}
    \item (\textbf{data efficiency}) How does the amount of labeled training data impact \method's effectiveness, and can it maintain robust effectiveness under varying degrees of data scarcity?
    \item (\textbf{time and memory efficiency})
    What is the performance of \method in terms of time efficiency during training and inference, and how much memory overhead does the cache incur during inference?
    \item (\textbf{configuration impact}) How does the performance of \method vary under  different configurations, including ablations of base models, RAG, and the cache, as well as alternative LLM choices?
\end{compactenum}
RQ1 investigates the overall effectiveness of \method for the \task, comprising two sub-RQs. With RQ1.1, we aim to demonstrate the strengths of \method when only limited labeled data is available. Specifically, we compare baselines trained with full datasets against \method trained with much smaller subsets.  With RQ1.2, we aim to highlight the distinct advantage of \method in handling gradually increasing instabilities in \task. To this end, we assess the effectiveness of \method and baselines on \slad and under the influence of varying ratios of instability in both log templates and sequence levels. RQ2 focuses on data efficiency by training \method and baselines on progressively larger subsets of \unstableadfa (e.g., containing 50, 500, 1000, 1500, and 2000 training samples). Due to computational constraints, we cannot perform data efficiency analysis on all datasets. Hence, we prioritize our most challenging dataset \unstableadfa for this analysis. RQ3 investigates the time efficiency of \method during training and inference. While employing LLM-based approaches (e.g., \method) may enhance effectiveness, it often comes at the cost of increased training and inference time. This question aims to evaluate the trade-offs between effectiveness and time efficiency, providing practical insights for those considering the use of LLMs in similar tasks. Additionally, RQ3 examines the memory efficiency of \method’s cache mechanism during inference, demonstrating its scalability in resource-constrained environments. Lastly, RQ4 involves exploring the impact of different configurations of \method. 
Specifically, we assess how the exclusion of the cache $C$, the exclusion of RAG in context-enriched prompting, and the choices of base models (e.g., removing some base models from the ensemble or replacing Mistral with other LLMs), affect the overall effectiveness or efficiency of \method.

\subsection{Experiment Setup}\label{sec:experimentalsetup}
\begin{table*}[htbp]
\caption{Overview of Datasets}
\centering
\resizebox{0.9\linewidth}{!}{ 
\begin{threeparttable}[htbp]
\resizebox{\linewidth}{!}{
\begin{tabular}{c@{\hspace{0.5\tabcolsep}}c@{\hspace{0.8\tabcolsep}}c@{\hspace{0.7\tabcolsep}}c@{\hspace{0.5\tabcolsep}}c@{\hspace{0.5\tabcolsep}}c@{\hspace{0.5\tabcolsep}}c@{\hspace{0.5\tabcolsep}}c@{\hspace{0.5\tabcolsep}}c@{\hspace{0.5\tabcolsep}}c}
\toprule
\multirow{2}{*}{\textbf{Name}} & \multirow{2}{*}{\textbf{Sys}} 
& \multirow{2}{*}{\makecell{\textbf{\#Log}\\\textbf{Messages}}} & \multirow{2}{*}{\textbf{\makecell{\#Anomalous\\Messages}}} & \multirow{2}{*}{\textbf{\makecell{\#Sessions}}} & \multirow{2}{*}{\textbf{\makecell{\#Log \\Templates}}} & \multicolumn{3}{c}{\textbf{Session Length}}\\
\cline{7-9}
&&&&&&\textbf{avg}&\textbf{min}&\textbf{max}\\
\midrule
ADFA & Linux & 2,747,550 &  317,388 (11.5\%) & 5,951 & 175 & 461.69 & 75 & 4,474 \\
\midrule
\multirow{2}{*}{LOGEVOL} & Hadoop 2 & 2,120,739 & 35,072 (1.6\%)& 333,699 & 319 & 6.35 & 1 & 1,963\\ 
& Hadoop 3 & 2,050,488 & 30,309 (1.4\%) & 343,013 & 313 & 5.97 & 1 & 1,818 \\
& Spark 2 & 931,960 & 1,702 (0.1\%)& 13,892 & 130 & 67.08 & 1 & 1125\\
& Spark 3 & 1,600,273 & 2,430 (0.1\%)& 21,232 & 134 & 75.37 & 1 & 1977\\
\midrule
HDFS & Hadoop & 11,110,850 & 284,818 (2.9\%)& 575,061 & 48 & 19.32 & 2 & 30\\
\bottomrule
\end{tabular}
}
\end{threeparttable}

}
\label{tab:datasets}
\end{table*}

\subsubsection{Datasets}\label{sec:datasets} We configured four datasets for \task from three public datasets, namely ADFA~\cite{ADFA}, LOGEVOL~\cite{evlog}, and HDFS~\cite{HDFS}. We exclude other popular benchmarks (e.g., BGL~\cite{bgl-spirit}, Thunderbird~\cite{bgl-spirit}, and Spirit~\cite{bgl-spirit}) because they contain only stable logs, and manually injecting instability is not feasible due to the lack of information about their annotation strategies.  

Table~\ref{tab:datasets} presents relevant statistics for these three datasets; column ``Sys'' indicates the system from which the logs were collected. ``\#Log Messages'', ``\#Anomalous Messages'', ``\#Sessions'', and ``\#Log Templates'' indicate the number of log messages, anomalous log messages, sessions, and unique log templates in each dataset, respectively. Column ``Session Length'' indicates the average, minimum, and maximum number of log messages in each session. We elaborate on each dataset next. 

\paragraph{ADFA}\label{sec:adfa} ~\citet{ADFA} created the Australian Defense Force Academy Linux Dataset by collecting Linux server operation logs and applying contemporary web attacks. ADFA comprises \num{2747550} log messages, i.e., Linux system calls in this context,  of which \num{317388} are anomalous (\SI{11.5}{\percent}).  The attacks applied to the system include the exploitation of a TIKI WIKI vulnerability using a Java-based Meterpreter (``java''), password brute-forcing with the Hydra tool (``hydra''), deploying a Linux Meterpreter payload via a poisoned executable (``meter''), leveraging a remote file inclusion vulnerability to deploy a C100 webshell (``web'') and creating privilege escalation by adding a superuser account with a poisoned executable (``adduser'').

\paragraph{LOGEVOL}\label{sec:logevol} ~\citet{evlog} introduced the LOGEVOL dataset captured from the real-world operations of Hadoop~2, Hadoop~3, Spark~2 and Spark~3 systems\footnote{Hadoop versions 2.10.2 and 3.3.3 are referred to as Hadoop~2 and~3, respectively, and Spark versions 2.4.0 and 3.0.3 are denoted as Spark~2 and Spark~3, respectively.}. All datasets were generated using HiBench~\cite{hibench} during the operation of 22 cloud computing tasks, such as sorting and classification~\cite{he2020loghub, lin2016log}. To capture real-world anomaly scenarios into logs, they injected 18 fault types into the system, including network fault, process suspension, process killing, and resource occupation.  The Hadoop~2 dataset consists of \num{2120739} log messages (including \SI{1.6}{\percent} anomalous) while the Hadoop~3 dataset is made up by \num{2050488} log messages 
(including \SI{1.4}{\percent} anomalous). Notably, 104 out of 303 (\SI{33.22}{\percent}) log templates from the Hadoop~3 dataset are novel and absent from the Hadoop~2 dataset, reflecting its instability and log template evolution. The Spark~2 dataset involves \num{931960} log messages, and the Spark~3 dataset is made up of \num{1600273} log messages. Compared to the Hadoop~2 and Hadoop~3 datasets, the proportion of anomalous logs in the Spark~2 and Spark~3 datasets is significantly lower, with both datasets having an anomaly rate of only \SI{0.1}{\percent}.

\paragraph{HDFS}\label{sec:hdfs} Hadoop Distributed File System (HDFS) logs~\cite{HDFS} were produced by running  MapReduce jobs on Amazon EC2 nodes, consisting of \num{11197954} log messages, of which \num{284818} (\SI{2.9}{\percent}) are anomalous.  The average number of log messages in a sequence is \num{19.32}. The total number of unique log templates is \num{48}. 
This dataset includes 11 types of anomalies, such as the deletion of a block that no longer exists or receiving a block that does not belong to any file. For a comprehensive description of the anomalies, we refer readers to the original paper~\cite{HDFS}.

\subsubsection{\task and \slad Configuration}
\label{sec:datasset-config}

Our experiments involve the evaluation of \method on both the \task and \slad, with \task being the primary focus of this paper and \slad serving as a baseline in RQ1 and RQ2. As mentioned in Section~\ref{sec:background}, \task is characterized by the disparity between the training and testing datasets, whereas \slad involves training and testing data drawn from the same distribution. Each dataset described in \S~\ref{sec:datasets} is configured to be used for both \slad and \task, consisting of a training dataset $\mathcal{D}^{\mathit{train}}$ and a testing dataset  $\mathcal{D}^{test}$. Additionally, a small, curated subset $\tilde{\mathcal{D}}^{train}$  is sampled from each full training dataset for the training of \method.  We provide details of the \slad and \task configurations on each dataset next, followed by configurations of $\tilde{\mathcal{D}}^{train}$ for \method. 

\noindent\textbf{\task Configuration.} As discussed in Section \ref{sec:background-anomalydetection}, unstable logs result from system or environmental evolution. To simulate these scenarios, we configured the unstable LOGEVOL dataset \unstablelogevol for system evolution and the unstable ADFA dataset \unstableadfa for environmental evolution. To further investigate the influence of different levels of instability, we include two synthesized datasets in our experiments, namely \synhdfs and \synlogevol. These datasets are created by injecting different levels of instability into the \hdfs and \logevol datasets, respectively. 
Table~\ref{tab:unstable-datasets} summarizes the \task configuration for each dataset and presents their statistics, including  the full training dataset size ($\mathcal{D}_\#^{train}$), \method's training dataset size ($\tilde{\mathcal{D}}_\#^{train}$), and the testing dataset size ($\mathcal{D}_\#^{test}$). We also report the duplication ratio, which quantifies data leakage, i.e., log sequences appearing in both training and testing datasets. As discussed in \S~\ref{sec:intro}, data leakage allows anomaly detectors to memorize the training data, resulting in artificially inflated effectiveness on the testing data. Hence, we addressed the data leakage issues identified by~\citet{boxi} through de-duplication in each configuration. Specifically, we removed log sequences from the testing dataset if they were already included in the training dataset. 

\begin{table*}[hbt!]
    \centering
    \caption{\task Configurations}
    \resizebox{\textwidth}{!}{
     \begin{tabular}{ccccccc}
\toprule
\multirow{2}{*}{\textbf{Dataset} } & \multicolumn{2}{c}{\textbf{Configuration}} & \multirow{2}{*}{\textbf{\makecell{Duplication\\Ratio}}} & \multicolumn{3}{c}{\textbf{\#Log Sequences}} \\
\cmidrule{2-3}
\cmidrule{5-7}
& \textbf{train} & \textbf{test} & & \textbf{$\mathcal{D}^{train}_\#$} & \textbf{$\tilde{\mathcal{D}}^{train}_\#$} & \textbf{$\mathcal{D}^{test}_\#$} \\
\midrule
\multirow{5}{*}{\unstableadfa}& ADFA $_{\mathit{w/o\ java}}$ & ADFA$_{\mathit{w/\ java}}$ & 0.32 &4786 & 1000 & 1165 \\ 
& ADFA$_{\mathit{w/o\ hydraSSH}}$ & ADFA$_{\mathit{w/\ hydraSSH}}$ & 0.31 & 4734 & 1000 & 1217 \\
& ADFA$_{\mathit{w/o\ hydraFTP}}$ & ADFA$_{\mathit{w/\ hydraFTP}}$ & 0.31 & 4748 & 1000 & 1203\\
& ADFA$_{\mathit{w/o\ meter}}$ & ADFA$_{\mathit{w/\ meter}}$ & 0.34 & 4835 & 1000 & 1116\\
& ADFA$_{\mathit{w/o\ web}}$ & ADFA$_{\mathit{w/\ web}}$ & 0.33 & 4792 & 1000 & 1159 \\
& ADFA$_{\mathit{w/o\ adduser}}$ & ADFA$_{\mathit{w/\ adduser}}$ & 0.33 & 4819 & 1000 & 1132\\
\midrule
\multirow{2}{*}{\unstablelogevol} & Hadoop 2  & Hadoop 3 & 0.84 &302312 & 8558& 34495 \\
\cmidrule{2-7}
& Spark 2 & Spark 3 & 0.50 & 11114 & 1134 &  4246 \\
\midrule
\multirow{12}{*}{\synlogevol} & \multirow{12}{*}{Spark 2} & Spark~2$_{5\%\_sequence}$ & 0.6 & \multirow{6}{*}{11114} & \multirow{6}{*}{1134} & \multirow{6}{*}{2778} \\
&  & Spark~2$_{10\%\_sequence}$ & 0.55 & & &  \\
&  & Spark~2$_{15\%\_sequence}$ & 0.50 & & &  \\
&  & Spark~2$_{20\%\_sequence}$ & 0.44 & & &  \\
&  & Spark~2$_{25\%\_sequence}$ & 0.37 & & &  \\
&  & Spark~2$_{30\%\_sequence}$ & 0.32 & & &  \\
\cmidrule{3-7}
&  & Spark~2$_{5\%\_template}$ & 0.54 & \multirow{6}{*}{11114} & \multirow{6}{*}{1134} & \multirow{6}{*}{2778}\\
&  & Spark~2$_{10\%\_template}$ & 0.45 & & &  \\
&  & Spark~2$_{15\%\_template}$ & 0.36 & & &  \\
&  & Spark~2$_{20\%\_template}$ & 0.28 & & &  \\
&  & Spark~2$_{25\%\_template}$ & 0.22 & & &  \\
&  & Spark~2$_{30\%\_template}$ & 0.18 & & &  \\
\midrule 
\multirow{4}{*}{\synhdfs } & \multirow{4}{*}{HDFS} & SynHDFS$_{5\%\_sequence}$ & 0.93 & \multirow{4}{*}{460048} & \multirow{4}{*}{5772} & \multirow{4}{*}{51000} \\
&  & SynHDFS$_{10\%\_sequence}$ & 0.88 & & & \\
&  & SynHDFS$_{20\%\_sequence}$ & 0.78 & & & \\
&  & SynHDFS$_{30\%\_sequence}$ & 0.69 & & & \\
\bottomrule
\end{tabular}

      }
    \label{tab:unstable-datasets}
\end{table*}

\paragraph{\unstableadfa} We derived six \task configurations by splitting ADFA based on attack types. Specifically, for each configuration, five out of six attack types are used for training (e.g.,  ADFA$_{\mathit{w/o\ java}}$, in column ``train'' in Table~\ref{tab:unstable-datasets}, represents training data containing all attack types except  the Java-based Meterpreter attack) and the remaining one for testing (e.g., ADFA$_{\mathit{w/\ java}}$, in the ``test'' column of Table~\ref{tab:unstable-datasets}, represents a testing dataset with only  the Java-based Meterpreter attack), simulating external changes in real-world scenarios where novel attack types emerge during operation. 
Consequently, we obtain six \task configurations for \unstableadfa involving  ADFA$_{\mathit{w/o\ java}}\rightarrow$ADFA$_{\mathit{w/\ java}}$, ADFA$_{\mathit{w/o\ hydraSSH}}\rightarrow$ADFA$_{\mathit{w/\ hydraSSH}}$, \newline ADFA$_{\mathit{w/o\ hydraFTP}}\rightarrow$ADFA$_{\mathit{w/\ hydraFTP}}$, ADFA$_{\mathit{w/o\ meter}}\rightarrow$ADFA$_{\mathit{w/\ meter}}$, ADFA$_{\mathit{w/o\ web}}\rightarrow$ADFA$_{\mathit{w/\ web}}$,  ADFA$_{\mathit{w/o\ adduser}}\rightarrow$ADFA$_{\mathit{w/\ adduser}}$, 
denoted as ``java'', ``hydraSSH'', ``hydraFTP'', ``meter'', ``web'', and ``adduser'' hereafter for brevity, respectively. 
As reported in Table \ref{tab:unstable-datasets}, the duplication ratio ranges from 0.31 to 0.34 in different training and testing dataset pairs, indicating that, without de-duplication, approximately \SI{31}{\percent} to \SI{34}{\percent} log sequences in the testing datasets are already included in the training datasets.

\paragraph{\unstablelogevol} The LOGEVOL dataset naturally captures software evolution, namely the transition from Hadoop~2 to Hadoop~3, as well as from Spark~2 to Spark~3. These transitions result in internal changes at both template and sequence levels (defined in \S~\ref{sec:background-logs}).  Hence, we use the Hadoop~2 and Spark~2 datasets for training and, correspondingly, the Hadoop~3 and Spark~3 datasets for testing. The duplication ratios for LOGEVOL Hadoop and LOGEVOL-Spark, as shown in Table~\ref{tab:unstable-datasets}, are 0.84 and 0.5, respectively; these values indicate that, without de-duplication,  84\% of Hadoop and 50\%  of Spark log sequences in the testing dataset are already included in the training dataset. 

\begin{figure*}[hbt!]
  \centering
  \begin{minipage}{0.4\textwidth}
    \centering
    \resizebox{\textwidth}{!}{
      \begin{tikzpicture}
[
    node distance = 5mm and 7mm,
    module/.style={%
        draw, rounded corners,
        minimum width=#1,
        minimum height=5mm,
        font=\linespread{1}\selectfont
        },
    module/.default=2cm,
    >=LaTeX,
 disc/.style = {shape=cylinder, draw, shape aspect=0.27,
                shape border rotate=90,
                text width=20mm, align=center, font=\linespread{1}\selectfont},
  mdl/.style = {shape=ellipse, aspect=3, draw},
  alg/.style = {draw, align=center, font=\linespread{1}\selectfont},
  alg2/.style = {draw, align=center, 
  minimum width=#1,
        minimum height=5mm,
        font=\linespread{1}\selectfont},
        alg2/.default=2cm
                    ]

    \node [alg2=2.6 cm, rounded corners, fill = gray!30!white, draw = gray!30!white] (s1) {Original Log Template};
    \node [alg2=1.5 cm,  below right= 2 mm and -26 mm of s1] (t1) {received block * from * dest: * };

    \node [alg2=2.6 cm, rounded corners, fill = gray!30!white, draw = gray!30!white, below right= 12mm and -24mm of s1] (s2) {Removing \small one word};
    \node [alg2=1.5 cm,  below right= 12 mm and -39 mm of t1] (t12) {received \sout{block} * from * dest: * };

     \node [alg2=2 cm, rounded corners, fill = gray!30!white, draw = gray!30!white, below right = 10 mm and -30mm of s2] (s3) {Adding \small one word};
    \node [alg2=1.5 cm,  below right = 27 mm and -39 mm of t1] (t13) {received block * from \textbf{IP} * dest: * };

     \node [alg2=3 cm, rounded corners, fill = gray!30!white, draw = gray!30!white, below right = 10 mm and -26mm of s3] (s4) {Replacing \small one word by another word};
    \node [alg2=1.5 cm, below right = 43 mm and -39 mm of t1] (t14) {\sout{received} \textbf{got} block * from * dest: * };

    \draw[-] (-1.2,-0.26)--(-1.2, -4.75);
    \draw[->] (-1.2, -4.75)--(-0.6, -4.75);
    \draw[->] (-1.2, -3.24)--(-0.6, -3.24);
    \draw[->] (-1.2, -1.7)--(-0.6, -1.7);

\end{tikzpicture}
    }
    \caption{Examples of Template-level Changes}
    \label{fig:synthetic-template}
  \end{minipage}
  \hfill
  \begin{minipage}{0.594\textwidth}
    \centering
    \resizebox{\textwidth}{!}{
      \begin{tikzpicture}
[
    node distance = 5mm and 7mm,
    module/.style={%
        draw, rounded corners,
        minimum width=#1,
        minimum height=5mm,
        font=\linespread{1}\selectfont
        },
    module/.default=2cm,
    >=LaTeX,
 disc/.style = {shape=cylinder, draw, shape aspect=0.27,
                shape border rotate=90,
                text width=20mm, align=center, font=\linespread{1}\selectfont},
  mdl/.style = {shape=ellipse, aspect=3, draw},
  alg/.style = {draw, align=center, font=\linespread{1}\selectfont},
  alg2/.style = {draw, align=center, 
  minimum width=#1,
        minimum height=5mm,
        font=\linespread{1}\selectfont},
        alg2/.default=2cm
                    ]

    \node [alg2=2.6 cm, rounded corners, fill = gray!30!white, draw = gray!30!white] (s1) {Original Log Sequence};
    \node [alg2=1.5 cm,  below= 2 mm of s1] (t1) {template 1};
    \node [alg2=1.5 cm,  right= 3 mm of t1] (t2) {template 2};
    \node [alg2=1.5 cm,  right= 3 mm of t2] (t3) {template 3};
    \node [alg2=1.5 cm,  right= 3 mm of t3] (t4) {template 4};

    \node [alg2=2.6 cm, rounded corners, fill = gray!30!white, draw = gray!30!white, below right= 12mm and -25mm of s1] (s2) {Removing \small a template};
    \node [alg2=1.5 cm,  below= 12 mm of t1] (t12) {template 1};
    \node [alg2=1.5 cm,  right= 3 mm of t12, draw = lightgray, text = lightgray] (t22) {template 2};
    \node [alg2=1.5 cm,  right= 3 mm of t22] (t32) {template 3};
    \node [alg2=1.5 cm,  right= 3 mm of t32] (t42) {template 4};

     \node [alg2=2.6 cm, rounded corners, fill = gray!30!white, draw = gray!30!white, below right = 10 mm and -31mm of s2] (s3) {Duplicating\small a template};
    \node [alg2=1.5 cm,  below= 10 mm of t12] (t13) {template 1};
    \node [alg2=1.5 cm, right= 3 mm of t13] (t23) {template 2};
    \node [alg2=1.5 cm, right= 3 mm of t23, fill = lightgray!30!white] (t33) {template 2};
    \node [alg2=1.5 cm, right= 3 mm of t33] (t43) {template 3};
    \node [alg2=1.5 cm, right= 3 mm of t43] (t53) {template 4};
    \node[fit=(t23)(t33), rounded corners, draw, inner
    sep=0.8mm, draw = lightgray, dashed, thick] (fit1) {};

    \node [alg2=2.6 cm, rounded corners, fill = gray!30!white, draw = gray!30!white, below right = 10 mm and -33mm of s3] (s4) {Shuffling a \small small subsequence};
    \node [alg2=1.5 cm, below= 10 mm of t13] (t14) {template 1};
    \node [alg2=1.5 cm, right= 3 mm of t14, fill = lightgray!30!white] (t24) {template 3};
    \node [alg2=1.5 cm, right= 3 mm of t24, fill = lightgray!30!white] (t34) {template 4};
    \node [alg2=1.5 cm, right= 3 mm of t34, fill = lightgray!30!white] (t44) {template 2};
    \node[fit=(t24) (t34) (t44), rounded corners, draw, inner
    sep=0.8mm, draw = lightgray, dashed, thick] (fit2) {};

    \draw[->] (t1)--(t2);
    \draw[->] (t2)--(t3);
    \draw[->] (t3)--(t4);

    \draw[->] (t12)--(t32);
    \draw[->] (t32)--(t42);

    \draw[->] (t13)--(t23);
    \draw[->] (t23)--(t33);
    \draw[->] (t33)--(t43);
    \draw[->] (t43)--(t53);

    \draw[->] (t14)--(t24);
    \draw[->] (t24)--(t34);
    \draw[->] (t34)--(t44);

    \draw[-] (-1.2,-0.26)--(-1.2, -4.75);
    \draw[->] (-1.2, -4.75)--(-0.7, -4.75);
    \draw[->] (-1.2, -3.24)--(-0.7, -3.24);
    \draw[->] (-1.2, -1.7)--(-0.7, -1.7);

\end{tikzpicture}
    }
    \caption{Examples of Sequence-level Changes}
    \label{fig:synthetic}
  \end{minipage}
\end{figure*}

\paragraph{\synlogevol}  The \task configurations in this dataset aim to simulate different levels of instability by applying internal changes of varying percentages to the \logevol dataset. \citet{evlog} injected log template/sequence-level changes into the \logevol Spark~2 dataset, with varying injection ratio of  \SI{5}{\percent}, \SI{10}{\percent}, \SI{15}{\percent}, \SI{20}{\percent}, \SI{25}{\percent}, and \SI{30}{\percent}. Figure \ref{fig:synthetic} demonstrates the three types of log sequence-level changes injected, firstly introduced by~\citet{LogRobust}, involving removing or duplicating a log template and shuffling a small subsequence. As shown in Figure \ref{fig:synthetic-template}, we introduce three types of log template-level changes, including adding, removing, or replacing a word in a log template. 
\citet{evlog} injected changes in the sequences in a way that sequence labels do not flip. 
The duplication ratio decreases from \num{0.6} to \num{0.18} as the testing set becomes more unstable.

\paragraph{\synhdfs} Similar to \synlogevol, we created four \task configurations for the \hdfs dataset, namely SynHDFS$_{5\%}$, SynHDFS$_{10\%}$, SynHDFS$_{20\%}$, and SynHDFS$_{30\%}$ by changing 5\%, 10\%, 20\%, and 30\% of log sequences in the HDFS dataset, respectively. The injection ratios are determined by following common practices in the literature~\cite{LogRobust}. Similar to \synlogevol, we are aware that such changes in log sequences can induce changes in their labels. We only apply sequence-level changes, excluding template-level changes due to the lack of implementation details reported by previous studies~\cite{LogRobust, hitanomaly, swisslog}.
At the sequence-level, to obviate the need for re-labeling, we applied changes only to log templates that are less likely to flip the labels of the entire log sequence. These log templates are identified by a strategy proposed by \citet{Xu2010}, which combines building a decision tree and manual examination. To reduce the cost of manual examination, we sampled and applied changes to a subset of the HDFS dataset instead of the full dataset.  Concretely, we randomly selected 50,000 normal and 1,000 anomalous log sequences, following the study by ~\citet{LogRobust}, to keep the anomaly percentage (\SI{2}{\percent}) close to that of the original HDFS dataset. The duplication ratio decreases from 0.93 to 0.69 as the testing set becomes more unstable.

~\\
\noindent\textbf{\slad Configuration.} For the ADFA dataset, we drew the training and testing data from the full dataset, containing all six types of anomalies. For \logevol Hadoop~2, Spark~2, and \hdfs, we adopt the same training datasets as in their \task configurations, whereas the testing datasets differ. While \task employs unstable testing data, \slad uses stable testing data collected from the same system as the training data, specifically from Hadoop~2, Spark~2, and HDFS operations, respectively. Notably, the testing dataset for \hdfs \slad configuration is the same as the one used in its \task configuration, a subset sampled from the original testing dataset, but without instability injection. This ensures consistency and a fair comparison between the \task and \slad configuration of \hdfs. Also, we did not configure LOGEVOL Hadoop~3 and Spark~3  for \slad as \task because the evolution information from Hadoop~3 and Spark~3 to other versions was not available. 

~\\
\noindent \textbf{Training Dataset Configuration for \method.} In RQ1, we compare \method and the baselines with their respective optimal settings. Baselines are trained on the full training datasets $\mathcal{D}^{\mathit{train}}$, following implementations in their original papers, whereas \method is trained on small subsets $\tilde{\mathcal{D}}^{train}$ randomly sampled from $\mathcal{D}^{\mathit{train}}$.  As reported in the second-to-last column ("$\tilde{\mathcal{D}}_\#^{train}$") of Table \ref{tab:unstable-datasets}, their data sizes are determined empirically for each dataset to achieve the optimal performance of \method. For small datasets with low duplication ratios such as \unstableadfa, we randomly selected \num{1000} log sequences from their full training dataset.  For larger datasets with high duplication ratios such as \unstablelogevol Hadoop, \unstablelogevol Spark, \synlogevol, and \synhdfs, unique anomalous log sequences are rare, accounting for only \SI{0.2}{\percent} to \SI{2}{\percent} of the full datasets, respectively. To maximize the use of these rare anomalous log sequences, we included all of them in the fine-tuning datasets and sampled \SI{20}{\percent} unique normal log sequences from all unique normal log sequences, preventing excessive duplication.

\subsubsection{Baselines}\label{sec:baseliens}

\begin{table*}
    \centering
    \captionsetup[table]{skip=0pt}
     \captionof{table}{Overview of Baselines}\label{tab:baselines}
    \footnotesize{
    \centering
     
\begin{threeparttable}[htbp]
\centering

\centering
\begin{tabular}{c@{\hspace{0.5\tabcolsep}}c@{\hspace{0.5\tabcolsep}}c@{\hspace{0.5\tabcolsep}}c@{\hspace{0.5\tabcolsep}}c@{\hspace{1.2\tabcolsep}}c}

\toprule
\textbf{\makecell{Learning Method}} & \textbf{Approach}  & \textbf{Parser} & \textbf{\makecell{Log Representation}} & \textbf{ML Method} & \textbf{Base Model}\\
\midrule
\multirow{4}{*}{Unsupervised} &PCA & Yes & Template ID & Traditional ML & PCA \\

&LogCluster & Yes & Template ID & Traditional ML & Clustering \\

&DeepLog & Yes & Template ID & Deep Learning & LSTM \\

&LogAnomaly & Yes & Template2Vec & Deep Learning & LSTM \\
\midrule
\makecell{Semi-supervised}&PLELog & Yes & \makecell{FastText and TF-IDF
}& Deep Learning & GRU \\
\midrule
\multirow{3}{*}{Supervised}&LogRobust & Yes & \makecell{FastText and TF-IDF} & Deep Learning & BiLSTM \\

&CNN & Yes & Logkey2vec & Deep Learning & CNN \\

&NeuralLog & No & BERT & Deep Learning & Transformer \\
&LightAD & Yes & \makecell{TemplateID} & \makecell{Traditional ML \& Deep Learning} & \makecell{KNN, DT, SLFN} \\
\bottomrule
\end{tabular}

\end{threeparttable}
     }
\end{table*}

We considered nine ML methods as baselines in this paper, including four unsupervised, one semi-supervised, and four supervised. Among these methods, LightAD~\cite{lightad} achieves the best performance on the \slad task. However, the leading approach for \task remains undetermined as different evaluation datasets are used in reported studies. Our choice of baselines is also determined by source code availability to ensure the reliability of the implementation. Consequently, we had to exclude models such as SwissLog~\cite{swisslog}, HitAnomaly~\cite{hitanomaly}, EvLog~\cite{evlog}, and LLMeLog~\cite{LLMeLog}. Our implementations are based on the code provided by~\citet{lightad}, ~\citet{le2022}, and~\citet{loglizer}. We have also not included LogPrompt~\cite{logprompt} in our evaluation since it relies on anomaly detection at the message level, ignoring sequential characteristics such as temporal dependencies, whereas our datasets are labeled at the sequence level. 

Table~\ref{tab:baselines} shows the main characteristics of the baselines; we provide a brief description in the following. 
\textit{Principal Component Analysis (PCA)}~\cite{Xu2010}, a dimensionality reduction method, converts logs into count vectors~\cite{countvector} and then uses the PCA algorithm to detect the label of log sequences by assigning them to either the normal or anomalous space.  In this paper, by PCA we refer to the PCA-based model introduced by~\citet{Xu2010} as an anomaly detector.
\textit{LogCluster}~\cite{lin2016log} clusters log sequences by computing the similarity of log representations to the centroid of normal logs.
\textit{DeepLog}~\cite{DeepLog} applies two layers of long short-term memories (LSTMs) in their network~\cite{LSTM} to predict the next event from a given log sequence and labels sequences as anomalous if the predicted log is different than the actual log template.
\textit{LogAnomaly}~\cite{LogAnomaly} has an architecture similar to DeepLog, but is further improved by adopting semantic embeddings for log templates and adding an attention layer between LSTM layers.
\textit{PLELog}~\cite{PLELog} is a semi-supervised strategy that uses normal data as well as a small subset of unlabeled data to train. First, it adopts a clustering method (HDBSCAN~\cite{hdbscan}) to probabilistically predict the labels of unlabeled data and then uses them to train an attention-based GRU~\cite{GRU} to detect anomalies. 

\textit{LogRobust}~\cite{LogRobust} uses a pre-trained word vectorizer (FastText~\cite{fasttextzip}) to extract semantic information from log templates and utilizes an attention-based BiLSTM model~\cite{BiLSTM} to detect anomalous log sequences.
\textit{CNN}~\cite{CNN} transforms an input log sequence into a trainable matrix and uses this matrix as input to train a Convolutional Neural Network~\cite{CNNdefinition,lecun2015deep} for log-based anomaly detection.
\textit{NeuralLog}~\cite{Neurallog} extracts the semantic meaning of raw log messages and represents them as semantic vectors, which are then used to detect anomalies through a transformer-based classification model~\cite{AttentionIA}. \textit{LightAD}~\cite{lightad} employes Bayesian method to select the most effective model from a heterogeneous pool of ML/DL algorithms---including KNN~\cite{KNN}, DT~\cite{Decisiontree}, and SLFN ~\cite{SLFN}---while simultaneously optimizing hyperparameters for the \slad task. To ensure fair comparisons, we adopted the same model pool and employed a small, held-out validation dataset to identify the optimal model for each dataset as instructed in~\citet{lightad}. The performance of the optimal model, evaluated on test data, serves as LightAD's reported effectiveness. 

\subsubsection{Evaluation Metrics and Statistical Testing}\label{sec:metrics} 
To provide a comprehensive evaluation, we assess \method in terms of effectiveness, data efficiency, and time efficiency. Further, we investigate the statistical significance of differences (from a point of view of effectiveness and time efficiency) on each dataset. 

~\\
\noindent \textbf{Effectiveness} To measure the effectiveness of \method, we use Precision, Recall, and F1-score as metrics. We consider TP (true positive) as the number of anomalies that are correctly detected by the model, FP (false positive) as the number of normal log sequences that are labeled as anomalous by the model, and FN (false negative) as the number of anomalous log sequences that the model fails to identify. Precision (P) is calculated by $\frac{\mathit{TP}}{\mathit{TP}+\mathit{FP}}$ as the percentage of true anomalies among all anomalies detected by the model. 
Recall (R) is the proportion of actual anomalies detected, computed by $\frac{\mathit{TP}}{\mathit{TP}+\mathit{FN}}$.
F1 score (F1) is the harmonic mean of Precision and Recall, i.e., $2*\frac{P*R}{P+R}$. 

~\\
\noindent \textbf{Data Efficiency} We define data efficiency to be the ability of a method to achieve accurate results while minimizing the use of labeled data for training.
Considering a training dataset with \datasize log sequences, we quantify the usage of labeled data using \uniquesize, which represents the number of unique log sequences. Each unique log sequence corresponds to a distinct pattern that requires annotation, meaning that a higher \uniquesize reflects greater labeling effort. 

A data-efficient method, such as \method, requires only a small subset of the full dataset for training, reducing the overall usage of labeled data. To compare data efficiency across different methods, we introduce the relative metric \uniquepercent as in Equation \ref{eq:label_cost_percent}, which measures the percentage of unique log sequences in the subset relative to the total unique log sequences in the full dataset (denoted by \uniquefullsize). 
\begin{equation}
    \label{eq:label_cost_percent}
    \mathcal{U}_\%=\frac{\mathcal{U}_\#}{\mathcal{U}_\#^{full}}
\end{equation}

To quantify the reduction in labeled data achieved by data-efficient methods compared to methods trained on full datasets, we define the labeled data usage reduction \lcreduction as in Equation \ref{eq:label_cost_reduction}. A higher \lcreduction indicates a greater reduction, demonstrating the superior data efficiency of the method under evaluation, and vice versa. 
\begin{equation}
    \label{eq:label_cost_reduction}
    \Delta \mathcal{U}_\%=1- \mathcal{U}_\%
\end{equation}
~\\
\noindent \textbf{Time Efficiency} We evaluate the time efficiency in terms of training and inference time for each model. For training time, we calculate the total training time taken for a model. For inference time, we calculate the average inference time for one input sequence in the testing set. 

~\\
\noindent \textbf{Memory Efficiency} We evaluate memory efficiency based on the additional memory required by the cache component during inference. Specifically, we measure the memory overhead introduced by the in-memory cache, which stores predictions for all previously seen unique sequences. To approximate this overhead, we compute the memory consumed by the cache in the structure of a dictionary after processing the entire test set. 
Since the cache grows incrementally with the number of unique sequences, this measurement represents its maximum size at the end of inference.

~\\
\noindent \textbf{Statistical Testing}
To mitigate the potential influence of randomness on our results, we repeat each experiment on each configuration 5 times and report the average performance across the runs. This ensures our analysis is robust and not unduly influenced by any single random sampling or stochastic training and fine-tuning, providing a reliable evaluation. We further perform Mann-Whitney U test as recommended in ~\cite{statistical_testing_guide} on each dataset, resulting in test group sizes of 30 for \unstableadfa (6 configurations), 10 for \unstablelogevol (2 configurations), 20 for \synhdfs (4 configurations), and 30 for \synlogevol (6 configurations).

The Mann-Whitney U test is a non-parametric statistical test that compares two methods, A and B, without any assumption of the data distribution. It computes a p-value, which indicates whether the observed performance difference is statistically significant. The null hypothesis presumes no significant distinction between performance A and B.  If the p-value falls below the commonly used threshold of \num{0.05}, we reject the null hypothesis and conclude that the difference is statistically significant. Conversely, if the p-value is greater than or equal to \num{0.05}, the difference is considered non-significant, meaning the observed difference could be due to randomness. 

\subsubsection{Other Settings. }
We conducted all experiments with on a cloud computing environment containing \num{28} CPU cores for computation, \num{2} $\times$ Nvidia L40S GPU devices, and \SI{256}{\giga\byte} RAM.

\subsection{Implementation}\label{sec:implementaitondetails}
In this section, we introduce the implementation details of preprocessing, \method, and the baselines. 

\subsubsection{Preprocessing} \label{sec:preprocessing-setting} Two primary steps of preprocessing are parsing and partitioning, as described in Section \ref{sec:GPT-basedAD}.

Log parsing is used to provide structured context for \method (as explained in~\S~\ref{subsec:preprocessing}) and log-parsing-based baselines such as LogRobust and CNN. 
For \unstableadfa, parsing is not required since each log message is a one-word system call. For datasets with evolving templates—\unstableadfa and \synlogevol—we follow their original authors’ practice and use the Prefix Graph parser~\cite{prefixgraph}. This parser does not require a training set and is more flexible in handling varying template lengths and substructures compared to fixed-depth approaches such as Drain~\cite{drain}. For \synhdfs, we use Drain following the common practice for this specific synthetic dataset~\cite{LogRobust, hitanomaly}\footnote{We acknowledge that on the original \hdfs dataset, more recent log parsers~\cite{lilac, LLMparser} demonstrated higher parsing effectiveness than Drain. However, as a recent study~\cite{khan_2024} demonstrated, there is no correlation between parsing accuracy and anomaly detection accuracy.}.\synhdfs exhibits instability at the sequence level, but its log templates remain stable; hence,  the limitations of Drain in handling evolving templates do not apply in this case.
Since each log message in ADFA consists of a one-word system call, the subsequent RAG can easily associate a log message with its relevant information, such as the description of that system call.

Applying a smaller window over long sequences facilitates the localization of anomalies when logs are labeled at the message level. Hence,  we applied sliding window-based partitioning with a window size of \num{50} on LOGEVOL-Hadoop. In contrast, sliding window partitioning is not an option for long sequences in the ADFA, LOGEVOL-Spark, \synlogevol, and HDFS datasets due to the absence of message-level labels. For HDFS, most sessions are short, with only \SI{3.5}{\percent} sessions exceeding 30 templates.  We followed the implementation of~\citet{le2022} and truncated these long sessions to ensure that all sessions were within the 30-template limit.

\subsubsection{\method}\label{sec:flexlog-details} The implementation of \method mainly involves two key aspects, namely, the selection of base models for ensemble learning and the fine-tuning of the LLM base model. 
~\\
\noindent \textbf{Base Model Selection} \method combines multiple heterogeneous base models through ensemble learning, including ML/DL models and LLMs. Specifically, the ensemble in \method comprises three ML base models (KNN, DT, and SLFN) with one LLM base model (Mistral ~\cite{mistralai}). We selected KNN, DT, and SLFN due to their high effectiveness on \slad task as reported by \citet{lightad}. We selected the LLM base model (Mistral 22B) through the empirical evaluation of multiple open-source and closed-source LLMs. To elaborate, closed-source LLMs like GPT are considered state-of-the-art LLMs in various domains, albeit at a high cost~\cite{yao2024survey}. We experimented with a major version--- GPT-4o (\texttt{GPT-4o-turbo} version) --- based on OpenAI's recommendation in terms of performance~\cite{fine-tuning-format, openai-models}. In contrast, open-source LLMs incur no cost and offer more flexibility regarding fine-tuning and inference. Within the limit of our computing resources, we explored two open-source LLMs that have shown competitive performance to closed-source LLMs~\cite{llama3.1,yang2024largelanguagemodelsperform}: LLama 3.1 8B and Mistral 22B.

\noindent \textbf{Fine-tuning LLMs} 
For open-source LLMs, we utilized 4-bits QLora~\cite{qlora} for fine-tuning, with the following configurations based on our preliminary experiments: rank=16, alpha=16, 
and batch=1. We fine-tuned Llama 3.1 8B and Mistral Small 22B using the Unsloth library due to its high efficiency~\cite{Unsloth}. The number of steps is empirically tuned for unique pairs of LLM and dataset separately, using grid-search and cross-validation; values range from 500 to 2500 in steps of 500. 
For closed-source LLM of GPT-4o, fine-tuning was accomplished through OpenAI API, which incurred an associated cost. Hyperparameters such as the number of epochs and batch size for fine-tuning GPT were optimized and automatically determined by OpenAI fine-tuning APIs on each dataset. 

LLMs, even after fine-tuning, tend to generate non-deterministic output, which threatens the reliability of \method for \task. To ensure reliable anomaly detection, we instruct the LLMs to generate the response with minimum temperature (e.g., 0.1 for Mistral ). In case the responses deviate from explicit labels (e.g., "normal", "anomalous", "0" or "1"), ambiguous responses trigger up to five regeneration attempts with progressively higher temperatures from 0.2 to 1, in steps of 0.2, to diversify outputs. If no valid label is parsed after all attempts, we classify the sequence as "normal" to reduce false positives, which might trigger alert fatigue and operational disruption unnecessarily. 

\noindent \textbf{Hyper-parameters settings of ML base models in \method} For the DT and SLFN base models, we use the default values provided by Scikit-learn Library across all datasets. Our preliminary experiments suggest that tuning these hyper-parameters with limited data often leads to overfitting and reduced effectiveness compared to the default settings. While further tuning could potentially improve \method's performance, we leave this for future work.   Specifically for DT, we set criterion to ``gini'', max\_depth to ``None'', and min\_samples\_split to 2. For SLFN, we set hidden\_layer\_sizes to 100, activation to ``relu'', solver=``adam'', and batch\_size to ``auto''.
For KNN, since the number of neighbors plays an important role in handling imbalanced datasets~\cite{10.1007/978-3-540-39964-3_62}, we empirically tuned it with grid search and cross-validation on limited training data.
We set the number of neighbors to 2 for ADFA and LOGEVOL Hadoop, and 1 for HDFS\footnote{Overall, due to the limited training data, we recommend using default parameters and tuning only those that are highly sensitive to imbalanced data, as anomalies are often rare in real-world datasets, using cross-validation. In the future, we plan to conduct more experiments to explore the correlation between the percentage of anomalous data and the optimal number of neighbors.}. For the extremely imbalanced datasets---LOGEVOL Spark and \synlogevol, which share the same training set---KNN performs poorly on the validation set, exhibiting significantly lower effectiveness compared to DT and SLFN. This aligns with known limitations of KNN on highly imbalanced datasets, where the majority class tends to dominate the predictions~\cite{knnimbalance}. Therefore, we excluded KNN from \method for LOGEVOL Spark and \synlogevol.

Lastly, for the ensemble strategy, we implemented a majority voting algorithm in which, in the event of a tie, the sequence is labeled as normal, as described in~\S~\ref{subsec:fsel}. Our preliminary results across all four datasets indicate that this strategy remains the most effective and straightforward compared to alternative ensemble learning approaches, including SNAIL~\cite{snail} and MetaFormer~\cite{metaformer} (see Appendix~\ref{sec:appendix-ensembling}).

\subsubsection{Baselines }
For baselines, we set hyper-parameters as reported in their original papers or suggested by their implementation packages. When hyper-parameters were not available in either of them, particularly for datasets such as ADFA, we empirically tuned the parameters by grid search with a cross-validation approach. For LightAD, we fine-tuned hyper-parameters using Bayesian optimization, available in their implementation code for KNN, DT, and SLFN.
LogAnomaly's representation model (\textsf{template2vec}) requires domain-specific antonyms and synonyms for training. This information is unavailable in its original paper and, thus, similar to the method previously adopted~\cite{le2022}, we used a pre-trained FastText model~\cite{fasttextzip} to compute the semantic vectors. 
As LOGEVOL Hadoop and SynHDFS training sets are too large to process on NeuralLog, we used a subset containing the first \num{200000} log messages, following the methodology adopted in prior work~\cite{le2022}.

\subsection{Data availability. }
The replication package, including our synthesized datasets, additional experiment results, and source code, is publicly available~\cite{replication-package}.

\section{Results}\label{sec:results}

\subsection{RQ1: Overall Effectiveness}\label{sec:results_rq1}

\begin{table}[htbp]
\centering
    \captionsetup[table]{skip=0pt}
\captionof{table}{ Statistics of training data for \method and baselines used in RQ1 on \unstableadfa, \logevol, \synlogevol and \synhdfs. 
\label{tab:labelcost-rq1} 
}
\begin{threeparttable}[t] \resizebox{0.85\linewidth}{!}{
        \begin{tabular}{@{\hspace{4\tabcolsep}}c@{\hspace{1.2\tabcolsep}}c@{\hspace{1.2\tabcolsep}}c@{\hspace{1.2\tabcolsep}}c@{\hspace{1.2\tabcolsep}}c@{\hspace{1.2\tabcolsep}}c@{\hspace{1.2\tabcolsep}}c@{\hspace{1.2\tabcolsep}}c@{\hspace{2\tabcolsep}}c@{\hspace{1.2\tabcolsep}}}
            \toprule
            \multirow{2}{*}{\textbf{Dataset}} & \multicolumn{2}{c}{\textbf{\datasize}} & \multicolumn{2}{c}{\textbf{\uniquesize}} & \multicolumn{2}{c}{\textbf{\uniquepercent}} & \multirow{2}{*}{\textbf{\lcreduction}}\\
            \cmidrule(lr){2-3} 
            \cmidrule(lr){4-5}
            \cmidrule(lr){6-7}
            &  Baselines & FlexLog & Baselines & FlexLog & Baselines & FlexLog & \\
            \midrule
            \unstableadfa & \num{ 4785 }&  \num{ 1000 }  & \num{  3181 }&   \num{ 686 } & 100 \% &   21.57\%& \SI{78.43} {\pp}   \\

            \midrule
            \unstablelogevol  & \num{ 313426 }&  \num{ 9692 }  & \num{  29809 }&   \num{ 9692 } & 100 \%   &   32.51\% & \SI{67.49}{\pp}     \\
            
            \midrule
            \synlogevol  &  \num{ 11114 }&  \num{ 1134 }  & \num{  4939 }&   \num{ 1134 }& 100 \%  &   22.96\%& \SI{77.04}{\pp}     \\
            
            \midrule
             \synhdfs  &  \num{ 460048 }&  \num{ 5772 }   & \num{  15545 }&   \num{ 5772 } & 100 \% &   37.13\% & \SI{62.87}{\pp}     \\
            
            \bottomrule
        \end{tabular}}
\end{threeparttable}
\end{table} 

\begin{table*}[t!]
\centering
    \captionsetup[table]{skip=0pt}
\captionof{table}{Effectiveness of \method and baselines for \task and \slad on the ADFA dataset \label{tab:adfa-rq1} 
}
     \begin{threeparttable}[ht!]
 \resizebox{\linewidth}{!}{
\begin{tabular}{@{\hspace{0.5\tabcolsep}}c@{\hspace{0.5\tabcolsep}}c@{\hspace{0.5\tabcolsep}}c@{\hspace{0.5\tabcolsep}}c@{\hspace{0.5\tabcolsep}}c@{\hspace{0.5\tabcolsep}}c@{\hspace{0.5\tabcolsep}}c@{\hspace{0.5\tabcolsep}}c@{\hspace{0.5\tabcolsep}}c@{\hspace{0.5\tabcolsep}}c@{\hspace{0.5\tabcolsep}}c@{\hspace{0.5\tabcolsep}}c@{\hspace{0.5\tabcolsep}}c@{\hspace{0.5\tabcolsep}}}
\toprule
\multirow{3}{*}{\textbf{Data}} & \multirow{3}{*}{\textbf{Unstable}}& \multirow{3}{*}{\textbf{M}} & \multicolumn{1}{c}{\textbf{limited data}} & \multicolumn{9}{c}{\textbf{full training set}}\\
\cmidrule(lr){4-4} 
\cmidrule(lr){5-13}
&&& \multicolumn{5}{c}{\textbf{Supervised}} & \textbf{Semi-S} & \multicolumn{4}{c}{\textbf{Unsupervised}}\\
\cmidrule(lr){4-8} 
\cmidrule(lr){9-8}
\cmidrule(lr){10-13}
& & & \method & LightAD & NeuralLog& LogRobust& CNN & PLELog & LogAnomaly& DeepLog& LogCluster& PCA\\
\midrule
 \multirow{3}{*}{ADFA}& \multirow{3}{*}{No} &  P & \num{0.708} & \num{0.820} & \num{0.538} & \num{0.718}&\num{0.666} & \num{0.736} & \num{0.357} & \num{0.334}  & \num{0.255} & \num{0.196}  \\ 

& & R & \num{0.894}  & \num{0.814} & \num{0.602} & \num{0.708} & \num{0.842} & \num{0.216}&\num{0.430} & \num{0.523}  &\num{0.907} & \num{0.139}  \\ 

&  & F1 & \num{.791} & \textbf{\num{0.817}}& \num{0.568} & \num{0.713} &\num{0.744}& \num{0.334} & \num{0.390} & \num{0.408}  & \num{0.398} & \num{0.162}  \\ 
\midrule
\midrule
 \multirow{3}{*}{adduser} & \multirow{3}{*}{Yes}& P & \num{0.619} &\num{0.673}& \num{0.303} & \num{0.711}& \num{0.547} & \num{0.507}  & \num{0.208} & \num{0.198} & \num{0.217}& \num{0.139} \\

& & R& \num{0.857} &\num{0.786} & \num{0.651} & \num{0.415}&\num{0.775}&\num{0.281} & \num{0.483} & \num{0.562}  & \num{0.725} & \num{0.202}  \\ 

 & & F1& \num{0.718} & \textbf{\num{0.725}}& \num{0.412} & \num{0.524} &\num{0.641} & \num{0.361}& \num{0.291} & \num{0.292}  & \num{0.334} & \num{0.165}  \\ 
\cmidrule{1-13}

 \multirow{3}{*}{hydraFTP}& \multirow{3}{*}{Yes}& P & \num{0.686}  & \num{0.780} & \num{0.532} & \num{0.612} &\num{0.882} & \num{0.247}& \num{0.345} & \num{0.411}  & \num{0.268} & \num{0.139}  \\ 

& & R & \num{0.913} & \num{0.731} & \num{0.306} & \num{0.306} &\num{0.653}& \num{0.897} & \num{0.650} & \num{0.581}  & \num{0.950} & \num{0.133}  \\ 

&  & F1 & \textbf{\num{0.784}} & \num{0.754} & \num{0.388} & \num{0.408}&\num{0.750} & \num{0.388} & \num{0.451} & \num{0.481}  & \num{0.418} & \num{0.144}  \\ 
\cmidrule{1-13}

 \multirow{3}{*}{hydraSSH}& \multirow{3}{*}{Yes}& P & \num{0.657}   &\num{0.833}& \num{0.298} & \num{0.656} & \num{0.882}&\num{0.522} & \num{0.408} & \num{0.390}  & \num{0.299} & \num{0.121}  \\ 

& & R & \num{0.804}  &  \num{0.635}& \num{0.479} & \num{0.262} & \num{0.653}&\num{0.433} & \num{0.583} & \num{0.554}  & \num{0.999} & \num{0.121}  \\ 

&  & F1 & \textbf{\num{0.723}} & \num{0.666} & \num{0.368} & \num{0.374} & \num{0.750}&\num{0.473} & \num{0.480} & \num{0.458}  & \num{0.461} & \num{0.140}  \\ 
\cmidrule{1-13}

 \multirow{3}{*}{java}& \multirow{3}{*}{Yes}& P & \num{0.634}  &\num{0.699} & \num{0.366} & \num{0.695} & \num{0.752}&\num{0.371} & \num{0.357} & \num{0.269}  & \num{0.213} & \num{0.169}  \\ 

& & R & \num{0.650} &\num{0.590} & \num{0.849} & \num{0.457} & \num{0.549}&\num{0.513} & \num{0.516} & \num{0.459}  & \num{0.959} & \num{0.157}  \\ 

&  & F1 & \num{0.642} & \num{0.639} & \num{0.511} & \num{0.558} & \num{0.635}&\num{0.430} & \num{0.422} & \num{0.339}  & \num{0.348} & \num{0.158}  \\ 
\cmidrule{1-13}

 \multirow{3}{*}{web}& \multirow{3}{*}{Yes}& P & \num{ 0.639} & \num{0.788} & \num{0.330} & \num{0.583} &\num{0.786} & \num{0.799} & \num{0.254} & \num{0.335}  & \num{0.186} & \num{0.204}  \\ 

& & R & \num{ 0.709} &  \num{0.487} & \num{0.741} & \num{0.395} &\num{0.513}& \num{0.136} & \num{0.530} & \num{0.373}  & \num{0.829}  &\num{0.191} \\ 

&  & F1 & \textbf{\num{0.672}}  &\num{0.602} & \num{0.457}   & \num{0.449} & \num{0.621}& \num{0.233} &\num{0.343} & \num{0.353} & \num{0.304}  & \num{0.197} \\ 
\cmidrule{1-13}

 \multirow{3}{*}{meter}& \multirow{3}{*}{Yes}& P & \num{ 0.564} &\num{0.710} & \num{0.312} & \num{0.745} & \num{0.642}&\num{0.569} & \num{0.147} & \num{0.178}  & \num{0.235} & \num{0.210}  \\ 

& & R & \num{0.859} & \num{0.732}& \num{0.889}  & \num{0.577} & \num{0.799} & \num{0.163}&\num{0.425} & \num{0.436} & \num{0.943}  & \num{0.139}   \\ 

&  & F1 & \textbf{\num{0.682}}  & \num{0.679} & \num{0.461} & \num{0.651} & \num{0.711}&\num{0.253} & \num{0.218} & \num{0.253}  & \num{0.175} & \num{0.241}  \\ 
\cmidrule{1-13}

 \multirow{3}{*}{average} & \multirow{3}{*}{Yes}& P &\num{0.633}& \num{0.747}& \num{0.357}& \num{0.667}& \num{0.748}& \num{0.503}& \num{0.286}& \num{0.297}& \num{0.236}& \num{0.164}\\ 

& & R &\num{0.799}& \num{0.660}& \num{0.652}& \num{0.402}& \num{0.657}& \num{0.404}& \num{0.531}& \num{0.494}& \num{0.901}& \num{0.157}\\ 

&  & F1 & \textbf{\num{0.704}}& \num{0.677} & \num{0.433}*& \num{0.494}*& \num{0.685} & \num{0.356}*& \num{0.368}*& \num{0.363}*& \num{0.340}*& \num{0.174}*
\\ 
\bottomrule
\end{tabular}
}
\begin{tablenotes}
         \item [*] \footnotesize{\method yields a significant higher F1-score than compared baseline.}
        \end{tablenotes}
\end{threeparttable}
\end{table*} 

\begin{table*}[htbp]
\centering
    \captionsetup[table]{skip=0pt}
\captionof{table}{Effectiveness of \method and baselines for \task and \slad on the LOGEVOL dataset \label{tab:crossversion-rq1}
}
    \begin{threeparttable}[htbp]
 \resizebox{\linewidth}{!}{

\begin{tabular}{@{\hspace{0.5\tabcolsep}}c@{\hspace{0.5\tabcolsep}}c@{\hspace{0.8\tabcolsep}}c@{\hspace{0.5\tabcolsep}}c@{\hspace{0.5\tabcolsep}}c@{\hspace{0.5\tabcolsep}}c@{\hspace{0.5\tabcolsep}}c@{\hspace{0.5\tabcolsep}}c@{\hspace{0.5\tabcolsep}}c@{\hspace{0.5\tabcolsep}}c@{\hspace{0.5\tabcolsep}}c@{\hspace{0.5\tabcolsep}}c@{\hspace{0.5\tabcolsep}}c@{\hspace{0.5\tabcolsep}}}
\toprule
\multirow{3}{*}{\textbf{Data}}& \multirow{3}{*}{\textbf{Unstable}} & \multirow{3}{*}{\textbf{M}} & \multicolumn{1}{c}{\textbf{limited data}} & \multicolumn{9}{c}{\textbf{full training set}}\\
\cmidrule(lr){4-4} 
\cmidrule(lr){5-13}
&&& \multicolumn{5}{c}{\textbf{supervised}} & \textbf{Semi-S} & \multicolumn{4}{c}{\textbf{Unsupervised}}\\
\cmidrule(lr){4-8} 
\cmidrule(lr){9-9}
\cmidrule(lr){10-13}
& & & \method & LightAD & NeuralLog& LogRobust& CNN & PLELog & LogAnomaly& DeepLog& LogCluster& PCA\\
\midrule
 \multirow{3}{*}{Hadoop$_{2\rightarrow2}$}& \multirow{3}{*}{No}& P & \num{0.999}  & \num{0.999} & \num{0.997} & \num{0.984} & \num{0.997}& \num{0.648}& \num{0.263} & \num{0.384} & \num{0.952}& \num{0.267} \\ 

& &R & \num{0.986}  &  \num{0.994}  & \num{0.986} & \num{0.976} & \num{0.997}& \num{0.888}& \num{0.616} & \num{0.352} & \num{0.320}& \num{0.867} \\ 

 & &F1 & \num{0.993} & \textbf{\num{0.997} } & \num{0.992} & \num{0.980} & \textbf{\num{0.997}}& \num{0.749}& \num{0.368} & \num{0.367} & \num{0.479}& \num{0.408}  \\ 
\midrule

 \multirow{3}{*}{Spark$_{2\rightarrow2}$}& \multirow{3}{*}{No}& P &  \num{0.999}   &\num{0.999} & \num{0.999} & \num{0.941} & \num{0.999}& \num{0.172} & \num{0.501} & \num{0.512}& \num{0.771} & \num{0.072}\\ 

& &R & \num{0.969}  & \num{0.939} & \num{0.636} & \num{0.969} & \num{0.878}& \num{0.129} & \num{0.393} & \num{0.443}& \num{0.818} & \num{0.471} \\ 

 && F1 & \textbf{\num{0.984}} & \num{0.968} & \num{0.777} & \num{0.952} & \num{0.935}& \num{0.243} & \num{0.441} & \num{0.475}& \num{0.794} & \num{0.125}\\ 
\midrule
\midrule
 \multirow{3}{*}{Hadoop$_{2\rightarrow3}$}&\multirow{3}{*}{Yes}&  P  & \num{0.998}  &\num{0.998} & \num{0.914} & \num{0.905} & \num{0.992}& \num{0.626}& \num{0.221} & \num{0.384} & \num{0.510}& \num{0.225}  \\ 

& &R& \num{0.965} &  \num{0.963}  & \num{0.984} & \num{0.950} & \num{0.968}& \num{0.819}& \num{0.522} & \num{0.352} & \num{0.371}& \num{0.898} \\ 

& & F1 & \textbf{\num{0.982}}  &\num{0.980}  & \num{0.948} & \num{0.927} & \num{0.980}& \num{0.709}& \num{0.310} & \num{0.367} & \num{0.430}& \num{0.360}  \\ 
\midrule

 \multirow{3}{*}{Spark$_{2\rightarrow3}$}&\multirow{3}{*}{Yes}&  P & \num{0.999}  & \num{0.981} & \num{0.916} & \num{0.696} & \num{0.992}& \num{0.105} & \num{0.141} & \num{0.08}& \num{0.347} & \num{0.061} \\ 

& &R & \num{0.805} & \num{0.708} & \num{0.766} & \num{0.832} & \num{0.736}& \num{0.377} & \num{0.458} & \num{0.606}& \num{0.805} & \num{0.484}  \\ 

 & &F1 &\textbf{\num{0.892}}  & \num{0.829} & \num{0.834} & \num{0.757} & \num{0.840}& \num{0.165} & \num{0.216} & \num{0.122}& \num{0.485} & \num{0.108} \\ 
\midrule
 \multirow{3}{*}{\makecell{Average \\ \footnotesize (Spark$_{2\rightarrow3}$,\\ \footnotesize Hadoop$_{2\rightarrow3}$)}}& \multirow{3}{*}{Yes}& P & \num{0.998}& \num{0.99}& \num{0.915}& \num{0.786}& \num{0.992}& \num{0.366}& \num{0.181}& \num{0.232}& \num{0.428}& \num{0.143}\\ 

& &R &\num{0.871}& \num{0.83}& \num{0.875}& \num{0.863}& \num{0.852}& \num{0.887}& \num{0.49}& \num{0.479}& \num{0.588}& \num{0.691} \\ 

& &F1 & \textbf{\num{0.928}}& \num{0.898}*& \num{0.891}*& \num{0.833}*& \num{0.910}*& \num{0.437}*& \num{0.263}*& \num{0.244}*& \num{0.458}*& \num{0.234}* \\ 
\bottomrule
\end{tabular}
}
\begin{tablenotes}
         \item [*] \footnotesize{\method yields a significant higher F1-score than baseline.}
        \end{tablenotes}
\end{threeparttable}

\end{table*} 

~\\
RQ1 investigates the effectiveness of \method trained with limited data and compares it to baselines trained with full datasets on \task and \slad. This comparison between \method and baselines is particularly less advantageous for \method, as it is trained with less data than the baselines\footnote{The effectiveness of baselines when trained on the same limited data as \method is reported in Appendix~\ref{sec:appendix-limited-data}.}. 
To prevent inflated effectiveness caused by data leakage, test data is de-duplicated from the training data, as explained in~\S~\ref{subsec:fsel}; a comparison of model performance evaluated on the test set with and without de-duplication is also provided in Appendix~\ref{sec:appendix-duplication}.

Table~\ref{tab:labelcost-rq1} reports the configurations of training datasets for \method and baselines, including the training dataset size (\datasize), the number of unique log sequences ($\mathcal{U}_\#$), the percentage of unique log sequences relative to the total unique log sequences in the full datasets (\uniquepercent) and the reduction in percentage points in labeled data achieved by \method (\lcreduction); please refer to \S~\ref{sec:metrics} for details of these metrics. In the rest of this section, we first report the evaluation of the overall effectiveness of \method and baselines on two real-world unstable datasets, \unstablelogevol and \unstableadfa (\S~\ref{sec:rq1_real}). To further analyze the influence of instability, we conducted controlled experiments on two synthesized datasets,  \synlogevol and \synhdfs, where varying levels of instability are systematically introduced (\S~\ref{sec:rq1_synthesized}).

\subsubsection{Effectiveness on Real-world Datasets}
\label{sec:rq1_real}
Table~\ref{tab:adfa-rq1} and Table~\ref{tab:crossversion-rq1} report the effectiveness of \method on two real-world datasets, \logevol and ADFA, respectively. Column \emph{``Data''} specifies different dataset configurations, including \slad alongside with various \task settings. For instance, Hadoop$_{2\rightarrow3}$ represents that logs produced from Hadoop 2 and Hadoop 3 are used as training and testing datasets, respectively.  Column ``Unstable'' indicates whether the data configuration is unstable or not. Column \emph{``M''} reports the effectiveness metrics introduced in \S~\ref{sec:metrics}. We recall in this table that \method is trained with only \emph{``limited data''} while all baselines are all trained with the \emph{``full training dataset''}, including \emph{``supervised''}, \emph{``semi-supervised''} and \emph{``unsupervised''} baselines.

As shown in Table~\ref{tab:adfa-rq1}, in the \slad of the ADFA dataset, \method achieves an F1 score of \num{0.791}, second only to LightAD (\num{0.817}) among all methods. However, in the \task, where log instability arises from environmental changes, i.e., the introduction of novel attack types, \method outperforms all baselines on \num{5} out of \num{6} configurations. The only exception occurs in the  \emph{adduser configuration}, where LightAD surpasses \method with a small margin of \SI{0.7}{\pp} ($72.5\%-71.8\%$). Overall, \method yields the highest average F1 score of \num{0.704} in \task configurations, followed by CNN (\num{0.685}) and LightAD (\num{0.677}). A Mann-Whitney U test indicates that the differences between the average F1 scores of \method and LightAD, as well as the differences between \method and CNN, are statistically insignificant, whereas \method outperforms all other baselines significantly. Moreover, when moving from \slad to \task, the most effective baseline in \slad (LightAD) experiences an average F1 score decrease of \SI{14}{\pp}, whereas the decrease with \method remains below \SI{9}{\pp}, alleviating the impact of unstable logs on anomaly detection effectiveness.

Notably, \method achieves such high effectiveness on the ADFA dataset with only limited training data, while baselines like LightAD and CNN are trained with full datasets. As shown in Table~\ref{tab:labelcost-rq1}, \method's fine-tuning datasets contain only \SI{21.57}{\percent}  of unique log sequences relative to the total unique log sequences in the full ADFA datasets, each requiring a dedicated label. This translates to a reduction of \SI{78.43}{\pp} in usage of labeled data while achieving state-of-the-art effectiveness on the ADFA dataset. 

We observe similar results on the \unstablelogevol dataset, where instability is introduced due to system evolution, e.g., software system version updates. As shown in Table \ref{tab:crossversion-rq1}, we evaluated two \slad configurations (Hadoop$_{2\rightarrow2}$ and Spark$_{2\rightarrow2}$) and two \task configurations (Hadoop$_{2\rightarrow3}$ and Spark$_{2\rightarrow3}$). In the Hadoop$_{2\rightarrow2}$ \slad configuration, all supervised approaches achieve a high F1 score ($\geq 0.980$), including \method, LightAD, NeuralLog, LogRobust, and CNN. In the other \slad configuration Spark$_{2\rightarrow2}$, \method surpass baselines, reaching an F1 score of \num{0.984} by \SI{1.4}{\pp} from the top baseline, LightAD. In the Hadoop$_{2\rightarrow3}$ and Spark$_{2\rightarrow3}$ \task configurations, \method yields  F1 scores of  \num{0.982} and \num{0.892}, respectively, remaining the best approach compared to all the baselines. On average, \method outperforms all the baselines in terms of F1 score, with a minimum margin of \SI{1.8}{\pp} ($0.928-0.910$). Additionally, when moving from \slad to \task, the most effective baseline in \slad (LightAD) experiences an average F1 score decrease of \SI{10}{\pp}, whereas the decrease with \method remains below \SI{7}{\pp}, alleviating once again the impact of unstable logs on anomaly detection effectiveness.

Similar to the \unstableadfa dataset, \method achieves such high effectiveness on the \logevol dataset by training on a relatively limited dataset instead of the entire \logevol training dataset. Specifically, as depicted in Table \ref{tab:labelcost-rq1}, \method's fine-tuning dataset contain only \SI{32.51}{\percent} of unique log sequences relative to the total unique log sequences in the full \logevol datasets, indicating a reduction in usage of labeled data of \SI{67.49}{\pp}.

\subsubsection{Impact of Log Instability}
\label{sec:rq1_synthesized}
As detailed in \S~\ref{sec:datasset-config}, we conducted comprehensive experiments on two synthesized datasets, \synhdfs and \synlogevol, to analyze the impact of instability at both the log sequence and log template levels. Specifically, \synlogevol exhibits instability at both the sequence and template levels, whereas \synhdfs is characterized solely by sequence-level instability.

\noindent\textbf{Instability at Log Sequence Level.}
Table~\ref{tab:injectionrateresults_synhdfs} presents the performance of \method and baseline approaches on \synhdfs with sequence-level instability. As the injection ratio of changes increases,
\method consistently achieves the highest F1 score across all unstable configurations, demonstrating its robustness to varying levels of instability. On average, \method outperforms all baselines, exceeding the top-performing baseline, LightAD, by a margin of \SI{1.2}{\pp} in F1 score. A Mann-Whitney test reveals that the difference in F1 scores between \method and LightAD is statistically insignificant, indicating comparable performance between the two. However, unlike all baselines---including LightAD---which are trained on the full dataset, \method is trained on only a small subset comprising \SI{37.13}{\percent} of unique log sequences (Table \ref{tab:labelcost-rq1}). This means that \method achieves similar effectiveness and robustness while reducing the usage of labeled data by \SI{62.87}{\pp}. The effectiveness of \method and baselines on \synlogevol under varying log sequence instability demonstrated similar results to \synhdfs that we have provided in Appendix~\ref{sec:appendix-synevol}. 

\noindent\textbf{Instability at Log Template Level.}
Table~\ref{tab:injectionrateresults_templates_synevol} reports the effectiveness of \method compared with baselines on \synlogevol under varying template-level changes. Once again, \method achieves the highest precision, recall and F1 score across all injection ratios, outperforming the top-performing baseline---LightAD---by \SI{1.9}{\pp} ($96.3\%-94.4\%$) in terms of average F1 score while reducing usage of labeled data by \SI{77.04}{\pp}. A Mann-Whitney U test suggests this difference is statistically significant. Similar to sequence-level changes in \synlogevol, increasing template-level instability negatively impacts the effectiveness of semi-supervised and unsupervised approaches, whereas supervised methods---\method, LightAD, NeuralLog, LogRobust, and CNN---are relatively robust in terms of effectiveness across instability levels. This is likely because many template modifications in \synlogevol involve minor textual variations, such as inserting, removing, or replacing short tokens, which have minimal impact on \task, especially in long log sequences with up to \num{1818} templates. For instance, a log template stating ``\texttt{SecurityManager: authentication disabled; ui acls disabled; users with view permissions: Set(root); groups with view permissions: Set(); users with modify permissions: Set(root); ...}'' was modified by inserting \texttt{"so"} before \texttt{"users with modify permissions"}, a minor change that has limited impact on \task effectiveness.

\begin{table*}[tb]
\centering
\captionsetup[table]{skip=0pt}
\captionof{table}{Effectiveness of \method and baselines under different sequence-level injection ratios on \synhdfs. 
\label{tab:injectionrateresults_synhdfs}}
\footnotesize{%
\begin{threeparttable}[htbp]
 \resizebox{\linewidth}{!}{
\begin{tabular}{@{\hspace{0.5\tabcolsep}}c@{\hspace{0.5\tabcolsep}}c@{\hspace{0.8\tabcolsep}}c@{\hspace{0.5\tabcolsep}}c@{\hspace{0.5\tabcolsep}}c@{\hspace{0.5\tabcolsep}}c@{\hspace{0.5\tabcolsep}}c@{\hspace{0.5\tabcolsep}}c@{\hspace{0.5\tabcolsep}}c@{\hspace{0.5\tabcolsep}}c@{\hspace{0.5\tabcolsep}}c@{\hspace{0.5\tabcolsep}}c@{\hspace{0.5\tabcolsep}}c@{\hspace{0.5\tabcolsep}}}
\toprule
\multirow{3}{*}{\textbf{Data}}& \multirow{3}{*}{\textbf{Unstable}} & \multirow{3}{*}{\textbf{M}} & \multicolumn{1}{c}{\textbf{limited data}} & \multicolumn{9}{c}{\textbf{full training set}}\\
\cmidrule(lr){4-4} 
\cmidrule(lr){5-13}
&&& \multicolumn{5}{c}{\textbf{supervised}} & \textbf{Semi-S} & \multicolumn{4}{c}{\textbf{Unsupervised}}\\
\cmidrule(lr){4-8} 
\cmidrule(lr){9-9}
\cmidrule(lr){10-13}
& & & \method & LightAD & NeuralLog& LogRobust& CNN & PLELog & LogAnomaly& DeepLog& LogCluster& PCA\\
\midrule
 \multirow{3}{*}{0\%} & \multirow{3}{*}{No}& P & \num{0.954}  & \num{0.9763} & \num{0.949} & \num{0.957} & \num{0.933} & \num{0.630} & \num{0.243} & \num{0.725} & \num{0.999} & \num{0.924}  \\ 

& &R & \num{0.999} & \num{0.990} & \num{0.985} & \num{0.980} & \num{0.951} & \num{0.966} & \num{0.971} & \num{0.927} & \num{0.346} & \num{0.667}\\ 

 && F1 & \num{0.976}  & \textbf{\num{0.983}} & \num{0.966} & \num{0.969} & \num{0.942} & \num{0.763} & \num{0.389} & \num{0.814} & \num{0.514}& \num{0.762} \\ 
\midrule
\midrule
 \multirow{3}{*}{5\%}& \multirow{3}{*}{Yes}& P & \num{0.953} &\num{0.957} & \num{0.944} & \num{0.995} & \num{0.932} & \num{0.602} & \num{0.236} & \num{0.678} & \num{0.986} & \num{0.976} \\ 

& &R & \num{0.995}  & \num{0.985} & \num{0.985} & \num{0.979} & \num{0.952} & \num{0.554} & \num{0.961} & \num{0.894} & \num{0.346}& \num{0.667}\\ 

 & &F1 & \textbf{\num{0.974}}  & \num{0.971} & \num{0.964} & \num{0.878} & \num{0.944} & \num{0.577} & \num{0.380} & \num{0.771} & \num{0.512} & \num{0.792} \\ 
\midrule

 \multirow{3}{*}{10\%}& \multirow{3}{*}{Yes}& P & \num{0.954} &  \num{0.966} & \num{0.914} & \num{0.692} & \num{0.933} & \num{0.404} & \num{0.239} & \num{0.649} & \num{0.999} & \num{0.225} \\ 

&& R & \num{0.995} & \num{0.980} & \num{0.980} & \num{0.974} & \num{0.951} & \num{0.884} & \num{0.975} & \num{0.927} & \num{0.350} & \num{0.673}\\ 

 && F1 & \textbf{\num{0.974} } & \num{0.973} & \num{0.946} & \num{0.809} & \num{0.942} & \num{0.555} & \num{0.385} & \num{0.764} &\num{0.519} & \num{0.337}\\ 
\midrule

 \multirow{3}{*}{20\%}& \multirow{3}{*}{Yes}& P & \num{0.949} & \num{0.943} & \num{0.906} & \num{0.560} & \num{0.933} & \num{0.345} & \num{0.482} & \num{0.644} & \num{0.949} & \num{0.158}\\ 

&& R & \num{0.995} & \num{0.966} & \num{0.980} & \num{0.956} & \num{0.951} & \num{0.798} & \num{0.538} & \num{0.889} & \num{0.360} & \num{0.694}\\ 

 && F1 & \textbf{\num{0.971}} & \num{0.954} & \num{0.942} & \num{0.707} & \num{0.942} & \num{0.482} & \num{0.509} & \num{0.747} & \num{0.522} & \num{0.258} \\ 
\midrule
 \multirow{3}{*}{30\%}& \multirow{3}{*}{Yes}& P & \num{0.940} & \num{0.925} & \num{0.896} & \num{0.489} & \num{0.929} & \num{0.243} & \num{0.464} & \num{0.548} & \num{0.915}& \num{0.137}\\ 

&& R & \num{0.985}  & \num{0.956} & \num{0.970} & \num{0.951} & \num{0.947} & \num{0.877} & \num{0.572} & \num{0.903} & \num{0.365}& \num{0.718}\\ 

 && F1 & \textbf{\num{0.964}}  & \num{0.940} & \num{0.931} & \num{0.646} & \num{0.938} & \num{0.381} & \num{0.512} & \num{0.682} & \num{0.522}& \num{0.230}\\ 
\midrule
 \multirow{3}{*}{Average}& \multirow{3}{*}{Yes}&  P& \num{0.949}& \num{0.948}& \num{0.915}& \num{0.684}& \num{0.932}& \num{0.398}& \num{0.355}& \num{0.63}& \num{0.962}& \num{0.374} \\ 

& &R &\num{0.992}& \num{0.972}& \num{0.979}& \num{0.965}& \num{0.95}& \num{0.778}& \num{0.762}& \num{0.903}& \num{0.355}& \num{0.688} \\ 

 & &F1 &\textbf{\num{0.971}}& \num{0.959}& \num{0.946}*& \num{0.760}*& \num{0.942}*& \num{0.499}*& \num{0.446}*& \num{0.741}*& \num{0.519}*& \num{0.404}* \\

\bottomrule
\end{tabular}
}
\begin{tablenotes}
         \item [*] \footnotesize{\method yields a significant higher F1-score than compared baseline.}
        \end{tablenotes}
\end{threeparttable}

}
\end{table*}

\begin{table*}[!ht]
\centering
\captionsetup[table]{skip=0pt}
\captionof{table}{Effectiveness of \method and baselines under different template-level injection ratios on \synlogevol. 
\label{tab:injectionrateresults_templates_synevol}}
\footnotesize{%
\begin{threeparttable}[!ht]
 \resizebox{\linewidth}{!}{
\begin{tabular}{@{\hspace{0.5\tabcolsep}}c@{\hspace{0.5\tabcolsep}}c@{\hspace{0.8\tabcolsep}}c@{\hspace{0.5\tabcolsep}}c@{\hspace{0.5\tabcolsep}}c@{\hspace{0.5\tabcolsep}}c@{\hspace{0.5\tabcolsep}}c@{\hspace{0.5\tabcolsep}}c@{\hspace{0.5\tabcolsep}}c@{\hspace{0.5\tabcolsep}}c@{\hspace{0.5\tabcolsep}}c@{\hspace{0.5\tabcolsep}}c@{\hspace{0.5\tabcolsep}}c@{\hspace{0.5\tabcolsep}}}
\toprule
\multirow{3}{*}{\textbf{Data}}& \multirow{3}{*}{\textbf{Unstable}} & \multirow{3}{*}{\textbf{M}} & \multicolumn{1}{c}{\textbf{limited data}} & \multicolumn{9}{c}{\textbf{full training set}}\\
\cmidrule(lr){4-4} 
\cmidrule(lr){5-13}
&&& \multicolumn{5}{c}{\textbf{supervised}} & \textbf{Semi-S} & \multicolumn{4}{c}{\textbf{Unsupervised}}\\
\cmidrule(lr){4-8} 
\cmidrule(lr){9-9}
\cmidrule(lr){10-13}
& & & \method & LightAD & NeuralLog& LogRobust& CNN & PLELog & LogAnomaly& DeepLog& LogCluster& PCA\\
\midrule
 \multirow{3}{*}{0\%}& \multirow{3}{*}{No}& P&  \num{0.999}   &\num{0.999} & \num{0.999} & \num{0.941} & \num{0.999}& \num{0.172} & \num{0.501} & \num{0.512}& \num{0.771} & \num{0.072}\\ 

& &R & \num{0.969}  & \num{0.939} & \num{0.636} & \num{0.969} & \num{0.878}& \num{0.129} & \num{0.393} & \num{0.443}& \num{0.818} & \num{0.471} \\ 

 && F1 & \textbf{\num{0.984}} & \num{0.968} & \num{0.777} & \num{0.952} & \num{0.935}& \num{0.243} & \num{0.441} & \num{0.475}& \num{0.794} & \num{0.125}\\ 
\midrule
\midrule
 \multirow{3}{*}{5\%} & \multirow{3}{*}{Yes}& P & \num{0.999} & \num{0.999} & \num{0.999} & \num{0.943} & \num{0.999} & \num{0.127} & \num{0.351} & \num{0.282} & \num{0.651} & \num{0.062}  \\ 

& & R & \num{0.970} & \num{0.941} & \num{0.647} & \num{0.971} & \num{0.882} & \num{0.315} & \num{0.393} & \num{0.382} & \num{0.823} & \num{0.500}  \\ 

 & & F1 & \textbf{\num{0.985}} & \num{0.969} & \num{0.785} & \num{0.956} & \num{0.937} & \num{0.181} & \num{0.440} & \num{0.325} & \num{0.727} & \num{0.111} \\ 
\midrule

 \multirow{3}{*}{10\%}& \multirow{3}{*}{Yes}&P & \num{0.999} & \num{0.999} & \num{0.999} & \num{0.921} & \num{0.937} & \num{0.120} & \num{0.260} & \num{0.181} & \num{0.622} & \num{0.054}  \\ 

& &R & \num{0.942} & \num{0.914} & \num{0.600} & \num{0.969} & \num{0.857} & \num{0.316} & \num{0.371} & \num{0.400} & \num{0.800} & \num{0.457} \\ 

 && F1 & \textbf{\num{0.970}} & \num{0.955} & \num{0.750} & \num{0.944} & \num{0.895} & \num{0.174} & \num{0.305} & \num{0.250} & \num{0.700} & \num{0.097}  \\ 
\midrule

 \multirow{3}{*}{15\%}& \multirow{3}{*}{Yes} & P & \num{0.999} & \num{0.999} & \num{0.999} & \num{0.944} & \num{0.916} & \num{0.117} & \num{0.265} & \num{0.180} & \num{0.604} & \num{0.054} \\ 

& &R & \num{0.921} & \num{0.894} & \num{0.605} & \num{0.918} & \num{0.891} & \num{0.315} & \num{0.447} & \num{0.447} & \num{0.763} & \num{0.447} \\ 

 && F1 & \textbf{\num{0.958}} & \num{0.944} & \num{0.754} & \num{0.931} & \num{0.904} & \num{0.171} & \num{0.333} & \num{0.257} & \num{0.674} & \num{0.097}  \\ 
\midrule
 \multirow{3}{*}{20\%}& \multirow{3}{*}{Yes}& P  & \num{0.999} & \num{0.999} & \num{0.916} & \num{0.943} & \num{0.906} & \num{0.112} & \num{0.180} & \num{0.095} & \num{0.457} & \num{0.047} \\ 

& &R & \num{0.941} & \num{0.911} & \num{0.647} & \num{0.951} & \num{0.852} & \num{0.315} & \num{0.382} & \num{0.411} & \num{0.794} & \num{0.470} \\ 

 & &F1  & \textbf{\num{0.969}} & \num{0.953} & \num{0.758} & \num{0.946} & \num{0.878} & \num{0.165} & \num{0.245} & \num{0.155} & \num{0.580} & \num{0.085}  \\ 
\midrule

 \multirow{3}{*}{25\%}& \multirow{3}{*}{Yes}&P & \num{0.999} & \num{0.999} & \num{0.999} & \num{0.908} & \num{0.916} & \num{0.127} & \num{0.188} & \num{0.100} & \num{0.507} & \num{0.051}  \\ 

& &R& \num{0.904} & \num{0.850} & \num{0.625} & \num{0.954} & \num{0.846} & \num{0.239} & \num{0.400} & \num{0.400} & \num{0.800} & \num{0.475}  \\ 

 & &F1 & \textbf{\num{0.950}} & \num{0.918} & \num{0.769} & \num{0.930} & \num{0.880} & \num{0.166} & \num{0.256} & \num{0.160} & \num{0.621} & \num{0.093} \\ 
\midrule

 \multirow{3}{*}{30\%}& \multirow{3}{*}{Yes}& P  & \num{0.973} & \num{0.999} & \num{0.931} & \num{0.923} & \num{0.947} & \num{0.121} & \num{0.180} & \num{0.096} & \num{0.438} & \num{0.053}  \\ 

& &R & \num{0.925} & \num{0.857} & \num{0.642} & \num{0.954} & \num{0.857} & \num{0.242} & \num{0.404} & \num{0.476} & \num{0.761} & \num{0.476} \\ 

 & &F1& \textbf{\num{0.948}} & \num{0.923} & \num{0.760} & \num{0.938} & \num{0.900} & \num{0.161} & \num{0.250} & \num{0.160} & \num{0.556} & \num{0.095}  \\ 

\midrule

 \multirow{3}{*}{Average}& \multirow{3}{*}{Yes}& P  &\num{0.995}& \num{0.999}& \num{0.974}& \num{0.93}& \num{0.937}& \num{0.121}& \num{0.237}& \num{0.156}& \num{0.547}& \num{0.053} \\ 

& &R & \num{0.934}& \num{0.894}& \num{0.628}& \num{0.953}& \num{0.864}& \num{0.29}& \num{0.399}& \num{0.419}& \num{0.79}& \num{0.471}\\ 

 & &F1& \textbf{\num{0.963}}& \num{0.944}*& \num{0.763}*& \num{0.941}*& \num{0.899}*& \num{0.17}*& \num{0.305}*& \num{0.218}*& \num{0.643}*& \num{0.096}*  \\ 
\bottomrule
\end{tabular}
}
\begin{tablenotes}
         \item [*] \footnotesize{\method yields a significant higher F1-score than compared baseline.}
        \end{tablenotes}
\end{threeparttable}

}
\end{table*}

\begin{tcolorbox}[compactbox]
The answer to RQ1 is that \method achieves \textbf{state-of-the-art effectiveness} on both real-world and synthesized datasets while exhibiting \textbf{high data efficiency} (with a reduction in labeled data usage ranging between \SI{62.87}{\pp} and \SI{78.43}{\pp}). On synthesized datasets, \method remains effective under up to \SI{30}{\percent} sequence- and template-level instability.
\end{tcolorbox}

\subsection{RQ2: Data Efficiency on the \unstableadfa Dataset}
\label{sec:results_rq2}
\begin{table*}[htbp]
\centering
    \captionsetup[table]{skip=0pt}
\captionof{table}{Statistics of the sampled subsets of \unstableadfa
\label{tab:adfa-sample}}
    \begin{threeparttable}[htbp]
 \resizebox{\linewidth}{!}{
\begin{tabular}{@{\hspace{\tabcolsep}}c@{\hspace{\tabcolsep}}c@{\hspace{\tabcolsep}}c@{\hspace{\tabcolsep}}c@{\hspace{\tabcolsep}}c@{\hspace{\tabcolsep}}c@{\hspace{\tabcolsep}}c@{\hspace{\tabcolsep}}c@{\hspace{\tabcolsep}}c@{\hspace{\tabcolsep}}c@{\hspace{\tabcolsep}}c@{\hspace{\tabcolsep}}c@{\hspace{\tabcolsep}}c@{\hspace{\tabcolsep}}c@{\hspace{\tabcolsep}}c@{\hspace{\tabcolsep}}c@{\hspace{\tabcolsep}}c@{\hspace{\tabcolsep}}c@{\hspace{\tabcolsep}}c@{\hspace{\tabcolsep}}c@{\hspace{\tabcolsep}}c@{\hspace{\tabcolsep}}c@{\hspace{\tabcolsep}}c@{\hspace{\tabcolsep}}c@{\hspace{\tabcolsep}}}
\toprule
\textbf{\datasize} & \multicolumn{3}{c}{\textbf{adduser}} & \multicolumn{3}{c}{\textbf{hydraFTP}} & \multicolumn{3}{c}{\textbf{hydraSSH}} & \multicolumn{3}{c}{\textbf{java}} & \multicolumn{3}{c}{\textbf{web}} & \multicolumn{3}{c}{\textbf{meter}}\\
\cmidrule(lr){2-4} 
\cmidrule(lr){5-7}
\cmidrule(lr){8-10}
\cmidrule(lr){11-13}
\cmidrule(lr){14-16}
\cmidrule(lr){17-19}
\\[-2ex]
& \uniquepercent & \uniqueabnormal & \uniquenormal & \uniquesize & \uniqueabnormal & \uniquenormal & \uniquesize & \uniqueabnormal & \uniquenormal & \uniquesize & \uniqueabnormal & \uniquenormal & \uniquesize & \uniqueabnormal & \uniquenormal & \uniquesize & \uniqueabnormal & \uniquenormal\\
\midrule

 50 & \num{ 1.43 }& \num{ 3.88 }& \num{ 0.81 }
& \num{ 1.46 }& \num{ 4.37 }& \num{ 0.81 }
& \num{ 1.37 }& \num{ 4.46 }& \num{ 0.70 }
& \num{ 1.44 }& \num{ 4.08 }& \num{ 0.81 }
& \num{ 1.34 }& \num{ 4.05 }& \num{ 0.70 }
& \num{ 1.30 }& \num{ 3.77 }& \num{ 0.66 }
\\
\midrule

 500 & \num{ 12.23 }& \num{ 32.34 }& \num{ 7.19 }
& \num{ 12.50 }& \num{ 36.18 }& \num{ 7.23 }
& \num{ 12.62 }& \num{ 36.96 }& \num{ 7.30 }
& \num{ 11.95 }& \num{ 32.84 }& \num{ 6.98 }
& \num{ 12.42 }& \num{ 34.19 }& \num{ 7.19 }
& \num{ 12.46 }& \num{ 31.11 }& \num{ 7.66 }
\\
\midrule
 1000 & \num{ 21.78 }& \num{ 53.65 }& \num{ 13.80 }
& \num{ 21.45 }& \num{ 57.16 }& \num{ 13.53 }
& \num{ 21.76 }& \num{ 58.03 }& \num{ 13.82 }
& \num{ 21.55 }& \num{ 54.08 }& \num{ 13.84 }
& \num{ 21.30 }& \num{ 53.80 }& \num{ 13.49 }
& \num{ 21.81 }& \num{ 52.41 }& \num{ 13.91 }
\\
\midrule
 1500 & \num{ 29.10 }& \num{ 67.49 }& \num{ 19.47 }
& \num{ 29.15 }& \num{ 70.97 }& \num{ 19.82 }
& \num{ 29.23 }& \num{ 72.32 }& \num{ 19.79 }
& \num{ 29.17 }& \num{ 70.09 }& \num{ 19.47 }
& \num{ 29.49 }& \num{ 69.85 }& \num{ 19.79 }
& \num{ 28.77 }& \num{ 67.22 }& \num{ 18.80 }
\\
\midrule
 2000 & \num{ 35.48 }& \num{ 78.38 }& \num{ 24.72 }
& \num{ 35.04 }& \num{ 82.69 }& \num{ 24.41 }
& \num{ 35.96 }& \num{ 82.32 }& \num{ 25.81 }
& \num{ 36.13 }& \num{ 79.90 }& \num{ 25.71 }
& \num{ 35.30 }& \num{ 79.57 }& \num{ 24.66 }
& \num{ 35.11 }& \num{ 76.43 }& \num{ 24.43 }
\\
\bottomrule
\end{tabular}
}
\end{threeparttable}
\end{table*} 

RQ2 focuses on the data efficiency analysis of \method and baselines. As mentioned in \S~\ref{sec:rqs}, computational constraints preclude us from investigating the data efficiency on all the datasets. Hence, we focus on \unstableadfa in this RQ since it includes six diverse configurations, enabling a robust evaluation under different real-world \task scenarios. As shown in RQ1, these diverse configurations make it the most challenging dataset in our experiments in terms of detection effectiveness.
Specifically, for each configuration, we randomly sample five training subsets (50, 500, 1000, 1500, and 2000 samples)  to simulate different levels of data scarcity.

Table~\ref{tab:adfa-sample} reports statistics for the sampled subsets, including the percentage of unique log sequences (\uniquepercent), which is calculated by dividing the number of unique log sequences in each subset by the total number of unique log sequences in the full training dataset. This percentage represents the proportion of labeling effort required for each subset compared to full dataset annotation. The table further provides the percentages of unique log sequences for each class ( \uniqueabnormal and \uniquenormal). 
Notably, the smallest subset ($\mathcal{D}_\#=50$) contains $<1.5\%$ of unique log sequences, representing extreme data scarcity scenarios where the availability of labeled data is highly limited.  

Figure~\ref{fig:data-efficiency} depicts the effectiveness of \method and top baselines trained on the sampled subsets of \unstableadfa. Less competitive baselines, such as PLELog, LogAnomaly, DeepLog, LogCluster, and PCA, are not illustrated in the figure for brevity. The dashed horizontal line in each figure represents the state-of-the-art results on each dataset, produced by LightAD trained with full datasets. 

Table~\ref{tab:adfa-delta} highlights the improvement of \method by calculating F1 score differences in percentage points between \method and baselines. We also report Mann-Whitney U Test results for F1 comparisons across datasets. A ``$*$'' symbol is appended to the average F1 score if \method significantly outperforms the baseline in terms of F1 score. 

Overall, \method consistently achieves superior performance in terms of average F1 scores compared to the baselines when trained with the same amount of labeled data. As shown in Table~\ref{tab:adfa-delta}, the average F1-score improvements over the strongest baseline (i.e., \method minus LightAD) for $\mathcal{D}_\#=50$, $500$, $1000$, $1500$, and $2000$ are \SI{2}{\pp}, \SI{13}{\pp}, \SI{10}{\pp}, \SI{8}{\pp}, and \SI{7}{\pp}, respectively.  Mann-Whitney U tests show the significance of all the improvements, except for the extreme data scarcity scenario with $\mathcal{D}_\#=50$. Notably, even when compared to LightAD trained on the full dataset (represented by the dashed horizontal lines in all plots in Figure~\ref{fig:data-efficiency}), \method achieves higher or comparable performance with significantly less labeled data. In the remainder of this section, we elaborate on the performance of \method and baselines under different data scarcity scenarios. 
Under extreme data scarcity ($\mathcal{D}_\#=50$), we observe that all approaches exhibit poor performance in terms of F1 score, which are lower than \num{0.60}, respectively, across all six configurations of \unstableadfa. This limitation likely stems from insufficient training data diversity. As reported in the first row of Table~\ref{tab:adfa-sample}, the 50 samples selected for each configuration of \unstableadfa contain less than \SI{1.46}{\percent} of total unique log sequences in full datasets. The percentages of unique anomalous log sequences are also low, with a maximum of  \SI{4.46}{\percent} ( $\mathcal{D}_\#=50$, the \emph{java configuration}). Such low diversity and small size of the labeled datasets tend to be insufficient for the training of all approaches, including \method and baselines. 

Under less severe data scarcity ($\mathcal{D}_\#=500, 1000, 1500, 2000$), all methods exhibit improved effectiveness compared to the extreme label scarcity scenario ($\mathcal{D}_\#=50$). \method outperforms all baselines across all six configurations in terms of F1 score. As reported in Table~\ref{tab:adfa-delta}, the maximum percentage points against LightAD are observed at $\mathcal{D}_\#=500$, reaching \SI{13}{\pp} on average. The average difference values diminish to \SI{7}{\pp} as data size increases to 2000, underscoring \method's advantage with limited labeled data. Mann-Whitney U tests confirm the significance of all the differences ($\mathcal{D}_\#=500, 1000, 1500, 2000$).

\begin{figure*}[t!]
\centering
\resizebox{0.7\textwidth}{!}{
        \ref{named}
        }\\
\resizebox{0.45\textwidth}{!}{\begin{subfigure}{\textwidth}
        \centering
\begin{tikzpicture}[scale=1]
        \begin{axis}[
            width=\textwidth ,
            height=8cm,
            xlabel={Sample Size},
    ylabel={F1-score},
    xmin=0, xmax=2000,
    ylabel style = {align=center},
        ymin=0, ymax=1, 
    ytick={0, 0.2, 0.4, 0.6,0.8,1},
    mark options={scale=1.5},
    legend columns = -2,
    legend entries = {, \method, LightAD, NeuralLog, LogRobust, CNN } ,
    legend to name=named,
    legend style={text=black}
]
\addplot[
    color=black,
    domain=0:2000     , dashed
    ]
    {0.7254};
    
\addplot[
    color=blue,
    mark=pentagon,thick,
    ]
    coordinates {
    (50, 0.425)(500, 0.620)(1000, 0.718)(1500, 0.7000)(2000, 0.7238)};

\addplot[
    color={rgb:red,237;green,177;blue,32} ,
    mark=square,thick,
    ]
    coordinates {
    (50, 0.386)(500, 0.512)(1000, 0.587)(1500, 0.626)(2000, 0.626)};

\addplot[
    color=orange ,
    mark=triangle,thick,
    ]
    coordinates {
    (50, 0.20)(500, 0.29)(1000, 0.6)(1500, 0.58)(2000, 0.49)};

\addplot[
    color= {rgb:red,45;green,160;blue,45} ,
    mark=star,thick,
    ]
    coordinates {
    (50, 0.29)(500, 0.34)(1000, 0.39)(1500, 0.41)(2000, 0.51)};
    
\addplot[
    color={rgb:red,160;green,20;blue,47},
    mark=Mercedes star,thick,
   ]
    coordinates {
    (50, 0.29)(500, 0.41)(1000, 0.53)(1500, 0.54)(2000, 0.54)};

            \end{axis}

    \end{tikzpicture} 
        \caption{adduser}
    \end{subfigure}}
\resizebox{0.45\textwidth}{!}{\begin{subfigure}{\textwidth}
        \centering
\begin{tikzpicture}[scale=1]
        \begin{axis}[
            width=\textwidth ,
            height=8cm,
            xlabel={Sample Size},
    ylabel={F1-score},
    xmin=0, xmax=2000,
    ylabel style = {align=center},
        ymin=0, ymax=1, 
    ytick={0, 0.2, 0.4, 0.6,0.8,1},
    mark options={scale=1.5},
]
\addplot[
    color=black,
    domain=0:2000     , dashed
    ]
    {0.7548};

\addplot[
    color=blue,
    mark=pentagon,thick,
    ]
    coordinates {
    (50, 0.5570)(500, 0.7039)(1000, 0.7840)(1500, 0.77)(2000, 0.7493)};

\addplot[
    color={rgb:red,237;green,177;blue,32} ,
    mark=square,thick,
    ]
    coordinates {
    (50, 0.5364)(500, 0.512)(1000, 0.720)(1500, 0.7233)(2000, 0.7264)};

\addplot[
    color=orange ,
    mark=triangle,thick,
    ]
    coordinates {
    (50, 0.28)(500, 0.54)(1000, 0.64)(1500, 0.65)(2000, 0.56)};

\addplot[
    color= {rgb:red,45;green,160;blue,45} ,
    mark=star,thick,
    ]
    coordinates {
    (50, 0.27)(500, 0.41)(1000, 0.55)(1500, 0.57)(2000, 0.59)};
    
\addplot[
    color={rgb:red,160;green,20;blue,47},
    mark=Mercedes star,thick,
   ]
    coordinates {
    (50, 0.27)(500, 0.64)(1000, 0.70)(1500, 0.70)(2000, 0.68)};

            \end{axis}

    \end{tikzpicture}
        \caption{hydraFTP}
    \end{subfigure}}
\resizebox{0.45\textwidth}{!}{
\begin{subfigure}{\textwidth}
        \centering%
\begin{tikzpicture}[scale=1]
        \begin{axis}[
            width=\textwidth ,
            height=8cm,
            xlabel={Sample Size},
    ylabel={F1-score},
    xmin=0, xmax=2000,
    ylabel style = {align=center},
        ymin=0, ymax=1, 
    ytick={0, 0.2, 0.4, 0.6,0.8,1},
    mark options={scale=1.5},
]
\addplot[
    color=black,
    domain=0:2000     , dashed
    ]
    {0.7213};
    
\addplot[
    color=blue,
    mark=pentagon,thick,
    ]
    coordinates {
    (50,0.5176)(500, 0.6971)(1000, 0.7139)(1500, 0.6612)(2000, 0.746)};

\addplot[
    color={rgb:red,237;green,177;blue,32} ,
    mark=square,thick,
    ]
    coordinates {
    (50, 0.5085)(500,0.5368 )(1000, 0.6334)(1500, 0.5946)(2000, 0.626)};

\addplot[
    color=orange ,
    mark=triangle,thick,
    ]
    coordinates {
    (50, 0.17)(500, 0.6)(1000, 0.63)(1500, 0.63)(2000, 0.56)};

\addplot[
    color= {rgb:red,45;green,160;blue,45} ,
    mark=star,thick,
    ]
    coordinates {
    (50, 0.31)(500, 0.39)(1000, 0.42)(1500, 0.55)(2000, 0.65)};
    
\addplot[
    color={rgb:red,160;green,20;blue,47},
    mark=Mercedes star,thick,
   ]
    coordinates {
    (50, 0.33)(500, 0.6)(1000, 0.6)(1500, 0.65)(2000, 0.66)};

            \end{axis}

    \end{tikzpicture}
            

        \caption{hydraSSH}
    \end{subfigure}}
\resizebox{0.45\textwidth}{!}{
\begin{subfigure}{\textwidth}
        \centering
\begin{tikzpicture}[scale=1]
        \begin{axis}[
            width=\textwidth ,
            height=8cm,
            xlabel={Sample Size},
    ylabel={F1-score},
    xmin=0, xmax=2000,
    ylabel style = {align=center},
        ymin=0, ymax=1,
    ytick={0, 0.2, 0.4, 0.6,0.8,1},
    mark options={scale=1.5},
]
\addplot[
    color=black,
    domain=0:2000     , dashed
    ]
    {0.640};

\addplot[
    color=blue,
    mark=pentagon,thick,
    ]
    coordinates {
    (50, 0.3533)(500, 0.7525)(1000, 0.642)(1500, 0.6429)(2000, 0.6468)};
    
\addplot[
    color={rgb:red,237;green,177;blue,32} ,
    mark=square,thick,
    ]
    coordinates {
    (50, 0.3237)(500, 0.6443)(1000, 0.5649)(1500, 0.5787)(2000, 0.600)};

\addplot[
    color=orange ,
    mark=triangle,thick,
    ]
    coordinates {
    (50, 0.24)(500, 0.38)(1000, 0.57)(1500, 0.57)(2000, 0.51)};

\addplot[
    color= {rgb:red,45;green,160;blue,45} ,
    mark=star,thick,
    ]
    coordinates {
    (50, 0.27)(500, 0.36)(1000, 0.39)(1500, 0.49)(2000, 0.55)};
    
\addplot[
    color={rgb:red,160;green,20;blue,47},
    mark=Mercedes star,thick,
   ]
    coordinates {
    (50, 0.32)(500, 0.52)(1000, 0.54)(1500, 0.46)(2000, 0.53)};

\addplot[
    color=gray,
    mark=o,
    ]
    coordinates {
    (50, 1)(500, 1)(1000, 1)(1500, 1)(2000, 1)};

            \end{axis}

    \end{tikzpicture}
        \caption{java}
    \end{subfigure}}
\resizebox{0.45\textwidth}{!}{\begin{subfigure}{\textwidth}
        \centering
\begin{tikzpicture}[scale=1]
        \begin{axis}[
            width=\textwidth ,
            height=8cm,
            xlabel={Sample Size},
    ylabel={F1-score},
    xmin=0, xmax=2000,
    ylabel style = {align=center},
        ymin=0, ymax=1, 
    ytick={0, 0.2, 0.4, 0.6,0.8,1},
    mark options={scale=1.5},
]
\addplot[
    color=black,
    domain=0:2000     , dashed
    ]
    {0.679};
    
\addplot[
    color=blue,
    mark=pentagon,thick,
    ]
    coordinates {
    (50, 0.2018)(500, 0.6383)(1000, 0.6667)(1500, 0.6784)(2000, 0.6860)};

\addplot[
    color={rgb:red,237;green,177;blue,32} ,
    mark=square,thick,
    ]
    coordinates {
    (50, 0.2379)(500, 0.5299)(1000, 0.5205)(1500, 0.5520)(2000, 0.6272)};

\addplot[
    color=orange ,
    mark=triangle,thick,
    ]
    coordinates {
    (50, 0.15)(500, 0.24)(1000, 0.43)(1500, 0.37)(2000, 0.46)};

\addplot[
    color= {rgb:red,45;green,160;blue,45} ,
    mark=star,thick,
    ]
    coordinates {
    (50, 0.15)(500, 0.3)(1000, 0.29)(1500, 0.34)(2000, 0.41)};
    
\addplot[
    color={rgb:red,160;green,20;blue,47},
    mark=Mercedes star,thick,
   ]
    coordinates {
    (50, 0.18)(500, 0.33)(1000, 0.42)(1500, 0.52)(2000, 0.63)};

            \end{axis}

    \end{tikzpicture}
        \caption{ meter}
    \end{subfigure}}
\resizebox{0.45\textwidth}{!}{\begin{subfigure}{\textwidth}
        \centering
\begin{tikzpicture}[scale=1]
        \begin{axis}[
            width=\textwidth ,
            height=8cm,
            xlabel={Sample Size},
    ylabel={F1-score},
    xmin=0, xmax=2000,
    ylabel style = {align=center},
        ymin=0, ymax=1, 
    ytick={0, 0.2, 0.4, 0.6,0.8,1},
    mark options={scale=1.5},
]
\addplot[
    color=black,
    domain=0:2000     , dashed
    ]
    {0.602};
    
\addplot[
    color=blue,
    mark=pentagon,thick,
    ]
    coordinates {
    (50, 0.4534)(500, 0.5957)(1000, 0.6726)(1500, 0.6966)(2000, 0.6310)};

\addplot[
    color={rgb:red,237;green,177;blue,32} ,
    mark=square,thick,
    ]
    coordinates {
    (50, 0.3730)(500, 0.5081)(1000, 0.5571)(1500, 0.5704)(2000, 0.5447)};

\addplot[
    color=orange ,
    mark=triangle,thick,
    ]
    coordinates {
    (50, 0.04)(500, 0.31)(1000, 0.32)(1500, 0.46)(2000, 0.46)};

\addplot[
    color= {rgb:red,45;green,160;blue,45} ,
    mark=star,thick,
    ]
    coordinates {
    (50, 0.25)(500, 0.37)(1000, 0.37)(1500, 0.38)(2000, 0.46)};
    
\addplot[
    color={rgb:red,160;green,20;blue,47},
    mark=Mercedes star,thick,
   ]
    coordinates {
    (50, 0.41)(500, 0.52)(1000, 0.54)(1500, 0.50)(2000, 0.52)};
   
            \end{axis}

    \end{tikzpicture} 

        \caption{web}
    \end{subfigure}}
 \caption{F1-scores of \method and top four baselines (LightAD, Neurallog, LogRobust, and CNN) trained on varying data scarcity level on six different \task configurations (adduser, hydraFTP, hydraSSH, meter, java, and web) of \unstableadfa.\label{fig:data-efficiency}}
 \end{figure*}

\begin{table*}[htbp]
\centering

    \captionsetup[table]{skip=0pt}
    \captionof{table}{F1 score differences (in percentage points) and statistical testing results when comparing \method to baseline methods on the \unstableadfa dataset. 
    \label{tab:adfa-delta}}
    \centering
     \begin{threeparttable}[htbp]
 \resizebox{\linewidth}{!}{
\begin{tabular}{@{\hspace{0.5\tabcolsep}}c@{\hspace{0.5\tabcolsep}}c@{\hspace{0.5\tabcolsep}}c@{\hspace{0.5\tabcolsep}}c@{\hspace{0.5\tabcolsep}}c@{\hspace{0.5\tabcolsep}}c@{\hspace{0.5\tabcolsep}}c@{\hspace{0.5\tabcolsep}}c@{\hspace{0.5\tabcolsep}}c@{\hspace{0.5\tabcolsep}}c@{\hspace{0.5\tabcolsep}}c@{\hspace{0.5\tabcolsep}}}
\toprule
\textbf{Configuration} &\textbf{\datasize}& \multicolumn{4}{c}{\textbf{Supervised}} & \textbf{Semi-S} &  \multicolumn{4}{c}{\textbf{Unsupervised}}\\
\cmidrule(lr){3-6}
\cmidrule(lr){7-7}
\cmidrule(lr){8-11}
\\[-2ex]
& & LightAD & NeuralLog & LogRobust & CNN & PLELog & LogAnomaly & DeepLog & LogCluster & PCA \\
\midrule

\multirow{5}{*}{adduser} &50 &\SI{4}{  } & \SI{\fpeval{ 42 - 20} }{  } & \SI{\fpeval{ 42 - 29} }{  } & \SI{\fpeval{42 - 29 } }{  } & \SI{\fpeval{ 42 - 39} }{  } & \SI{\fpeval{42- 32 } }{  } & \SI{\fpeval{ 42 - 32} }{  }  & \SI{\fpeval{42 - 19}}{  } & \SI{\fpeval{42 - 23 } }{  }\\
&500&\SI{11}{  } & \SI{\fpeval{62-29 } }{  } & \SI{\fpeval{62 -34} }{  } & \SI{\fpeval{62-41 } }{  } & \SI{\fpeval{62 -38} }{  } & \SI{\fpeval{62-26 } }{  } & \SI{\fpeval{62 -18} }{  }  & \SI{\fpeval{62 -20} }{  } & \SI{\fpeval{62 -7} }{  }\\
&1000&\SI{13}{  } & \SI{\fpeval{71-60 } }{  } & \SI{\fpeval{71 -39} }{  } & \SI{\fpeval{71-53 } }{  } & \SI{\fpeval{71-49 } }{  } & \SI{\fpeval{71 -31} }{  } & \SI{\fpeval{71-26 } }{  }  & \SI{\fpeval{71 -22} }{  } & \SI{\fpeval{71-7 } }{  }\\
&1500&\SI{7}{  } & \SI{\fpeval{70 - 58 } }{  } & \SI{\fpeval{70 -41} }{  } & \SI{\fpeval{70-54 } }{  } & \SI{\fpeval{70 -30} }{  } & \SI{\fpeval{70 -23} }{  } & \SI{\fpeval{70-23 } }{  }  & \SI{\fpeval{70 -23} }{  } & \SI{\fpeval{70-6 } }{  } \\
&2000&\SI{10}{  } & \SI{\fpeval{72-49 } }{  } & \SI{\fpeval{72-51 } }{  } & \SI{\fpeval{72 -54} }{  } & \SI{\fpeval{72-49 } }{  } & \SI{\fpeval{72 -25} }{  } & \SI{\fpeval{72-22 } }{  }  & \SI{\fpeval{72 -24} }{  } & \SI{\fpeval{72-5 } }{  }  \\
 
\midrule  
\multirow{5}{*}{hydraFTP} &50  &\SI{2}{  } & \SI{\fpeval{55-28 } }{  } & \SI{\fpeval{55-27 } }{  } & \SI{\fpeval{55-35 } }{  } & \SI{\fpeval{55-35 } }{  } & \SI{\fpeval{55 -33} }{  } & \SI{\fpeval{55-33 } }{  }  & \SI{\fpeval{55 -29 } }{  } & \SI{\fpeval{55 -17} }{  } \\
&500&\SI{19}{  } & \SI{\fpeval{70 -54} }{  } & \SI{\fpeval{70-41 } }{  } & \SI{\fpeval{70-64 } }{  } & \SI{\fpeval{70 -22} }{  } & \SI{\fpeval{70-37 } }{  } & \SI{\fpeval{70 -35} }{  }  & \SI{\fpeval{70-32 } }{  } & \SI{\fpeval{70 -12} }{  } \\
&1000&\SI{6}{  } & \SI{\fpeval{78-64 } }{  } & \SI{\fpeval{78 -55} }{  } & \SI{\fpeval{78-70 } }{  } & \SI{\fpeval{78-46 } }{  } & \SI{\fpeval{78 -36} }{  } & \SI{\fpeval{78-36 } }{  }  & \SI{\fpeval{78-34 } }{  } & \SI{\fpeval{78 -14} }{  }  \\
&1500&\SI{5}{  } & \SI{\fpeval{77 -65} }{  } & \SI{\fpeval{77 -57} }{  } & \SI{\fpeval{77-70 } }{  } & \SI{\fpeval{77-30 } }{  } & \SI{\fpeval{77 -39} }{  } & \SI{\fpeval{77-38 } }{  }  & \SI{\fpeval{77-35 } }{  } & \SI{\fpeval{77-14 } }{  } \\
&2000&\SI{3}{  } & \SI{\fpeval{75 -56} }{  } & \SI{\fpeval{75-59 } }{  } & \SI{\fpeval{75-68 } }{  } & \SI{\fpeval{75-47 } }{  } & \SI{\fpeval{75 -40} }{  } & \SI{\fpeval{75 -41} }{  }  & \SI{\fpeval{75 -38} }{  } & \SI{\fpeval{75 -11} }{  } \\
\midrule

\multirow{5}{*}{hydraSSH} &50  &\SI{1}{  } & \SI{\fpeval{52 -17} }{  } & \SI{\fpeval{52 -31} }{  } & \SI{\fpeval{52 -33} }{  } & \SI{\fpeval{52 -46} }{  } & \SI{\fpeval{52-35 } }{  } & \SI{\fpeval{52-33 } }{  }  & \SI{\fpeval{52 -30} }{  } & \SI{\fpeval{52 -24} }{  } \\
&500&\SI{16}{  } & \SI{\fpeval{70 -60} }{  } & \SI{\fpeval{70 -39} }{  } & \SI{\fpeval{70-60 } }{  } & \SI{\fpeval{70-30 } }{  } & \SI{\fpeval{70 -38} }{  } & \SI{\fpeval{70 -41} }{  }  & \SI{\fpeval{70 -34} }{  } & \SI{\fpeval{70-18 } }{  }\\
&1000&\SI{8}{  } & \SI{\fpeval{71 -63 } }{  } & \SI{\fpeval{71 -42} }{  } & \SI{\fpeval{71-60 } }{  } & \SI{\fpeval{71-15 } }{  } & \SI{\fpeval{71 -34} }{  } & \SI{\fpeval{71 -40} }{  }  & \SI{\fpeval{71 -35} }{  } & \SI{\fpeval{71 -18} }{  }\\
&1500&\SI{7}{  } & \SI{\fpeval{67 - 63 } }{  } & \SI{\fpeval{67 -55} }{  } & \SI{\fpeval{67-65 } }{  } & \SI{\fpeval{67 -32 } }{  } & \SI{\fpeval{67 -36} }{  } & \SI{\fpeval{67 -37} }{  }  & \SI{\fpeval{67 -37} }{  } & \SI{\fpeval{67-15 } }{  } \\
&2000&\SI{12}{  } & \SI{\fpeval{74-56 } }{  } & \SI{\fpeval{74 -65} }{  } & \SI{\fpeval{74-66 } }{  } & \SI{\fpeval{74 -41 } }{  } & \SI{\fpeval{74 -40} }{  } & \SI{\fpeval{74- 43} }{  }  & \SI{\fpeval{74-37 } }{  } & \SI{\fpeval{74 -17} }{  }\\

\midrule

\multirow{5}{*}{java} &50 &\SI{3}{  } & \SI{\fpeval{35-24 } }{  } & \SI{\fpeval{35-27 } }{  } & \SI{\fpeval{35 -32} }{  } & \SI{\fpeval{35 -20} }{  } & \SI{\fpeval{35 -30} }{  } & \SI{\fpeval{35 -34} }{  }  & \SI{\fpeval{35-24 } }{  } & \SI{\fpeval{35-28 } }{  } \\
&500&\SI{11}{  } & \SI{\fpeval{75 -38} }{  } & \SI{\fpeval{75-36} }{  } & \SI{\fpeval{75 -52} }{  } & \SI{\fpeval{75-24 } }{  } & \SI{\fpeval{75-34 } }{  } & \SI{\fpeval{75 -28} }{  }  & \SI{\fpeval{75-27 } }{  } & \SI{\fpeval{75-27 } }{  }\\
&1000&\SI{6}{  } & \SI{\fpeval{64-57 } }{  } & \SI{\fpeval{64-39 } }{  } & \SI{\fpeval{64 -54} }{  } & \SI{\fpeval{64 -14} }{  } & \SI{\fpeval{64 -35} }{  } & \SI{\fpeval{64 -29 } }{  }  & \SI{\fpeval{64-28 } }{  } & \SI{\fpeval{64-17 } }{  } \\
&1500&\SI{6}{  } & \SI{\fpeval{64 -57 } }{  } & \SI{\fpeval{64-49 } }{  } & \SI{\fpeval{64 -46} }{  } & \SI{\fpeval{64-18 } }{  } & \SI{\fpeval{64 -35} }{  } & \SI{\fpeval{64 -29} }{  }  & \SI{\fpeval{64-29 } }{  } & \SI{\fpeval{64 -16} }{  } \\
&2000&\SI{5}{  } & \SI{\fpeval{65-51 } }{  } & \SI{\fpeval{65-55} }{  } & \SI{\fpeval{65-53 } }{  } & \SI{\fpeval{65-32 } }{  } & \SI{\fpeval{65-36 } }{  } & \SI{\fpeval{65-29 } }{  }  & \SI{\fpeval{65-29 } }{  } & \SI{\fpeval{65-14 } }{  }\\
\midrule

 \multirow{5}{*}{meter}&50&\SI{-4}{  } & \SI{\fpeval{20-15 } }{  } & \SI{\fpeval{20-15 } }{  } & \SI{\fpeval{20 -18} }{  } & \SI{\fpeval{20-13 } }{  } & \SI{\fpeval{20-20 } }{  } & \SI{\fpeval{20-19 } }{  }  & \SI{\fpeval{20-15 } }{  } & \SI{\fpeval{20 -10} }{  } \\
&500&\SI{11}{  } & \SI{\fpeval{63 -24} }{  } & \SI{\fpeval{63 -30} }{  } & \SI{\fpeval{63 -33} }{  } & \SI{\fpeval{63-12 } }{  } & \SI{\fpeval{63-18 } }{  } & \SI{\fpeval{63 -17} }{  }  & \SI{\fpeval{63-16 } }{  } & \SI{\fpeval{63-10 } }{  }  \\
&1000&\SI{15}{  } & \SI{\fpeval{66-43 } }{  } & \SI{\fpeval{66-29 } }{  } & \SI{\fpeval{66 -42} }{  } & \SI{\fpeval{66-8 } }{  } & \SI{\fpeval{66 -16} }{  } & \SI{\fpeval{66 -14} }{  }  & \SI{\fpeval{66 -17} }{  } & \SI{\fpeval{66-1 } }{  } \\
&1500&\SI{13}{  } & \SI{\fpeval{67-37 } }{  } & \SI{\fpeval{67 -34} }{  } & \SI{\fpeval{67 -52} }{  } & \SI{\fpeval{67-14 } }{  } & \SI{\fpeval{67 -17} }{  } & \SI{\fpeval{67-15 } }{  }  & \SI{\fpeval{67-19 } }{  } & \SI{\fpeval{67 -4} }{  } \\
&2000&\SI{6}{  } & \SI{\fpeval{68-46 } }{  } & \SI{\fpeval{68 -41} }{  } & \SI{\fpeval{68 -63} }{  } & \SI{\fpeval{68-15 } }{  } & \SI{\fpeval{68 -20} }{  } & \SI{\fpeval{68-22 } }{  }  & \SI{\fpeval{68 -22} }{  } & \SI{\fpeval{68-1 } }{  }\\
\midrule

\multirow{5}{*}{web} &50&\SI{8}{  } & \SI{\fpeval{45-4 } }{  } & \SI{\fpeval{45-25 } }{  } & \SI{\fpeval{45 -41} }{  } & \SI{\fpeval{45 -31} }{  } & \SI{\fpeval{45 -32} }{  } & \SI{\fpeval{45-32 } }{  }  & \SI{\fpeval{45-23 } }{  } & \SI{\fpeval{45 -25} }{  } \\
&500&\SI{9}{  } & \SI{\fpeval{60-31 } }{  } & \SI{\fpeval{60-37 } }{  } & \SI{\fpeval{60-52 } }{  } & \SI{\fpeval{60 -43} }{  } & \SI{\fpeval{60-23 } }{  } & \SI{\fpeval{60 -24} }{  }  & \SI{\fpeval{60-26 } }{  } & \SI{\fpeval{60 -13} }{  } \\
&1000&\SI{12}{  } & \SI{\fpeval{67-32 } }{  } & \SI{\fpeval{67-37 } }{  } & \SI{\fpeval{67-54 } }{  } & \SI{\fpeval{67-32 } }{  } & \SI{\fpeval{67 -27} }{  } & \SI{\fpeval{67-27 } }{  }  & \SI{\fpeval{67-26 } }{  } & \SI{\fpeval{67 -7} }{  } \\
&1500&\SI{13}{  } & \SI{\fpeval{70-46 } }{  } & \SI{\fpeval{70 -38} }{  } & \SI{\fpeval{70 -50} }{  } & \SI{\fpeval{70 -25} }{  } & \SI{\fpeval{70 -27} }{  } & \SI{\fpeval{70 -28} }{  }  & \SI{\fpeval{70-26 } }{  } & \SI{\fpeval{70 -12} }{  } \\
&2000&\SI{9}{  } & \SI{\fpeval{63 -46} }{  } & \SI{\fpeval{63 -46} }{  } & \SI{\fpeval{63 -52 } }{  } & \SI{\fpeval{63-21 } }{  } & \SI{\fpeval{63-31 } }{  } & \SI{\fpeval{63 -32} }{  }  & \SI{\fpeval{63-26 } }{  } & \SI{\fpeval{63 -10} }{  } \\
\midrule
\midrule
\multirow{5}{*}{average} &50 &\SI{2}{  } & \SI{\fpeval{23 } }{  }*& \SI{\fpeval{18 } }{  }*& \SI{\fpeval{10 } }{  }* & \SI{\fpeval{11 } }{  }* & \SI{\fpeval{11 } }{  }* & \SI{\fpeval{11 } }{  }*  & \SI{\fpeval{18 } }{  }* & \SI{\fpeval{20 } }{  }* \\
&500 &\SI{13}{  }* & \SI{\fpeval{27 } }{  } & \SI{\fpeval{30 } }{  } & \SI{\fpeval{16 } }{  }* & \SI{\fpeval{38 } }{  }* & \SI{\fpeval{37 } }{  }* & \SI{\fpeval{39 } }{  }*  & \SI{\fpeval{41 } }{  }* & \SI{\fpeval{52 } }{  }* \\
&1000&\SI{10}{  }* & \SI{\fpeval{16 } }{  }* & \SI{\fpeval{29 } }{  }* & \SI{\fpeval{14 } }{  }* & \SI{\fpeval{42 } }{  }* & \SI{\fpeval{40 } }{  }* & \SI{\fpeval{41 } }{  }*  & \SI{\fpeval{42 } }{  }* & \SI{\fpeval{59 } }{  }*  \\
&1500&\SI{8}{  }* & \SI{\fpeval{15 } }{  }* & \SI{\fpeval{23 } }{  }* & \SI{\fpeval{13 } }{  }* & \SI{\fpeval{44 } }{  }* & \SI{\fpeval{40 } }{  }* & \SI{\fpeval{41 } }{  }*  & \SI{\fpeval{41 } }{  }* & \SI{\fpeval{58 } }{  }* \\
&2000&\SI{7}{  }* & \SI{\fpeval{19 } }{  }* & \SI{\fpeval{17 } }{  }* & \SI{\fpeval{10 } }{  }* & \SI{\fpeval{35 } }{  }* & \SI{\fpeval{37 } }{  }* & \SI{\fpeval{38 } }{  }*  & \SI{\fpeval{40 } }{  }* & \SI{\fpeval{60 } }{  }*  \\

\bottomrule
\end{tabular}
}
\begin{tablenotes}
         \item [*] \method yields a significant higher F1-score than compared baseline.
         \end{tablenotes}
\end{threeparttable}
\end{table*} 

\begin{tcolorbox}[compactbox]
The answer to RQ2 is that \method outperforms baselines in F1 scores under \textbf{varying data scarcity levels} of \unstableadfa except at \datasize$=50$, where all methods perform poorly due to insufficient data. 
\end{tcolorbox}
\subsection{RQ3: Time and Memory Efficiency }\label{sec:results_rq3}

\begin{table*}[htbp]
\centering
    \captionsetup[table]{skip=0pt}
\captionof{table}{Training (T) and Inference (I) time of \method and the baselines (in seconds) on \unstableadfa, \unstablelogevol, \synhdfs, and \synlogevol. \label{tab:time-efficiency}}
\begin{threeparttable}[t]
 \resizebox{\linewidth}{!}{
\begin{tabular}{@{\hspace{0.5\tabcolsep}}c@{\hspace{0.5\tabcolsep}}c@{\hspace{0.5\tabcolsep}}c@{\hspace{0.5\tabcolsep}}c@{\hspace{0.5\tabcolsep}}c@{\hspace{0.5\tabcolsep}}c@{\hspace{0.8\tabcolsep}}c@{\hspace{0.5\tabcolsep}}c@{\hspace{0.5\tabcolsep}}c@{\hspace{0.5\tabcolsep}}c@{\hspace{0.5\tabcolsep}}c@{\hspace{0.5\tabcolsep}}c@{\hspace{0.5\tabcolsep}}c@{\hspace{0.5\tabcolsep}}c@{\hspace{0.5\tabcolsep}}c@{\hspace{0.5\tabcolsep}}c@{\hspace{0.5\tabcolsep}}c@{\hspace{0.5\tabcolsep}}}
\toprule
 & \multirow{2}{*}{\textbf{Config}}& \multirow{2}{*}{\textbf{M}} & \multicolumn{9}{c}{\textbf{Supervised}} & \textbf{Semi-S} & \multicolumn{4}{c}{\textbf{Unsupervised}}\\
\cmidrule(lr){4-12} 
\cmidrule(lr){13-13}
\cmidrule(lr){14-17}
& & & \method & Mistral & KNN & DT & SLFN & LightAD & NeuralLog& LogRobust& CNN & PLELog & LogAnomaly& DeepLog& LogCluster& PCA\\
\midrule
\multicolumn{17}{c}{\unstableadfa}\\
\midrule
& \multirow{2}{*}{adduser}& T & \num{16161} &\num{16160} & \num{2} & $<$\num{0.001} & $<$\num{0.001} & \num{5} & \num{922} & \num{221} & \num{271} & \num{285} & \num{276} & \num{184} & \num{4} & \num{0.017} \\ 

& & I & \num{0.836} & $<$\num{0.001} & $<$\num{0.001} & \num{0.005}& \num{0.842} & \num{0.012} & \num{0.664} & \num{0.234} & \num{0.164} & \num{0.211} & \num{0.025} & \num{0.012} & $<$\num{0.001} & $\ll 0.001$ \\  
\midrule

&  \multirow{2}{*}{hydraFTP}& T & \num{15906} & \num{15906} & \num{1} & $<$\num{0.001} & $<$\num{0.001}  & \num{4} & \num{920} & \num{220} & \num{269} & \num{285} & \num{276} & \num{181} & \num{4} & \num{0.017} \\ 

& & I & \num{0.818} & \num{0.813} & $<$\num{0.001} & $<$\num{0.001} & \num{0.005}& \num{0.014} & \num{0.635} & \num{0.229} & \num{0.165} & \num{0.211} & \num{0.024} & \num{0.011} & $<$\num{0.001} & $\ll 0.001$ \\  
\midrule

 & \multirow{2}{*}{hydraSSH}& T & \num{14147}& \num{14146} & \num{1} & $<$\num{0.001} & $<$\num{0.001} & \num{5} & \num{923} & \num{220} & \num{269} & \num{285} & \num{275} & \num{183} & \num{4} & \num{0.017} \\ 

&& I & \num{0.832}& \num{0.832} & $<$\num{0.001} & $<$\num{0.001} & \num{0.004} & \num{0.014} & \num{0.655} & \num{0.232} & \num{0.164} & \num{0.211} & \num{0.025} & \num{0.012} & $<$\num{0.001} & $\ll 0.001$ \\  
\midrule

 &\multirow{2}{*}{java}& T & \num{13933} & \num{13932} & \num{1} & $<$\num{0.001} & $<$\num{0.001} & \num{6} & \num{921} & \num{220} & \num{270} & \num{285} & \num{275} & \num{180} & \num{4} & \num{0.017} \\ 

&& I & \num{0.875} & \num{0.870} & $<$\num{0.001} & $<$\num{0.001} & \num{0.005}& \num{0.014} & \num{0.673} & \num{0.236} & \num{0.164} & \num{0.211} & \num{0.025} & \num{0.011} & $<$\num{0.001} & $\ll 0.001$ \\  
\midrule

 &\multirow{2}{*}{web}& T & \num{14079}& \num{14078} & \num{1} & $<$\num{0.001} & $<$\num{0.001} & \num{5} & \num{918} & \num{220} & \num{270} & \num{285} & \num{276} & \num{181} & \num{4} & \num{0.017}\\ 

&& I & \num{0.864} & \num{0.860} & $<$\num{0.001} & $<$\num{0.001} & \num{0.004} & \num{0.013} & \num{0.679} & \num{0.231} & \num{0.165} & \num{0.211} & \num{0.025} & \num{0.012} & $<$\num{0.001} & $\ll 0.001$ \\  
\midrule

 &\multirow{2}{*}{meter} & T & \num{15324}& \num{15323} & \num{1} & $<$\num{0.001} & $<$\num{0.001}  & \num{3} & \num{921} & \num{220} & \num{269} & \num{285} & \num{276} & \num{182} & \num{4} & \num{0.017} \\ 

&& I & \num{0.888} & \num{0.885}& $<$\num{0.001} & $<$\num{0.001} & \num{0.003}& \num{0.014} & \num{0.682} & \num{0.238} & \num{0.165} & \num{0.211} & \num{0.025} & \num{0.012} & $<$\num{0.001} & $\ll 0.001$ \\  
\midrule
\multicolumn{17}{c}{\unstablelogevol}\\
\midrule
 &\multirow{2}{*}{ Hadoop}& T & \num{4031}& \num{4031}& \num{6} & $<$\num{0.001} & $<$\num{0.001}  & \num{30} & \num{1550} & \num{178} & \num{316} & \num{48} & \num{1340} & \num{1020} & \num{21} & \num{0.180} \\ 

& & I & \num{0.435}& \num{0.414} & \num{0.003} & \num{0.001} & \num{0.017} & \num{0.067} & \num{0.275} & \num{0.078} & \num{0.077} & \num{0.072} & \num{0.004} & \num{0.001} & $<$\num{0.001} & $\ll 0.001$ \\  
\midrule

 & \multirow{2}{*}{Spark}& T & \num{23706}  & \num{23706} & \textit{N/A} & $\ll$\num{0.001} & $\ll$\num{0.001} & \num{0.2} & \num{712} & \num{232} & \num{136} & \num{220} & \num{944} & \num{514} & \num{9} & \num{0.015} \\ 

& & I & \num{0.844}  & \num{0.838}& \textit{N/A} & $<$\num{0.001} & \num{0.006}& \num{0.038} & \num{0.564} & \num{0.082} & \num{0.072} & \num{0.006} & \num{0.007} & \num{0.003} & $< 0.001$ & $\ll 0.001$ \\  
\midrule
\multicolumn{17}{c}{\synhdfs}\\
\midrule
 & \multirow{2}{*}{average}& T & \num{13587}& \num{13583} & \num{4} & $<$\num{0.001} & $<$\num{0.001}  & \num{22} & \num{1260} & \num{293} & \num{355} & \num{42} & \num{976} & \num{1110} & \num{20} & \num{0.067} \\ 

& & I & \num{0.771} & \num{0.771} & $\ll$\num{0.001} & $\ll$\num{0.001} & $<$\num{0.001} & \num{0.005} & \num{0.259} & \num{0.008} & \num{0.016} & \num{0.007} & \num{0.004} & \num{0.002} & $< 0.001$ & $\ll 0.001$ \\  
\midrule
\multicolumn{17}{c}{\synlogevol}\\
\midrule
 & \multirow{2}{*}{average}& T & \num{23706}& \num{23706} & \textit{N/A} & $<$\num{0.001} & $<$\num{0.001} & \num{0.2} & \num{712} & \num{232} & \num{136} & \num{220} & \num{944} & \num{514} & \num{9} & \num{0.015} \\ 

& & I & \num{0.988} & \num{0.979} & \textit{N/A} & $<$\num{0.001} & \num{0.009} & \num{0.038} & \num{0.703} & \num{0.116} & \num{0.076} & \num{0.008} & \num{0.013} & \num{0.004} & $< 0.001$ & $\ll 0.001$ \\  
\bottomrule
\end{tabular}
}
\end{threeparttable}

\end{table*} 
\subsubsection{Time Efficiency}
Table~\ref{tab:time-efficiency} reports the training and inference time of \method and the baselines across different datasets. Column ``M'' denotes the metric, where ``T'' represents the training time of full datasets (in seconds) and ``I'' represents the average inference time of one log sequence (in seconds). Specifically, we include \method's base models---one LLM (Mistral) and three ML models (KNN, DT, and SLFN), which are introduced in~\S~\ref{sec:flexlog-details}. The reported training and inference times represent the total time aggregated across these models (without considering parallel computation).

\method requires substantially more training time than the baselines, with a maximum of \num{23706} seconds (\num{6} hours and \num{35} minutes) on the \unstablelogevol Spark and \synlogevol datasets. In contrast, the maximum training time of the baselines is recorded as only \num{1550} seconds when training NeuralLog on the \unstablelogevol Hadoop dataset.  However, since training is a one-time cost, such a long training time is generally acceptable. 

\method's average inference time, per log sequence, remains below 1 second, ranging from \num{0.771} on \synhdfs to \num{0.988} on \synlogevol. While this is significantly higher than traditional ML and DL baselines---some of which make one prediction in milliseconds---it reflects the inherent trade-off between model complexity and efficiency. Simpler models such as PCA and LogCluster achieve near instantaneous inference but suffer from low detection effectiveness, failing to detect critical anomalies (\S~\ref{sec:results_rq1}). The inference time of \method is acceptable in user-oriented monitoring systems, where inference times on the order of seconds are generally tolerable. Examples include cloud service monitoring  (e.g., AWS CloudWatch ~\cite{diagboya2021infrastructure}, Microsoft Azure Monitor~\cite{collier2015microsoft}), where system logs are analyzed to detect performance issues; IT infrastructure monitoring (e.g., Prometheus~\cite{turnbull2018monitoring}),  where server and network health metrics are updated periodically; and industrial IoT monitoring, where manufacturing systems and smart grids detect equipment failures. In these cases, inference time or response time of an anomaly detector within seconds still allows for timely interventions. In contrast, latency-sensitive domains, such as high-frequency trading systems, demand millisecond-level inference time, as decisions must be made within microseconds to capitalize on market fluctuation~\cite{bello2024deep}. In such scenarios, the inference time of \method, which is on the order of seconds, is not acceptable. 

We remark that although \method is not as efficient as traditional methods, its superior effectiveness justifies its application in scenarios where reliability is paramount. In anomaly detection, false negatives---undetected anomalies---can have far greater consequences than minor delays in inference time.  For instance, in high-stakes environments like cybersecurity or critical infrastructure monitoring, failing to detect an anomaly or an intrusion attempt could lead to data breaches, financial losses, or system-wide failures. While simpler models offer faster inference, they often sacrifice effectiveness, leading to unacceptable numbers of false positives and false negatives. 

To further explain the time cost of \method, Table~\ref{tab:time-efficiency} shows the training and inference time of its base models, which include one LLM (Mistral) and three ML models (KNN, DT, and SLFN). We find that Mistral requires substantially more training time and average inference time per sequence compared to the three ML base models, accounting for the majority of time needed for \method. Specifically, while Mistral's training and inference time remain under 7 hours and \SI{1}{\s}, respectively, all ML base models take less than \SI{6}{\s} for training and less than \SI{0.001}{\s}  for inference per log sequence. This discrepancy is primarily due to Mistral's significantly large number of trainable parameters. As a result, Mistral has the most impact on the time efficiency of \method compared to ML models. However, with parallel computing and more powerful hardware, the efficiency of LLMs can easily be significantly improved, thereby enhancing the overall efficiency of \method.

\subsubsection{Memory Efficiency of \method's Cache Mechanism}\label{subsubsec:memoryefficiency}\method leverages a caching mechanism to improve efficiency during inference by storing predictions for previous log sequences, as described in~\S~\ref{subsec:cache}. The cache improves \method's time efficiency by reducing repetitive computation, although it introduces extra memory overhead. 
To evaluate the memory efficiency of this caching component, we report the memory usage, defined in~\S~\ref{sec:metrics}, during inference across our datasets. For~\unstableadfa, which contains the longest sequences with an average length of 461 templates, \method consumes between \SI{17.16}{\mega\byte} and \SI{19.60}{\mega\byte} of memory. For \unstablelogevol, which has the largest number of log sequences, the memory usage remains below \SI{4}{\mega\byte}---specifically, \SI{1.75}{\mega\byte} for Hadoop and \SI{3.5}{\mega\byte} for Spark. For our synthetic datasets, memory usage also stays below  \SI{1}{\mega\byte}, averaging \SI{576}{\kilo\byte} and \SI{2.88}{\mega\byte} for \synhdfs\ and \synlogevol, respectively, across different injection ratios. Overall, this level of memory usage demonstrates the scalability potential of \method’s caching mechanism, especially when balanced against the time savings presented as part of the answer to RQ4 (\S~\ref{subsubsec:ablation_cache}). Regarding scalability in systems with extremely long or highly diverse sequences, we note that the cache’s \textit{delete} function helps keep memory consumption under control (see~\S~\ref{subsec:cache} for details). Further analysis of memory efficiency will depend on the specific memory resources available in the monitoring system during inference.

\begin{tcolorbox}[compactbox]
The answer to RQ3 is that while \method is \textbf{not the most time-efficient} in \task inference, it processes each log sequence within \textbf{\SI{1}{\s} on average}. 
\method's cache memory remains \textbf{below \SI{4}{\mega\byte}} for most datasets (up to \SI{19.6}{\mega\byte} for \unstableadfa), confirming its \textbf{memory efficiency}. 
\end{tcolorbox}

\subsection{RQ4: Configuration Impact}\label{sec:results_rq4}
In this section, we investigate the impact of  \method's different ablation configurations, such as excluding the cache mechanism (\S~\ref{subsubsec:ablation_cache}), excluding RAG (\S~\ref{subsubsec:ablation_rag}) and different choices of base models in ensemble learning (\S~\ref{subsubsec:ablation_ensemble}). Additionally, we investigate the impact of alternative LLMs in the configuration of \method (\S~\ref{subsubsec:alternative-llms}) to highlight the advantages of using Mistral Small as the LLM component. 
\subsubsection{Impact of Cache-empowered Inference}
\label{subsubsec:ablation_cache}
\begin{table*}[htbp]
\centering
    \captionsetup[table]{skip=0pt}
\captionof{table}{\method vs. \method w/o cache --- Comparisons of inference time per log sequence (in seconds) on  \unstableadfa, \unstablelogevol, \synlogevol, and \synhdfs \label{tab:flexlog-cache}}
     \begin{threeparttable}[htbp]
 \resizebox{\linewidth}{!}{
\begin{tabular}{@{\hspace{1.5\tabcolsep}}c@{\hspace{1.5\tabcolsep}}c@{\hspace{1.5\tabcolsep}}c@{\hspace{1.5\tabcolsep}}c@{\hspace{1.5\tabcolsep}}c@{\hspace{1.5\tabcolsep}}c@{\hspace{1.5\tabcolsep}}c@{\hspace{1.5\tabcolsep}}c@{\hspace{1.5\tabcolsep}}c@{\hspace{1.5\tabcolsep}}c@{\hspace{1.5\tabcolsep}}c@{\hspace{1.5\tabcolsep}}c@{\hspace{1.5\tabcolsep}}c@{\hspace{1.5\tabcolsep}}}
\toprule
\multirow{2}{*}{\textbf{Config}} &\multicolumn{7}{c}{\textbf{\unstableadfa}} & \multicolumn{3}{c}{\textbf{\unstablelogevol}} & \multicolumn{1}{c}{\textbf{\synlogevol}} & \multicolumn{1}{c}{\textbf{\synhdfs}}\\
\cmidrule(lr){2-8}
\cmidrule(lr){9-11}
\cmidrule(lr){12-12}
\cmidrule(lr){13-13}
& adduser & hydraFTP & hydraSSH & java & meter & web & average& Hadoop & Spark & average & average & average\\
\midrule

 \multirow{1}{*}{\method} & \num{0.842} & \num{0.818} & \num{0.832}& \num{0.875} & \num{0.864} & \num{0.888} &\textbf{\num{0.853}} & \num{0.435} & \num{0.844} &\textbf{ \num{0.628}} & \textbf{\num{0.941}} & \textbf{\num{0.771}} \\
 \midrule
 \multirow{1}{*}{\method w/o cache} & \num{0.896} & \num{0.861} & \num{0.898} & \num{0.916} & \num{0.909} & \num{0.952} & \num{0.905} & \num{0.793} & \num{1.086} & \num{0.940}  & \num{0.988}  & \num{0.794} \\
 \midrule
 \midrule
Difference (\si{\s}) & -0.054 & -0.043 & -0.066 & -0.041 & -0.045 & -0.064 & -0.052$^*$ & -0.358 & -0.242 & -0.312$^*$ & -0.047$^*$ & -0.023\\
\bottomrule
\end{tabular}
}
\begin{tablenotes}
         \item [$*$] \footnotesize{\method yields a significant lower inference time than \method without cache.}
        \end{tablenotes}
\end{threeparttable}

\end{table*} 
\method maintains a cache $C$ to avoid redundant predictions incurred by recurring log sequences, thus improving inference efficiency. To assess the impact of $C$ on inference time, we compare \method and \method without cache (denoted by ``w/o cache'') in terms of inference time across four datasets: \unstableadfa, \unstablelogevol, \synlogevol, and \synhdfs. The results of this comparison are reported in Table ~\ref{tab:flexlog-cache}. Column ``Config'' indicates the subjects under comparison --- \method and \method w/o cache; the following four columns ``\unstableadfa'', ``\unstablelogevol'', ``\synlogevol'' and ``\synhdfs'' report the results of inference time per log sequence (in seconds) for each respective dataset. For the real-world datasets (\unstableadfa and \unstablelogevol), we report inference time for each configuration. In contrast, for the synthetic datasets (\synlogevol and \synhdfs), we provide only the average inference time across configurations. This is because different configurations of \synlogevol and \synhdfs vary in the injection ratio of changes, which has minimal impact on inference time, leading to similar inference time across configurations. Hence, we report only the average inference time on synthesized datasets for brevity.  The last row ``Difference'' indicates the difference between 
 the inference time of \method and that for \method w/o cache. Similar to RQ1 (\S~\ref{sec:results_rq1}), we conducted Mann-Whitney U tests on each dataset to assess the statistical significance of the differences in inference times; the symbol ``*'' indicates the inference time of \method is significantly lower than that of \method w/o cache.

We observe that the introduction of the cache consistently reduces the inference time across all datasets. The reduction on two real-world datasets---\unstableadfa and \unstablelogevol---and \synlogevol are more pronounced than \synhdfs, with reductions of \num{0.052}, \num{0.312}, and \num{0.047}, respectively. Mann-Whitney U tests confirm these reductions are statistically significant, whereas the reduction on \synhdfs is smaller (\num{ 0.023}) and statistically not significant. However, as discussed in \S~\ref{sec:results_rq3}, the practical impact of caching is limited in user-oriented monitoring systems, where inference times below 1 second are generally considered acceptable. While caching improves efficiency, the reduction in inference time might be too small to make a noticeable difference in such systems. 

That said, we remark that the impact of caching previously seen log sequences depends on how frequently identical log sequences reappear in real-world systems. In our experiments, \SI{6}{\percent}, \SI{43}{\percent}, \SI{2.7}{\percent}, and \SI{2.2}{\percent} of log sequences appear more than once in the testing dataset of \unstableadfa, \unstablelogevol, \synlogevol, and \synhdfs, respectively. Hence, the average reduction of inference time on each dataset, ordered from the greatest to the least, is as follows: \unstablelogevol (\num{-0.312} seconds), \unstableadfa (\num{-0.052} seconds), \synlogevol (\num{-0.047} seconds), and \synhdfs (\num{-0.023} seconds). 
In practice, in many operational environments, such as distributed cloud computing frameworks (e.g., Hadoop~\cite{hadoop} and Spark~\cite{spark}) and security monitoring systems, certain types of log sequences, such as scheduled job reports and system diagnostics, occur repeatedly over time. In these cases, the cache mechanism of \method can significantly improve inference efficiency, but its impact depends on the extent of redundant computations and the acceptable latency requirements of the system. 

\subsubsection{Impact of RAG}
\label{subsubsec:ablation_rag}
\begin{table*}[htbp]
\centering
    \captionsetup[table]{skip=0pt}
\captionof{table}{\method vs. \method w/o RAG --- Comparisons in terms of F1 score (in percentage points) on the \unstableadfa dataset. \label{tab:adfa-rag}}
     \begin{threeparttable}[htbp]
 \resizebox{\linewidth}{!}{
\begin{tabular}{@{\hspace{1.5\tabcolsep}}c@{\hspace{1.5\tabcolsep}}c@{\hspace{1.5\tabcolsep}}c@{\hspace{1.5\tabcolsep}}c@{\hspace{1.5\tabcolsep}}c@{\hspace{1.5\tabcolsep}}c@{\hspace{1.5\tabcolsep}}c@{\hspace{1.5\tabcolsep}}c@{\hspace{1.5\tabcolsep}}}
\toprule
\multirow{1}{*}{\textbf{Config}}  &\multicolumn{1}{c}{\textbf{adduser}} & \multicolumn{1}{c}{\textbf{hydraFTP}} & \multicolumn{1}{c}{\textbf{hydraSSH}} & \multicolumn{1}{c}{\textbf{java}} & \multicolumn{1}{c}{\textbf{meter}} & \multicolumn{1}{c}{\textbf{web}} & \textbf{Average}\\
\midrule

 \multirow{1}{*}{\method}
& \num{71.8} & \num{78.4} & \num{72.3} & \num{64.2} & \num{68.2} & \num{67.2} & \textbf{\num{70.4}}  \\
\midrule

 \multirow{1}{*}{\method w/o RAG}
 &\num{68.8} & \num{70.5} & \num{64.8} & \num{62.2} & \num{65.9} & \num{64.1} & 66.0\\
\midrule
\midrule
Difference (pp) & 3.0 & 7.9 & 7.5 & 2.0 & 2.3 & 3.1 & 4.4$^*$\\
\bottomrule
\end{tabular}
}
\begin{tablenotes}
         \item [$*$] \footnotesize{\method yields a significant higher F1-score than \method without RAG.}
        \end{tablenotes}
\end{threeparttable}
\end{table*} 
\method employs RAG to retrieve context for log sequences, providing relevant information that enriches the prompts. In our experiments, only \unstableadfa has available contextual information, specifically Linux system call descriptions. Hence, RAG is only applied in the experiments on the \unstableadfa dataset. To assess the impact of RAG, we compare the F1 scores of \method and \method without RAG (denoted by ``w/o RAG'') on \unstableadfa. The results of this comparison are reported in Table~~\ref{tab:adfa-rag}. Column ``Config'' represents the subjects under comparison --- \method and \method w/o RAG. Column ``adduser'', ``hydraFP'', ``hydraSSH'', ``java'', ``meter'', and ``web'' reports the F1 scores of six different configurations of \unstableadfa; configuration details are described in \S~\ref{sec:datasset-config}. The last column ``Average'' indicates the average F1 score across all six configurations; the last row ``Difference (\si{pp})'' reports the difference between the F1 scores of \method and those of \method w/o RAG. We performed a Mann-Whitney U test to assess the significance of the differences; however,  testing on individual configurations was not feasible since we repeated the experiments on each configuration 5 times, which does not provide sufficient statistical power. Instead, a Mann-Whitney U test was run across all six configurations, increasing the test sample size to 30.  

The results in Table~\ref{tab:adfa-rag} show that \method outperforms \method w/o RAG on all six configurations of \unstableadfa, with differences ranging from \SI{2}{\pp} to \SI{7.9}{\pp}. On average, RAG improves the F1 score of \task from \num{0.660} to \num{0.704} (\SI{4.4}{\pp}). A Mann-Whitney U test confirms this improvement is statistically significant. We conclude that the RAG component of \method is effective in improving F1 scores on \unstableadfa. This suggests that RAG could similarly improve performance on other datasets that have contextually rich log information. For example, in cloud computing platforms (e.g., AWS or Azure), where logs often include metadata such as instance IDs, resource types, or service configurations, RAG could enhance the effectiveness of \task by providing context that helps identify patterns or anomalies related to specific resources or services.

\subsubsection{Impact of Base Model Choices in Ensemble Learning}
\label{subsubsec:ablation_ensemble}
\method employs an ensemble of four base models for \task, leveraging diverse perspectives to enhance predictive effectiveness. As described in \S~\ref{sec:implementaitondetails}, the ensemble in \method includes one LLM (Mistral) and three ML models (KNN, DT, and SLFN), which were selected empirically due to their superior performance across different datasets. In this section, we present the results of different base model choices on \unstableadfa, \unstablelogevol, \synhdfs, and \synlogevol, including ablation studies (e.g., removing individual base models and removing all ML base models) and replacing Mistral with alternative LLMs (Llama 3.1 and GPT-4o). 

\noindent\textbf{Ablation Study} Table~\ref{tab:ablation} reports ablation studies of \method on all \task configurations. Column ``Config'' denotes the configuration of Flexlog; for instance, ``w/o SLFP'' represents \method without SLFP base model, and ``w/o ML'' represents \method without the three ML models (KNN, DT, and SLFN), resulting in standalone Mistral.
To evaluate the individual contribution of each base model, we first assessed four configurations of \method, each obtained by removing a single base model: \method w/o Mistral, w/o KNN, w/o SLFN, and w/o DT. The results in Table ~\ref{tab:ablation} indicate that removing Mistral leads to the most significant drop in F1 scores, with a maximum reduction of \SI{6.7}{\pp} on \unstableadfa adduser. Removing any of the ML models (KNN, DT, or SLFN) also results in reduced F1 scores across all configurations in all the datasets.
To assess the statistical significance of these reductions, we conducted Mann-Whitney U tests on each dataset. The results confirm that the decreases in F1 scores are statistically significant in all cases, except for \method w/o KNN, w/o DT, and w/o SLFN on \unstablelogevol. This highlights the effectiveness of each ML base model in contributing to \method’s overall effectiveness. Further, we assess the contribution of all ML models by excluding all three ML models from \method, leaving only Mistral for predictions. This resulted in decreased F1 scores across most datasets, except for \unstablelogevol Spark and \synlogevol, where ``w/o ML'' outperformed \method by \SI{7}{\pp} ($96.2\%-89.2\%$) and \SI{0.8}{\pp} ($97.9\%-97.1\%$), respectively. Mann-Whitney U tests show that \method significantly outperforms ``w/o ML'' in terms of F1 score on \unstableadfa and \synhdfs. The F1 score difference on \synlogevol between \method and ``w/o ML'' is statistically not significant, whereas on \unstablelogevol Spark, "w/o ML" achieves a significantly higher F1 score than \method. These results align with the findings from removing individual ML base models, further suggesting that the ML base models contribute negatively to \method's performance on \unstablelogevol Spark and 
, leading to ``w/o ML'' outperforming the ensemble.  

A likely explanation for the better performance of \method without any ML base models on \unstablelogevol Spark and \synlogevol is these datasets' extreme class imbalance. As detailed in \S~\ref{sec:datasets}, \unstablelogevol Spark and \synlogevol share the same training dataset, sampled from LogEvol Spark 2, which is highly imbalanced: only \SI{16}{\percent} of log sequences in the sampled training dataset are anomalous, compared to \SI{50}{\percent} in \unstableadfa and \unstablelogevol Hadoop, and \SI{57}{\percent} in \synhdfs. It is well known that traditional ML models, such as KNN, SLFN, and DT, struggle with highly imbalanced datasets~\cite{imbalanceddata}. Since \method integrates these ML models, its performance is negatively affected.

To mitigate data imbalance, we experimented with both down-sampling and over-sampling techniques~\cite{mohammed2020machine}. Down-sampling by reducing normal logs to match the number of anomalous logs led to a decrease of \SI{5}{\pp} in F1 score. For over-sampling, we applied several standard strategies, including duplicating anomalous logs, adding small perturbations to representation vectors, and using the Synthetic Minority Oversampling Technique (SMOTE)~\cite{smote} on representation vectors. However, none of these approaches improved the effectiveness of the ML models. The key challenge lies in the representation of log sequences as count vectors. Unlike continuous feature spaces where interpolation can generate plausible synthetic samples, count vector representations encode categorical relationships, making common over-sampling techniques such as SMOTE~\cite{Chawla_2002} prone to producing unrealistic or noisy data. In contrast, standalone Mistral (``w/o ML'') remains robust in predictive effectiveness, leveraging its pretraining on vast and diverse corpora to mitigate data imbalance. Based on these findings, we \emph{recommend} using standalone Mistral instead of the full \method when dealing with extremely imbalanced training datasets, eliminating the negative effect of poorly trained ML models. 

For future improvements, more advanced data augmentation techniques, such as Generative Adversarial Networks (GANs), could be used to generate synthetic log sequences directly, rather than modifying their count vector representations. This avoids the issue of unrealistic or noisy data caused by over-sampling in a discrete feature space, making the synthetic data more useful for training ML models on imbalanced datasets like \unstablelogevol Spark. As a result, this could potentially improve \method’s overall performance.

\noindent \textbf{Alternative LLMs. }\label{subsubsec:alternative-llms} Table~\ref{tab:LLM-alternative} reports the F1-score of \method with three LLM choices. Column ``Config'' represents different configurations of \method, wherein Mistral is replaced by Llama 3.1 8B (``\textit{$Mistral \rightarrow Llama$}'') and GPT-4o (``\textit{$Mistral \rightarrow GPT$}''). Column ``Source'' indicates whether the employed LLM is open-source or closed-source. Experimental results indicate that GPT-4o and Mistral achieve comparable effectiveness, both consistently outperforming Llama across all configurations and datasets. Mann-Whitney U tests at the dataset level confirm that the F1 score differences between \method and `` \textit{$Mistral \rightarrow GPT$}'' are not significant across all four datasets. However, \method significantly outperforms ``\textit{$Mistral \rightarrow Llama$}'' on \unstableadfa, \unstablelogevol, and \synhdfs, while the difference on \synlogevol is not statistically significant. These findings underscore the effectiveness of Mistral for \task, demonstrating performance on par with the closed-source GPT-4o while avoiding the financial cost associated with API invoking. Thus, we recommend Mistral as a cost-effective yet competitive base model for \method.  
\begin{table*}[t!]
\centering
    \captionsetup[table]{skip=0pt}
\captionof{table}{Ablation studies of \method on all unstable datasets. \label{tab:ablation}}
    \begin{threeparttable}[htbp]
 \resizebox{\linewidth}{!}{
\begin{tabular}{@{\hspace{0.5\tabcolsep}}c@{\hspace{0.5\tabcolsep}}c@{\hspace{0.8\tabcolsep}}c@{\hspace{0.8\tabcolsep}}c@{\hspace{0.8\tabcolsep}}c@{\hspace{0.8\tabcolsep}}c@{\hspace{0.8\tabcolsep}}c@{\hspace{0.8\tabcolsep}}c@{\hspace{0.8\tabcolsep}}c@{\hspace{0.8\tabcolsep}}c@{\hspace{0.8\tabcolsep}}c@{\hspace{0.8\tabcolsep}}c@{\hspace{0.8\tabcolsep}}c@{\hspace{0.8\tabcolsep}}c@{\hspace{0.8\tabcolsep}}}
\toprule
\multirow{2}{*}{\textbf{Config}} & \multicolumn{7}{c}{\textbf{\unstableadfa}} & \multicolumn{3}{c}{\textbf{\unstablelogevol}} & \multirow{1}{*}{\textbf{\synhdfs}} & \multirow{1}{*}{\textbf{\synlogevol}} \\
\cmidrule(lr){2-8} 
\cmidrule(lr){9-11}
\cmidrule(lr){12-12}
\cmidrule(lr){13-13}
& adduser& hydraFTP  & hydraSSH & java & meter & web & average & Hadoop & Spark & average & average & average\\
\midrule
\method & \textbf{\num{0.718}} & \textbf{\num{0.784}} &  \textbf{\num{0.723}} &  \textbf{\num{0.642}} &  \textbf{\num{0.682}} &  \textbf{\num{0.672}} & \textbf{\num{0.704}} & \num{0.982} &  \num{0.892} & \num{0.937}& \textbf{\num{0.972}} & \num{0.971}\\ 
\midrule
\midrule
w/o Mistral & \num{0.645} & \num{0.769} & \num{0.692} &  \num{0.628} & \num{0.639} &  \num{0.616} &  \num{0.664}* &  \num{0.975} &  \num{0.841} &  \num{0.908}* &  \num{0.939}* & \num{0.936}*\\ 
\midrule
w/o KNN & \num{0.683} & \num{0.768} & \num{0.654} & \num{0.621} & \num{0.651} & \num{0.579} &  \num{0.659}* &  \num{0.980} &  \textit{N/A} &  \num{0.980} &  \num{0.945}* & \textit{N/A}\\ 
\midrule
w/o DT & \num{0.667} & \num{0.749} & \num{0.690} & \num{0.640} & \num{0.654} & \num{0.623} &  \num{0.670}* &  \num{0.973} &  \num{0.875} &  \num{0.924} &  \num{0.948}* &\num{0.959}*\\ 
\midrule
w/o SLFN & \num{0.677} & \num{0.691} & \num{0.613} & \num{0.641} & \num{0.647} & \num{0.556} & \num{0.637}* & \num{0.978} &  \num{0.871} &  \num{0.924} &  \num{0.934}* &  \num{0.948}*\\ 
\midrule

w/o ML & \num{0.579} & \num{0.591} & \num{0.630} & \num{0.628} & \num{0.674} & \num{0.571} & \num{0.612}* &  \textbf{\num{0.998}} &  \textbf{\num{0.962}} &  \textbf{\num{0.980}}$^\dagger$ &  \num{0.928}* & \textbf{\num{0.979}}\\ 
\bottomrule
\end{tabular}
}

\begin{tablenotes}
         \item [$*$] \footnotesize{\method yields a significant higher F1-score than the ablation configuration.}
         \item [$\dagger$] \footnotesize{\method yields a significant lower F1-score than the ablation configuration.}
          \item [\textit{N/A}] \footnotesize{Not applicable. KNN is excluded from \method on the \synhdfs and \unstablelogevol datasets (see~\S~\ref{sec:flexlog-details})}
        \end{tablenotes}
\end{threeparttable}
\end{table*} 

\begin{table*}[ht!]
\centering
    \captionsetup[table]{skip=0pt}
\captionof{table}{F1 scores of using alternative LLMs in \method. \label{tab:LLM-alternative}}
     \begin{threeparttable}[ht!]
 \resizebox{\linewidth}{!}{
\begin{tabular}{@{\hspace{1.5\tabcolsep}}c@{\hspace{1.5\tabcolsep}}c@{\hspace{1.5\tabcolsep}}c@{\hspace{1.5\tabcolsep}}c@{\hspace{1.5\tabcolsep}}c@{\hspace{1.5\tabcolsep}}c@{\hspace{1.5\tabcolsep}}c@{\hspace{1.5\tabcolsep}}c@{\hspace{1.5\tabcolsep}}c@{\hspace{1.5\tabcolsep}}c@{\hspace{1.5\tabcolsep}}c@{\hspace{1.5\tabcolsep}}c@{\hspace{1.5\tabcolsep}}c@{\hspace{1.5\tabcolsep}}c@{\hspace{1.5\tabcolsep}}}
\toprule
\multirow{1}{*}{\textbf{Config}} & \multirow{1}{*}{\textbf{Source}} &\multicolumn{7}{c}{\textbf{\unstableadfa}} & \multicolumn{3}{c}{\textbf{\unstablelogevol}}  & \multicolumn{1}{c}{\textbf{\synhdfs}} & \multicolumn{1}{c}{\textbf{\synlogevol}}\\

\cmidrule(lr){3-9}
\cmidrule(lr){10-12}
\cmidrule(lr){13-13}
\cmidrule(lr){14-14}

&& adduser & hydraFTP & hydraSSH & java & meter & web & average & Hadoop & Spark & average & average & average\\
\midrule

 \multirow{1}{*}{\method} & open &\num{0.718} & \num{0.784} &  \textbf{\num{0.723}} &  \num{0.642} &  \num{0.682} &  \num{0.672} & \num{0.704} & \num{0.982} &  \num{0.892} & \textbf{\num{0.937}}& \num{0.972} & \textbf{\num{0.971}}\\
 \midrule
 \midrule
 \multirow{1}{*}{\textit{$Mistral \rightarrow Llama$}}& open & \num{0.679} & \num{0.743} & \num{0.707} & \num{0.593} & \num{0.669} & \num{0.591} & \num{0.664}* & \num{0.941}&\num{0.850} & \num{0.895}*  & \num{0.949}*  & \num{0.970} \\
\midrule
 \multirow{1}{*}{\textit{$Mistral \rightarrow GPT$}} & closed &  \textbf{\num{0.721}} & \textbf{\num{0.786}} & \num{0.718} & \textbf{\num{0.648}} & \textbf{\num{0.690}} & \textbf{\num{0.696}} & \textbf{\num{0.710}} & \textbf{\num{0.981}} & \num{0.886} &\num{0.933}  & \textbf{\num{0.976}} & \textbf{\num{0.971}}\\
\bottomrule
\end{tabular}
}
\begin{tablenotes}
         \item [$*$] \footnotesize{\method yields a significant higher F1 score compared to using the alternative LLM.}
        \end{tablenotes}
\end{threeparttable}
\end{table*}

\begin{tcolorbox}[compactbox]
The answer to RQ4 is that the full \method configuration---combining \textbf{cache}, \textbf{RAG}, and an ensemble of \textbf{KNN, DT, SLFN, and Mistral}---performs best, with each component improving efficiency or effectiveness. 
Among LLMs, \textbf{Mistral} is a cost-effective choice, matching the performance of GPT-4o while outperforming Llama.
\end{tcolorbox}

\subsection{Discussion} \label{sec:discussion}
In this section, to guide AIOps engineers, we highlight the implications of our study for \task.

\subsubsection{Token Consumption of LLMs}
\label{subsubsec:token_consumption}
In the early times of leveraging attentive language models, language models accepted limited input tokens, such as a maximum of \num{512} input tokens for BERT~\cite{bert}, making it challenging to apply them to long log sequences. However, as language models expanded into LLMs, the maximum input token limit also increased, both for open-source and closed-source models. 
Table~\ref{tab:token-limit} reports the input token limits of the LLMs used in our work and compares them with the maximum token consumption observed for each dataset. Column ``Model'' denotes LLMs used in our experiments. Column ``Source'' indicates whether the LLM is open-sourced or closed-source. Column ``Input Token Limit'' reports the input token limit specific to each LLM. Column ``Maximum Input Token'' denotes the maximum token consumption of \method's prompts on each dataset. We note that these values vary across LLMs due to variations in their tokenization processes. Our results show that \method's token consumption remains well within input token limits for all LLMs, indicating that the challenge of input token limits has been alleviated and even eliminated for our datasets. The highest usage is only \SI{37.7}{\percent} (\num{24735} out of the maximum limit of \num{65536} tokens), with prompts used by GPT-4o for \unstableadfa.  Notably, in our experiments, we have included various log datasets with log sequences as long as \num{4474} templates (see Table~\ref{tab:datasets}), and yet LLMs effectively accommodate them. Moreover, recent advances in LLMs can potentially extend the token limit to one million tokens~\cite{qwen2.5-1m, qwen2.5},  making it no longer a hard limitation for using LLMs.

\begin{table*}[hbt!]
\centering    
\captionsetup[table]{skip=0pt}
\captionof{table}{Overview of token consumption of \unstableadfa, \unstablelogevol, \synlogevol, and \synhdfs on open-source and closed-source LLMs \label{tab:token-limit}}
    \begin{threeparttable}[htbp]
 \resizebox{\linewidth}{!}{
\begin{tabular}{@{\hspace{1.5\tabcolsep}}c@{\hspace{1.5\tabcolsep}}c@{\hspace{1.5\tabcolsep}}c@{\hspace{1.5\tabcolsep}}c@{\hspace{1.5\tabcolsep}}c@{\hspace{1.5\tabcolsep}}c@{\hspace{1.5\tabcolsep}}c@{\hspace{1.5\tabcolsep}}}
\toprule

\multirow{2}{*}{\textbf{Model}} &\multirow{2}{*}{\textbf{Accessibility}} & \multirow{2}{*}{\textbf{\makecell{Input\\Token Limit}}} & \multicolumn{4}{c}{\textbf{Maximum Input Token}}\\
\cmidrule(lr){4-7}
&&& \textbf{\unstableadfa} & \textbf{\unstablelogevol} & \textbf{\synhdfs} & \textbf{\synlogevol}\\
\midrule

 \multirow{1}{*}{\textit{Mistral Small}} & open & \num{128000} & \num{25848} & \num{28934} & \num{1243}& \num{16840} \\
 \midrule

 \multirow{1}{*}{\textit{Llama 3.1 8B}} & open & \num{128000} & \num{21371} & \num{24798} & \num{862}& \num{12533} \\
 \midrule

 \multirow{1}{*}{\textit{GPT-4o}} & closed & \num{65536 } & \num{21383} & \num{24735} & \num{860} &\num{12479} \\

\bottomrule
\end{tabular}
}
\end{threeparttable}
\end{table*}

\subsubsection{Error Analysis}
\label{subsubsec:discussion_ea}
In RQ4, we observed that removing any base model from \method generally decreases effectiveness. To better understand the role of the LLM base model (i.e., Mistral) and ML base models (i.e., KNN, DT, and SLFN), in this section, we analyze the log sequences that \method correctly classify but mis-classify when either Mistral or ML models are removed. We denote these mis-classified log sequences as \woMistralError and \woMLError, respectively. We investigate, for each dataset, whether \woMistralError and \woMLError differ in various characteristics to highlight the unique contribution of LLM and ML to the effectiveness of \method. 
Specifically, we focus on the unseen log templates in \woMistralError and \woMLError because unseen log templates pose a key challenge in \task. Traditional anomaly detectors often rely on predefined patterns and struggle with generalization; in contrast,  \method, by means of the integration of LLM, may be better at detecting anomalies involving novel log templates. By analyzing the misclassified cases in \woMistralError and \woMLError, we can determine whether the LLM (Mistral) or ML models (KNN, DT, and SLFN) contribute more to handling novel log templates, helping us understand their complementary strengths in \method.

Table~\ref{tab:error-analysis} reports the error analysis results across all four datasets: \unstableadfa, \unstablelogevol, \synlogevol, and \synhdfs. Column ``Data'' specifies the dataset; Column ``Config'' reports the configuration of each dataset; Column ``Condition'' shows the subject of the statistics, including the testing dataset, \woMistralError, and \woMLError. Column ``\# Sequences'' reports the number of sequences; column ``Sequence Length'' indicates the average, minimum, and maximum length of the sequences, denoted as ``avg'', ``min'', and ``max'', respectively. Column ``\% Anomaly'' reports the percentage of anomalous log sequences, and column ``\% Unseen Template'' denotes the percentages of log sequences that have at least one unseen log template. 
Overall, for each dataset, \woMistralError and \woMLError demonstrate different characteristics in terms of log sequence length, percentage of anomaly, and percentage of unseen templates, highlighting how Flexlog’s base models complement each other in an effective prediction.

In terms of unseen templates, in \unstablelogevol Hadoop and \synhdfs, we did not observe any in either \woMLError or \woMistralError. In \unstableadfa, \woMistralError contains \SI{7.38}{\percent} unseen log templates, much higher than that of \woMLError (\SI{1.89}{\percent}). Similarly, in \unstablelogevol Spark, the percentage of unseen log templates in \woMistralError reaches \SI{33.33}{\percent}, while \woMLError contains no unseen template-related mis-classifications. In \synlogevol, \SI{11.76}{\percent} of misclassifications in \woMistralError are related to unseen templates, higher than that in \woMLError (\SI{9.28}{\percent}). Overall, we observe a higher percentage of unseen templates in \woMistralError than \woMLError across most datasets. The higher percentage of unseen log templates in \woMistralError, across several datasets, indicates that Mistral plays a crucial role in handling unseen log templates. This highlights LLM's adaptability in recognizing new templates, making it particularly valuable for \task, where logs are unstable due to environment and system evolution. In contrast, perhaps unsurprisingly, ML models appear to rely more on established patterns, struggling with new templates.

\begin{table*}[ht!]
\centering    
\captionsetup[table]{skip=0pt}
\captionof{table}{Overview of Error Analysis of \unstableadfa, \unstablelogevol, \synlogevol, and \synhdfs when Mistral or ML models are removed from \method \label{tab:error-analysis}}
    \begin{threeparttable}[htbp]
 \resizebox{\linewidth}{!}{
\begin{tabular}{@{\hspace{0.8\tabcolsep}}c@{\hspace{0.8\tabcolsep}}c@{\hspace{0.8\tabcolsep}}c@{\hspace{0.8\tabcolsep}}c@{\hspace{0.8\tabcolsep}}c@{\hspace{0.8\tabcolsep}}c@{\hspace{0.8\tabcolsep}}c@{\hspace{0.8\tabcolsep}}c@{\hspace{0.8\tabcolsep}}c@{\hspace{0.8\tabcolsep}}}
\toprule
\multirow{2}{*}{\textbf{Data}}& \multirow{2}{*}{\textbf{Config}} &  \multirow{2}{*}{\textbf{Condition}}  & \multirow{2}{*}{\textbf{$\#$ Sequence}} & \multicolumn{3}{c}{\textbf{Sequence Length}}& \multirow{2}{*}{ \textbf{\% Anomaly}} & \multirow{2}{*}{\textbf{\makecell{\% Unseen\\ Template}}} \\
\cmidrule(lr){5-7} 
&& & & avg & min & max & & \\ 
\midrule
\multirow{3}{*}{\unstableadfa} & \multirow{3}{*}{all \task} & \fulltest & \num{5146} &  \num{523.98} & \num{75} & \num{4494} & \num{14.34} & \num{3.54} \\
& &\woMistralError& \num{149}  & \num{476.74}& \num{82} & \num{2513} & \num{0} & \num{7.38}\\
&  & \woMLError & \num{264} & \num{610.48} & \num{80} & \num{3089} & \num{11.74} & \num{1.89}\\
\midrule
\multirow{9}{*}{\unstablelogevol} & \multirow{3}{*}{Hadoop \task} & \fulltest & \num{5329} &  \num{13.41} & \num{1} & \num{50} & \num{9.8} & \num{74.37} \\
& &\woMistralError & \num{5}  & \num{41.8}& \num{9} & \num{50} & \num{0.0} & \num{0.0}\\
&  & \woMLError & \num{0} & \textit{N/A} & \textit{N/A} & \textit{N/A} & \textit{N/A} & \textit{N/A}\\
\cmidrule(ll){2-9} 
 & \multirow{3}{*}{Spark \task} & \fulltest & \num{4045} &  \num{81.70} & \num{1} & \num{1977 } & \num{1.78} & \num{38.07} \\
& & \woMistralError& \num{6}  & \num{ 15.33 }& \num{ 4} & \num{35} & \num{100} & \num{33.33}\\
&  &\woMLError & \num{1} & \num{26.00} & \num{26} & \num{26} & \num{100} & \num{0}\\
\cmidrule(ll){2-9} 
 & \multirow{3}{*}{all \task} & \fulltest & \num{9374} &  \num{42.88} & \num{1} & \num{1977} & \num{6.34} & \num{58.70} \\
& & \woMistralError & \num{11}  & \num{ 29.83}& \num{4} & \num{50} & \num{ 54.54} & \num{18.18}\\
&  & \woMLError & \num{1} & \num{26.00} & \num{26} & \num{26} & \num{100.00} & \num{0.00}\\
\midrule
\multirow{3}{*}{\synhdfs} & \multirow{3}{*}{all \task} & \fulltest & \num{3744} &  \num{26.91} & \num{10} & \num{57} & \num{22.22} & \num{0.43} \\
& & \woMistralError & \num{24}  & \num{40.08}& \num{35} & \num{48} & \num{0.0 } & \num{0.0}\\
&  & \woMLError & \num{92} & \num{32.43} & \num{19} & \num{52} & \num{5.43} & \num{0.0}\\
\midrule
\multirow{3}{*}{\synlogevol} & \multirow{3}{*}{all \task} &\fulltest & \num{15406} &  \num{151.07} & \num{1} & \num{1022} & \num{2.96} & \num{91.3} \\
& &\woMistralError & \num{34}  & \num{135.03}& \num{6} & \num{323} & \num{100} & \num{11.76}\\
&  & \woMLError & \num{23} & \num{39.04} & \num{4} & \num{177} & \num{0} & \num{9.28}\\
\bottomrule
\end{tabular}
}
\begin{tablenotes}
         \item [\textit{N/A}] \footnotesize{Not applicable as no prediction errors were observed in\\ this configuration}
        \end{tablenotes}
\end{threeparttable}

\end{table*}

\subsection{Threats to Validity} 

\label{sec:treadstovalidity}

\subsubsection{Internal Validity}~\label{sec:internal-validity}\noindent\textbf{Data Leakage from LLMs.} 
One potential threat to internal validity is data leakage, where test data may inadvertently overlap with training data. Although we performed de-duplication across all test datasets, the risk of data leakage cannot be entirely eliminated, as some test data may have been included in the pretraining corpus of LLMs. This could lead to inflated effectiveness. To mitigate this risk, we evaluated \method not only on publicly available datasets (ADFA, LOGEVOL, and \synlogevol), but also on the \synhdfs dataset, which we synthesized ourselves. In addition, the cut-off date of Mistral Small is August 2023, and LOGEVOL and \synlogevol were introduced after this date. Hence, these datasets cannot be part of its pretraining corpus, ensuring a more reliable assessment of \method’s effectiveness.
\newline

\noindent\textbf{Selection of Training Dataset Sizes.} Another threat to internal validity arises from the selection of training dataset sizes (\datasize) for evaluating data efficiency. Since \datasize directly affects the performance of \method and the baselines, an exhaustive evaluation across all possible values is infeasible. To address this limitation, we systematically experimented with multiple \datasize values, as detailed in \S~\ref{sec:results_rq2}, ranging from \num{50} to \num{2000}. This range allows us to provide a comprehensive analysis of data efficiency while remaining within our computational constraints.
\newline

\noindent\textbf{Selection of \method's Base Models.} Since \method is an ensemble learning-based approach, its effectiveness is significantly influenced by the selection of base models. While it is impractical to evaluate all possible model combinations, we carefully selected representative base models from the literature on log-based anomaly detection, including KNN, DT, SLFN, LightAD~\cite{boxi}, NeuralLog~\cite{Neurallog}, CNN~\cite{CNN}, LLaMA 3.1, and GPT-4o. Through extensive empirical evaluation, we tested multiple model combinations and ultimately selected KNN, DT, MLP, and Mistral due to their consistently high effectiveness and robustness across both real-world and synthesized datasets.
\newline

\noindent\textbf{Synthetic Datasets.} The synthetic datasets in our experiments are generated by injecting different levels of instability. One potential threat to internal validity lies in the realism of this injected instability and the validity of labels after injections. To mitigate this threat, we adopted datasets synthesized by experts and applied well-established injection strategies from the literature.  Specifically, the \synlogevol dataset, proposed by ~\citet{evlog}, was annotated by two experienced Spark developers and carefully reviewed after annotation. For~\synhdfs datasets, which we constructed following the \synlogevol approach, we applied only sequence-level changes that are unlikely to affect the overall label of a log sequence. To guide this process, we leveraged the decision tree analysis~\cite{HDFS} of the original dataset to identify low-impact templates. These precautions help maintain the reliability of the labels, despite the synthetic nature of the data.
\newline

\noindent\textbf{Consistency of Labels.} For faster inference, \method's cache mechanism reuses the prediction for previously seen, identical log sequences. However, the assumption that identical log sequences always correspond to the same label may not hold in extreme log evolution cases, particularly in security-critical domains. This assumption is nevertheless valid for all of our datasets, where each unique log sequence is consistently labeled.  To further mitigate this risk, the \textit{update} function in \method's cache enables label revisions by system administrators if changes are observed over time.
\newline

\noindent\textbf{Cascading Effect of Parsing Errors.}  The use of log parsing introduces a potential risk of parsing errors, which could negatively affect the effectiveness of \method. 
However, prior work by~\citet{khan_2024} reports no strong correlation between log parsing accuracy and anomaly detection accuracy, suggesting that minor parsing errors do not necessarily degrade the effectiveness of \method. 
Our evaluation shows that \method, as a parsing-based approach, achieves significantly higher effectiveness than NeuralLog and LightAD, both of which operate on raw logs. Specifically, LightAD uses some of the same base models as \method but applies them without log parsing. 
In \method, this risk is further mitigated through two design factors. First, we adopt widely used and reliable parsers 
to reduce the likelihood of significant parsing errors (see~\S~\ref{sec:preprocessing-setting}). Second, the architecture limits error propagation: predictions are aggregated via majority voting across multiple base models, preventing a single erroneous output from dominating the final decision. Additionally, the fine-tuned LLM in \method leverages contextual information, making it resilient to potential inconsistencies introduced by parsing.

\subsubsection{Conclusion Validity}~\label{sec:conclusion-validity} It is widely acknowledged that LLMs often produce non-deterministic responses even when provided with identical prompts, posing a potential threat to the conclusion validity of \method. To address this challenge, as detailed in~\S~\ref{sec:implementaitondetails}, we configured the LLMs to generate responses with minimal randomness, achieved by setting the temperature parameter to 0.  To further reduce the influence of randomness from our evaluation results, as discussed in~\S~\ref{sec:implementaitondetails}, we ran each experiment configuration five times and calculated the average value as the final result. We also conducted Mann-Whitney U test to assess the significance of effectiveness and efficiency differences between \method and baselines. To ensure sufficient statistical power, we performed statistical testing at the dataset level rather than the configuration level, since each configuration only has five samples. Aggregation at the dataset level ensures each test includes at least 10 samples.

\subsubsection{External Validity}
A potential threat to external validity is the impact of highly unstable logs on anomaly detection performance. While \method achieves state-of-the-art effectiveness, its performance, like that of other methods, may degrade when log instability is extreme. For example, the introduction of YARN for job management in Hadoop 2 resulted in substantial architectural modifications, leading to significant changes in its logs~\cite{vavilapalli2013apache}. Such drastic shifts pose challenges for all existing methods, potentially limiting their applicability in highly unstable logging environments. To address this concern, we evaluate \method and the baselines across diverse datasets that capture two primary sources of log instability: software evolution (\unstablelogevol) and environment change (\unstableadfa). Additionally, we conduct experiments on two synthesized datasets (\synhdfs and \synlogevol) that simulate instability levels ranging from \SI{0}{\percent} to \SI{30}{\percent}. Our findings indicate that while all methods experience performance degradation as instability increases, \method remains robust in terms of precision, recall and F1 score. With a minimum F1 score of \num{0.948} (\SI{30}{\percent} template level injections on \synlogevol), \method consistently outperforms all the baselines, demonstrating greater robustness in handling highly unstable logs. Nevertheless, further evaluation on a broader range of real-world systems is necessary to fully assess the limitations of its applicability.

\section{Related Work}\label{sec:related-works}
\subsection{Anomaly Detection on Unstable Logs}
Log-based anomaly detection has been extensively studied in the literature to enhance the dependability of software-intensive systems~\cite{le2022,boxi,LANDAUER2023100470}. However, only a few studies have investigated anomaly detection on unstable logs~\cite{LogRobust,swisslog, hitanomaly, evlog}, a common situation in practice. \citet{LogRobust} first identified such a challenge and proposed LogRobust (see~\S~\ref{sec:baseliens}) to leverage an attention-based Bi-LSTM as an anomaly detector. They also created a new unstable log dataset called Synthetic HDFS to evaluate the effectiveness and robustness of LogRobust. This inspired a number of follow-up works, including supervised \cite{hitanomaly,swisslog} and unsupervised approaches \cite{evlog}. 

Supervised approaches like HitAnomaly~\citep{hitanomaly} and SwissLog~\citep{swisslog} require training with a labeled dataset, encompassing both normal and anomalous data. SwissLog adopts the same architecture (i.e., Bi-LSTM) as LogRobust and aims to further improve it by incorporating time embeddings and Bert-based semantic embeddings. HitAnomaly, however, leverages a much larger model based on a hierarchical transformer architecture. The high complexity of the HitAnomaly model allows it to tackle not only the static parts of log messages but also dynamic parts, such as numerical values that have been masked in the log templates. Experiment results demonstrate the superiority of HitAnomaly on stable logs compared to LogRobust, while showing a robust performance for small injection ratios (under 20\%) in unstable logs and being outperformed by LogRobust from 20\% to 30\%. 

\citet{evlog} proposed EvLog, an unsupervised approach leveraging a multi-level semantics extractor and attention mechanism to identify anomalous log messages in unstable logs. 
Unlike \method, EvLog is a parser-free method to combat potential parsing errors in a dataset. However, by avoiding log parsing, EvLog may face challenges in generalizing to datasets with diverse or domain-specific log formats, as it relies solely on semantic extraction without leveraging structured context.
As part of the EvLog study, they also introduced two log datasets: \logevol and SYNEVOL (referred to as \synlogevol in our paper), which serve as valuable benchmarks for \task research, including our work. 

To summarize, compared to existing \task approaches, \method: 1) leverages the synergy of ML models and LLMs through ensemble learning 2) requires significantly less training data, reduces labeling cost, 
3) achieves state-of-the-art effectiveness across all datasets, outperforming baselines in terms of F1 scores consistently while being trained on limited labeled data.

\subsection{Application of LLMs to Log Analysis}

Over the past few years, LLMs have been widely adopted on different log-related software engineering tasks to enhance effectiveness and generalizability, including anomaly detection and log parsing ~\cite{logprompt, LLMparser, divlog}. 

\textit{Anomaly detection}. The application of LLMs in the field of anomaly detection started by leveraging BERT~\cite{bert} to capture contextual information of logs with semantic-based representations. 
LogBERT~\cite{logbert} utilizes BERT to learn the semantics of normal log messages and predicts an anomaly where the representation of log messages of an input sequence deviates from the distribution of normal log sequences.
\citet{Neurallog} proposed NeuralLog (discussed in~\S~\ref{sec:baseliens}). 
\citet{loggpt-gpt-2} introduced LogGPT, which leverages reinforcement learning to fine-tune GPT-2 for anomaly detection.
More recently, \citet{logprompt} proposed LogPrompt, which adopts LLMs such as GPT-3 and Vicuna \cite{vicuna2023} for online log parsing and anomaly detection via in-context learning; we discussed the reasons for not considering LogPrompt as baseline in~\S~\ref{sec:baseliens}.
\citet{LLMeLog} proposed LLMeLog, which leverages a BERT model fine-tuned with log sequences enriched with contextual information that was retrieved by GPT-3.5. Inspired by LLMeLog, we further explored RAG with various LLMs, including closed-source and open-source LLMs. Additionally, we equipped LLMs with a cache mechanism and ensemble learning to enhance their effectiveness and data efficiency. Note that we did not consider LLMeLog as a baseline in this work due to the unavailability of their replication package.

\textit{Log Parsing}. 
\citet{logparsingchatgpt} explored the in-context learning of ChatGPT~\cite{Ouyang2022TrainingLM} on log parsing and achieved promising results with zero-shot and few-shot prompts. \citet{divlog} proposed DivLog, another few-shot, in-context learning method that constructs prompts with five labeled examples for each target log template. DivLog explicitly optimizes the diversity of included examples using the Determinantal Point Process (DPP)~\cite{chen2018fastgreedymapinference}, reducing the potential biases in the examples by maximizing sample diversity. \citet{lilac} proposed a novel parser called LILAC, equipping LLMs with an adaptive cache to reduce the LLM query times and, consequently, the efficiency. Recently, \citet{selfLog} introduced a self-evolutionary LLM-based parser, which identifies new log templates by grouping history log messages. Fewer studies focus on fine-tuning LLMs for log parsing. \citet{LogPPT} introduced LogPPT to fine-tune RoBERTa for log parsing. In addition, an adaptive random sampling strategy was designed to select a small yet diverse training dataset. \citet{LLMparser} compared in-context learning and fine-tuning using open-source LLMs such as Flan-T5~\cite{chung2022scaling} and LLaMA~\cite{touvron2023llama} on log parsing.
\citet{YALP} introduced YALP, which leverages the capabilities of ChatGPT (gpt-3.5-turbo) in conjunction with traditional methods --- Longest Common Subsequence,without incorporating user labeling (zero-shot learning).
\citet{logparserllm} proposed LogParser-LLM, which essentially blends a prefix tree and an LLM-based
template extractor. This extractor parses log messages with different LLMs, including GPT-3.5-turbo, GPT-4, and Llama-2-13B, in either ICL or fine-tuning manner;  the highest results were obtained using GPT-4 with ICL.
\citet{logbatcher} introduced LogBatcher, which is a cost-effective LLM-based log parser based on GPT-3.5-Turbo. Similar to YALP, they control the cost of using closed-source LLM by storing inferred messages in a basic cache. Moreover, it does  not require any training by prompting the LLM with a group of high-diversity log messages to ensure that the LLM understands the diversity of the dataset in a zero-shot manner. 
Overall, the results of the above works align with our findings: the vast pretrained knowledge of LLMs enables data efficiency and robustness on unseen log templates. 

\section{Conclusion and future work}\label{sec:conclusion}
This paper proposed a novel approach, \method, for anomaly detection on unstable logs (\task), exploiting the synergy between Large Language Models (LLMs) and Machine Learning (ML) models via ensemble learning. \method incorporates four base models, one LLM (Mistral), and three ML models (KNN, DT, and SLFN), which are trained on limited stable logs, reducing the usage of labeled data significantly. To classify unstable logs, \method first processes unstable logs into log sequences through log parsing and partitioning. Then, \method employs Retrieval-Augmented Generation (RAG) to fetch relevant information (if available) for these sequences, constructing context-enriched prompts. The fine-tuned LLM processes these prompts, while ML models use the log sequences directly for prediction. Finally, \method combines the predictions of all the base models using majority voting to produce the final classification.  

Our extensive experiments on two real-world and two synthesized datasets show that \method achieves state-of-the-art effectiveness on all datasets while reducing the usage of labeled data by \SI{62.87}{\pp} to \SI{78.43}{\pp}, respectively. Further experiments on \unstableadfa with varying limited training data size demonstrate that \method maintains robust effectiveness under varying levels of data scarcity, except the extreme data scarcity scenario (\datasize$=50$), where all methods exhibit poor performance due to insufficient labeled data. \method outperforms the top baseline in terms of F1 by \SI{13}{\pp}  when the training dataset contains only 500 samples. However, experiments assessing time efficiency show that \method trades off some time efficiency for this effectiveness but still manages to keep the inference time below one second per log sequence. This suggests \method is applicable for most systems, except those with stringent latency requirements, such as high-frequency trading systems. Furthermore, in terms of memory efficiency, \method’s cache memory remains \textbf{below \SI{4}{\mega\byte}} for most datasets (up to \SI{19.6}{\mega\byte} for \unstableadfa)
confirming its \textbf{memory efficiency}. Finally, we conducted ablation studies on individual components of \method, as well as evaluating alternative LLM choices. We confirmed the significant contributions of the cache-based inference, RAG, and ensemble learning with Mistral as LLM and KNN, DT, and SLFN as ML models. 

In the future, we plan to further enhance the effectiveness of \method by exploring more powerful open-source LLMs as base models, such as DeepSeek R1~\cite{deepseek}, along with more advanced prompting techniques such as agent-based prompting ~\cite{llmagent}. Additionally, we wish to investigate different ensemble learning techniques that dynamically decide the optimal ensemble composition for a given log sequence. To enhance the generalizability of our caching mechanism, we plan to incorporate similarity-based retrieval in the future, allowing the model to infer labels from similar seen log sequences instead of identical ones. Furthermore, to address the dataset imbalance observed in certain datasets, we aim to explore techniques like Generative Adversarial Networks (GANs) for data augmentation, potentially improving the performance of ML-based models within the ensemble. 

\section*{Acknowledgments}

This work was partly supported by the Natural Sciences and Engineering Research Council of Canada (NSERC), through the Canada Research Chairs  and discovery programs, the Research Ireland grant 13/RC/2094-2,  the
  Luxembourg  National  Research Fund (FNR), grant reference C22/IS/17373407/LOGODOR.
The experiments conducted in this work were enabled in part by the computation support provided by the Digital Research Alliance of Canada.\footnote{\url{https://alliancecan.ca/en}}

\bibliographystyle{plainnat}
\bibliography{manuscript}

\begin{thebibliography}{106}
\providecommand{\natexlab}[1]{#1}
\providecommand{\url}[1]{\texttt{#1}}
\expandafter\ifx\csname urlstyle\endcsname\relax
  \providecommand{\doi}[1]{doi: #1}\else
  \providecommand{\doi}{doi: \begingroup \urlstyle{rm}\Url}\fi

\bibitem[Arcuri and Briand(2011)]{statistical_testing_guide}
Andrea Arcuri and Lionel Briand.
\newblock A practical guide for using statistical tests to assess randomized algorithms in software engineering.
\newblock In \emph{Proceedings of the 33rd International Conference on Software Engineering}, ICSE '11, page 1–10, New York, NY, USA, 2011. Association for Computing Machinery.
\newblock ISBN 9781450304450.
\newblock \doi{10.1145/1985793.1985795}.
\newblock URL \url{https://doi.org/10.1145/1985793.1985795}.

\bibitem[Bello et~al.(2024)Bello, Ige, and Ameyaw]{bello2024deep}
Halima~Oluwabunmi Bello, Adebimpe~Bolatito Ige, and Maxwell~Nana Ameyaw.
\newblock Deep learning in high-frequency trading: conceptual challenges and solutions for real-time fraud detection.
\newblock \emph{World Journal of Advanced Engineering Technology and Sciences}, 12\penalty0 (02):\penalty0 035--046, 2024.

\bibitem[Brown et~al.(2020)Brown, Mann, Ryder, Subbiah, Kaplan, Dhariwal, Neelakantan, Shyam, Sastry, Askell, et~al.]{brown2020language}
Tom Brown, Benjamin Mann, Nick Ryder, Melanie Subbiah, Jared~D Kaplan, Prafulla Dhariwal, Arvind Neelakantan, Pranav Shyam, Girish Sastry, Amanda Askell, et~al.
\newblock Language models are few-shot learners.
\newblock \emph{Advances in neural information processing systems}, 33:\penalty0 1877--1901, 2020.

\bibitem[Chawla et~al.(2002{\natexlab{a}})Chawla, Bowyer, Hall, and Kegelmeyer]{Chawla_2002}
N.~V. Chawla, K.~W. Bowyer, L.~O. Hall, and W.~P. Kegelmeyer.
\newblock Smote: Synthetic minority over-sampling technique.
\newblock \emph{Journal of Artificial Intelligence Research}, 16:\penalty0 321–357, June 2002{\natexlab{a}}.
\newblock ISSN 1076-9757.
\newblock \doi{10.1613/jair.953}.
\newblock URL \url{http://dx.doi.org/10.1613/jair.953}.

\bibitem[Chawla et~al.(2002{\natexlab{b}})Chawla, Bowyer, Hall, and Kegelmeyer]{smote}
Nitesh~V Chawla, Kevin~W Bowyer, Lawrence~O Hall, and W~Philip Kegelmeyer.
\newblock Smote: synthetic minority over-sampling technique.
\newblock \emph{Journal of artificial intelligence research}, 16:\penalty0 321--357, 2002{\natexlab{b}}.

\bibitem[Chen et~al.(2018)Chen, Zhang, and Zhou]{chen2018fastgreedymapinference}
Laming Chen, Guoxin Zhang, and Hanning Zhou.
\newblock Fast greedy map inference for determinantal point process to improve recommendation diversity, 2018.
\newblock URL \url{https://arxiv.org/abs/1709.05135}.

\bibitem[Chen et~al.(2004)Chen, Zheng, Lloyd, Jordan, and Brewer]{Decisiontree}
Mike~Y. Chen, Alice~X. Zheng, Jesper Lloyd, Michael~I. Jordan, and Eric Brewer.
\newblock Failure diagnosis using decision trees.
\newblock \emph{International Conference on Autonomic Computing, 2004. Proceedings.}, pages 36--43, 2004.
\newblock URL \url{https://api.semanticscholar.org/CorpusID:56849250}.

\bibitem[Chiang et~al.(2023)Chiang, Li, Lin, Sheng, Wu, Zhang, Zheng, Zhuang, Zhuang, Gonzalez, Stoica, and Xing]{vicuna2023}
Wei-Lin Chiang, Zhuohan Li, Zi~Lin, Ying Sheng, Zhanghao Wu, Hao Zhang, Lianmin Zheng, Siyuan Zhuang, Yonghao Zhuang, Joseph~E. Gonzalez, Ion Stoica, and Eric~P. Xing.
\newblock Vicuna: An open-source chatbot impressing gpt-4 with 90\%* chatgpt quality, March 2023.
\newblock URL \url{https://lmsys.org/blog/2023-03-30-vicuna/}.

\bibitem[Cho et~al.(2014)Cho, van Merrienboer, G{\"{u}}l{\c{c}}ehre, Bahdanau, Bougares, Schwenk, and Bengio]{GRU}
Kyunghyun Cho, Bart van Merrienboer, {\c{C}}aglar G{\"{u}}l{\c{c}}ehre, Dzmitry Bahdanau, Fethi Bougares, Holger Schwenk, and Yoshua Bengio.
\newblock Learning phrase representations using {RNN} encoder-decoder for statistical machine translation.
\newblock In Alessandro Moschitti, Bo~Pang, and Walter Daelemans, editors, \emph{Proceedings of the 2014 Conference on Empirical Methods in Natural Language Processing, {EMNLP} 2014, October 25-29, 2014, Doha, Qatar, {A} meeting of SIGDAT, a Special Interest Group of the {ACL}}, pages 1724--1734. {ACL}, 2014.
\newblock \doi{10.3115/V1/D14-1179}.
\newblock URL \url{https://doi.org/10.3115/v1/d14-1179}.

\bibitem[Chu et~al.(2021)Chu, Wang, Qi, Sun, Tao, and Liao]{prefixgraph}
Guohao Chu, Jiaqi Wang, Qi~Qi, Haiyang Sun, Shengtao Tao, and Jun Liao.
\newblock Prefix-graph: A versatile log parsing approach merging prefix tree with probabilistic graph.
\newblock In \emph{Proceedings of the 37th International Conference on Data Engineering (ICDE)}, pages 2411--2422, Virtual Event, 2021. IEEE.
\newblock URL \url{https://ieeexplore.ieee.org/document/9458609}.

\bibitem[Chung et~al.(2022)Chung, Hou, Longpre, Zoph, Tay, Fedus, Li, Wang, Dehghani, Brahma, Webson, Gu, Dai, Suzgun, Chen, Chowdhery, Castro-Ros, Pellat, Robinson, Valter, Narang, Mishra, Yu, Zhao, Huang, Dai, Yu, Petrov, Chi, Dean, Devlin, Roberts, Zhou, Le, and Wei]{chung2022scaling}
Hyung~Won Chung, Le~Hou, Shayne Longpre, Barret Zoph, Yi~Tay, William Fedus, Yunxuan Li, Xuezhi Wang, Mostafa Dehghani, Siddhartha Brahma, Albert Webson, Shixiang~Shane Gu, Zhuyun Dai, Mirac Suzgun, Xinyun Chen, Aakanksha Chowdhery, Alex Castro-Ros, Marie Pellat, Kevin Robinson, Dasha Valter, Sharan Narang, Gaurav Mishra, Adams Yu, Vincent Zhao, Yanping Huang, Andrew Dai, Hongkun Yu, Slav Petrov, Ed~H. Chi, Jeff Dean, Jacob Devlin, Adam Roberts, Denny Zhou, Quoc~V. Le, and Jason Wei.
\newblock Scaling instruction-finetuned language models, 2022.

\bibitem[Collier and Shahan(2015)]{collier2015microsoft}
Michael Collier and Robin Shahan.
\newblock \emph{Microsoft azure essentials-fundamentals of azure}.
\newblock Microsoft Press, 2015.

\bibitem[Creech and Hu(2013)]{ADFA}
Gideon Creech and Jiankun Hu.
\newblock Generation of a new ids test dataset: Time to retire the kdd collection.
\newblock In \emph{2013 IEEE Wireless Communications and Networking Conference (WCNC)}, pages 4487--4492, 2013.
\newblock \doi{10.1109/WCNC.2013.6555301}.

\bibitem[Dean and Ghemawat(2008)]{spark}
Jeffrey Dean and Sanjay Ghemawat.
\newblock Mapreduce: Simplified data processing on large clusters.
\newblock \emph{Communications of the ACM}, 51\penalty0 (1):\penalty0 107--113, 2008.
\newblock \doi{10.1145/1327452.1327492}.
\newblock URL \url{http://doi.acm.org/10.1145/1327452.1327492}.

\bibitem[DeepSeek-AI et~al.(2025)DeepSeek-AI, Guo, Yang, Zhang, Song, Zhang, Xu, Zhu, Ma, Wang, and Bi]{deepseek}
DeepSeek-AI, Daya Guo, Dejian Yang, Haowei Zhang, Junxiao Song, Ruoyu Zhang, Runxin Xu, Qihao Zhu, Shirong Ma, Peiyi Wang, and Xiao Bi.
\newblock Deepseek-r1: Incentivizing reasoning capability in llms via reinforcement learning, 2025.
\newblock URL \url{https://arxiv.org/abs/2501.12948}.

\bibitem[Deerwester et~al.(1990)Deerwester, Dumais, Furnas, Landauer, and Harshman]{countvector}
Scott Deerwester, Susan~T Dumais, George~W Furnas, Thomas~K Landauer, and Richard Harshman.
\newblock Indexing by latent semantic analysis.
\newblock \emph{Journal of the American society for information science}, 41\penalty0 (6):\penalty0 391--407, 1990.

\bibitem[Dettmers et~al.(2023)Dettmers, Pagnoni, Holtzman, and Zettlemoyer]{qlora}
Tim Dettmers, Artidoro Pagnoni, Ari Holtzman, and Luke Zettlemoyer.
\newblock Qlora: Efficient finetuning of quantized llms, 2023.
\newblock URL \url{https://arxiv.org/abs/2305.14314}.

\bibitem[Devlin et~al.(2019)Devlin, Chang, Lee, and Toutanova]{bert}
Jacob Devlin, Ming{-}Wei Chang, Kenton Lee, and Kristina Toutanova.
\newblock {BERT:} pre-training of deep bidirectional transformers for language understanding, 2019.
\newblock URL \url{https://doi.org/10.18653/v1/n19-1423}.

\bibitem[Diagboya(2021)]{diagboya2021infrastructure}
Ewere Diagboya.
\newblock \emph{Infrastructure Monitoring with Amazon CloudWatch: Effectively monitor your AWS infrastructure to optimize resource allocation, detect anomalies, and set automated actions}.
\newblock Packt Publishing Ltd, 2021.

\bibitem[Dong et~al.(2022)Dong, Li, Dai, Zheng, Ma, Li, Xia, Xu, Wu, Liu, et~al.]{dong2022survey}
Qingxiu Dong, Lei Li, Damai Dai, Ce~Zheng, Jingyuan Ma, Rui Li, Heming Xia, Jingjing Xu, Zhiyong Wu, Tianyu Liu, et~al.
\newblock A survey on in-context learning.
\newblock \emph{arXiv preprint arXiv:2301.00234}, 2022.

\bibitem[Du et~al.(2017)Du, Li, Zheng, and Srikumar]{DeepLog}
Min Du, Feifei Li, Guineng Zheng, and Vivek Srikumar.
\newblock {DeepLog: Anomaly detection and diagnosis from system logs through deep learning}, 2017.
\newblock ISSN 15437221.

\bibitem[Fix and Hodges(1989)]{KNN}
Evelyn Fix and Joseph~L. Hodges.
\newblock Discriminatory analysis - nonparametric discrimination: Consistency properties.
\newblock \emph{International Statistical Review}, 57:\penalty0 238, 1989.
\newblock URL \url{https://api.semanticscholar.org/CorpusID:120323383}.

\bibitem[Grattafiori et~al.(2024)Grattafiori, Dubey, Jauhri, Pandey, Kadian, Al-Dahle, and Letman]{llama3.1}
Aaron Grattafiori, Abhimanyu Dubey, Abhinav Jauhri, Abhinav Pandey, Abhishek Kadian, Ahmad Al-Dahle, and Aiesha Letman.
\newblock The llama 3 herd of models, 2024.
\newblock URL \url{https://arxiv.org/abs/2407.21783}.

\bibitem[Guo et~al.(2003)Guo, Wang, Bell, Bi, and Greer]{10.1007/978-3-540-39964-3_62}
Gongde Guo, Hui Wang, David Bell, Yaxin Bi, and Kieran Greer.
\newblock Knn model-based approach in classification.
\newblock In Robert Meersman, Zahir Tari, and Douglas~C. Schmidt, editors, \emph{On The Move to Meaningful Internet Systems 2003: CoopIS, DOA, and ODBASE}, pages 986--996, Berlin, Heidelberg, 2003. Springer Berlin Heidelberg.
\newblock ISBN 978-3-540-39964-3.

\bibitem[Guo et~al.(2021)Guo, Yuan, and Wu]{logbert}
Haixuan Guo, Shuhan Yuan, and Xintao Wu.
\newblock Logbert: Log anomaly detection via bert.
\newblock In \emph{2021 International Joint Conference on Neural Networks (IJCNN)}, pages 1--8, 2021.
\newblock \doi{10.1109/IJCNN52387.2021.9534113}.

\bibitem[Hadadi et~al.(2025)Hadadi, Xu, Bianculli, and Briand]{replication-package}
Fatemeh Hadadi, Qinghua Xu, Domenico Bianculli, and Lionel Briand.
\newblock {Replication Package}, 10 2025.
\newblock URL \url{https://figshare.com/s/2a8a747a5b6029ce93f4}.

\bibitem[Han et~al.(2023)Han, Yuan, and Trabelsi]{loggpt-gpt-2}
Xiao Han, Shuhan Yuan, and Mohamed Trabelsi.
\newblock Loggpt: Log anomaly detection via gpt.
\newblock In \emph{2023 IEEE International Conference on Big Data (BigData)}, pages 1117--1122, 2023.
\newblock \doi{10.1109/BigData59044.2023.10386543}.

\bibitem[Han et~al.(2024)Han, Gao, Liu, Zhang, and Zhang]{han2024parameterefficient}
Zeyu Han, Chao Gao, Jinyang Liu, Jeff Zhang, and Sai~Qian Zhang.
\newblock Parameter-efficient fine-tuning for large models: A comprehensive survey, 2024.

\bibitem[He et~al.(2024)He, Jia, Duan, Cai, Li, , and Huang]{LLMeLog}
M.~He, T.~Jia, C.~Duan, H~Cai, Y.~Li, , and G.~Huang.
\newblock Llmelog: An approach for anomaly detection based on llm-enriched log events.
\newblock In \emph{2024 IEEE 35th International Symposium on Software Reliability Engineering (ISSRE)}, Tsukuba, Japan, 2024. IEEE Computer Society.

\bibitem[He et~al.(2017)He, Zhu, Zheng, and Lyu]{drain}
Pinjia He, Jieming Zhu, Zibin Zheng, and Michael~R Lyu.
\newblock Drain: An online log parsing approach with fixed depth tree, 2017.

\bibitem[He et~al.(2016)He, Zhu, He, and Lyu]{loglizer}
Shilin He, Jieming Zhu, Pinjia He, and Michael~R. Lyu.
\newblock Experience report: System log analysis for anomaly detection, 2016.
\newblock URL \url{https://doi.org/10.1109/ISSRE.2016.21}.

\bibitem[He et~al.(2020)He, Zhu, He, and Lyu]{he2020loghub}
Shilin He, Jieming Zhu, Pinjia He, and Michael~R. Lyu.
\newblock {Loghub: A Large Collection of System Log Datasets Towards Automated Log Analytics}.
\newblock \emph{CoRR}, abs/2008.06448, 2020.
\newblock URL \url{https://arxiv.org/abs/2008.06448}.

\bibitem[He et~al.(2021)He, He, Chen, Yang, Su, and Lyu]{HeSurvey}
Shilin He, Pinjia He, Zhuangbin Chen, Tianyi Yang, Yuxin Su, and Michael~R. Lyu.
\newblock A survey on automated log analysis for reliability engineering.
\newblock \emph{ACM Comput. Surv.}, 54\penalty0 (6), jul 2021.
\newblock ISSN 0360-0300.
\newblock \doi{10.1145/3460345}.
\newblock URL \url{https://doi.org/10.1145/3460345}.

\bibitem[Hejazi and Singh(2013)]{hejazi2013one}
Maryamsadat Hejazi and Yashwant~Prasad Singh.
\newblock One-class support vector machines approach to anomaly detection.
\newblock \emph{Applied Artificial Intelligence}, 27\penalty0 (5):\penalty0 351--366, 2013.

\bibitem[Hochreiter and Schmidhuber(1997)]{LSTM}
Sepp Hochreiter and J\"{u}rgen Schmidhuber.
\newblock Long short-term memory.
\newblock \emph{Neural Comput.}, 9\penalty0 (8):\penalty0 1735–1780, nov 1997.
\newblock ISSN 0899-7667.
\newblock \doi{10.1162/neco.1997.9.8.1735}.
\newblock URL \url{https://doi.org/10.1162/neco.1997.9.8.1735}.

\bibitem[Hu et~al.(2021)Hu, Shen, Wallis, Allen-Zhu, Li, Wang, Wang, and Chen]{hu2021lora}
Edward~J. Hu, Yelong Shen, Phillip Wallis, Zeyuan Allen-Zhu, Yuanzhi Li, Shean Wang, Lu~Wang, and Weizhu Chen.
\newblock Lora: Low-rank adaptation of large language models, 2021.

\bibitem[Huang et~al.(2000)Huang, Chen, and Babri]{SLFN}
Guangbin Huang, Yan~Qiu Chen, and Haroon~Atique Babri.
\newblock Classification ability of single hidden layer feedforward neural networks.
\newblock \emph{IEEE transactions on neural networks}, 11 3:\penalty0 799--801, 2000.
\newblock URL \url{https://api.semanticscholar.org/CorpusID:1852628}.

\bibitem[Huang et~al.(2020)Huang, Liu, Fung, He, Zhao, Yang, and Luan]{hitanomaly}
Shaohan Huang, Yi~Liu, Carol Fung, Rong He, Yining Zhao, Hailong Yang, and Zhongzhi Luan.
\newblock Hitanomaly: Hierarchical transformers for anomaly detection in system log.
\newblock \emph{IEEE Transactions on Network and Service Management}, 17\penalty0 (4):\penalty0 2064--2076, 2020.
\newblock \doi{10.1109/TNSM.2020.3034647}.

\bibitem[Huo et~al.(2023)Huo, Lee, Su, Shan, Liu, and Lyu]{evlog}
Y.~Huo, C.~Lee, Y.~Su, S.~Shan, J.~Liu, and M.~R. Lyu.
\newblock Evlog: Identifying anomalous logs over software evolution.
\newblock In \emph{2023 IEEE 34th International Symposium on Software Reliability Engineering (ISSRE)}, pages 391--402, Los Alamitos, CA, USA, oct 2023. IEEE Computer Society.
\newblock \doi{10.1109/ISSRE59848.2023.00018}.
\newblock URL \url{https://doi.ieeecomputersociety.org/10.1109/ISSRE59848.2023.00018}.

\bibitem[Intel(2021)]{hibench}
Intel.
\newblock Hibench.
\newblock \url{https://github.com/Intel-bigdata/HiBench}, 2021.

\bibitem[Irugalbandara et~al.(2024)Irugalbandara, Mahendra, Daynauth, Arachchige, Dantanarayana, Flautner, Tang, Kang, and Mars]{10590016}
Chandra Irugalbandara, Ashish Mahendra, Roland Daynauth, Tharuka~Kasthuri Arachchige, Jayanaka Dantanarayana, Krisztian Flautner, Lingjia Tang, Yiping Kang, and Jason Mars.
\newblock Scaling down to scale up: A cost-benefit analysis of replacing openai's llm with open source slms in production.
\newblock In \emph{2024 IEEE International Symposium on Performance Analysis of Systems and Software (ISPASS)}, pages 280--291, 2024.
\newblock \doi{10.1109/ISPASS61541.2024.00034}.

\bibitem[Jiang et~al.(2023)Jiang, Sablayrolles, Mensch, Bamford, Chaplot, de~las Casas, Bressand, Lengyel, Lample, Saulnier, Lavaud, Lachaux, Stock, Scao, Lavril, Wang, Lacroix, and Sayed]{jiang2023mistral7b}
Albert~Q. Jiang, Alexandre Sablayrolles, Arthur Mensch, Chris Bamford, Devendra~Singh Chaplot, Diego de~las Casas, Florian Bressand, Gianna Lengyel, Guillaume Lample, Lucile Saulnier, Lélio~Renard Lavaud, Marie-Anne Lachaux, Pierre Stock, Teven~Le Scao, Thibaut Lavril, Thomas Wang, Timothée Lacroix, and William~El Sayed.
\newblock Mistral 7b, 2023.
\newblock URL \url{https://arxiv.org/abs/2310.06825}.

\bibitem[Jiang et~al.(2024)Jiang, Liu, Chen, Li, Huang, Huo, He, Gu, and Lyu]{lilac}
Zhihan Jiang, Jinyang Liu, Zhuangbin Chen, Yichen Li, Junjie Huang, Yintong Huo, Pinjia He, Jiazhen Gu, and Michael~R. Lyu.
\newblock Lilac: Log parsing using llms with adaptive parsing cache, 2024.

\bibitem[Joulin et~al.(2016)Joulin, Grave, Bojanowski, Douze, Jégou, and Mikolov]{fasttextzip}
Armand Joulin, Edouard Grave, Piotr Bojanowski, Matthijs Douze, Hérve Jégou, and Tomas Mikolov.
\newblock Fasttext.zip: Compressing text classification models, 2016.

\bibitem[Kabinna et~al.(2016)Kabinna, Shang, Bezemer, and Hassan]{logevolution}
Suhas Kabinna, Weiyi Shang, Cor-Paul Bezemer, and Ahmed~E. Hassan.
\newblock Examining the stability of logging statements.
\newblock In \emph{2016 IEEE 23rd International Conference on Software Analysis, Evolution, and Reengineering (SANER)}, volume~1, pages 326--337, 2016.
\newblock \doi{10.1109/SANER.2016.29}.

\bibitem[Kaur et~al.(2019)Kaur, Pannu, and Malhi]{imbalanceddata}
Harsurinder Kaur, Husanbir~Singh Pannu, and Avleen~Kaur Malhi.
\newblock A systematic review on imbalanced data challenges in machine learning: Applications and solutions.
\newblock \emph{ACM computing surveys (CSUR)}, 52\penalty0 (4):\penalty0 1--36, 2019.

\bibitem[Khan et~al.(2024)Khan, Shin, Bianculli, and Briand]{khan_2024}
Zanis~Ali Khan, Donghwan Shin, Domenico Bianculli, and Lionel~C. Briand.
\newblock Impact of log parsing on deep learning-based anomaly detection.
\newblock \emph{Empirical Software Engineering}, 29\penalty0 (6), August 2024.
\newblock ISSN 1573-7616.
\newblock \doi{10.1007/s10664-024-10533-w}.
\newblock URL \url{http://dx.doi.org/10.1007/s10664-024-10533-w}.

\bibitem[Kim(2014)]{CNNdefinition}
Yoon Kim.
\newblock Convolutional neural networks for sentence classification.
\newblock In Alessandro Moschitti, Bo~Pang, and Walter Daelemans, editors, \emph{Proceedings of the 2014 Conference on Empirical Methods in Natural Language Processing ({EMNLP})}, pages 1746--1751, Doha, Qatar, October 2014. Association for Computational Linguistics.
\newblock \doi{10.3115/v1/D14-1181}.
\newblock URL \url{https://aclanthology.org/D14-1181}.

\bibitem[Krasner(2021)]{krasner2021cost}
Herb Krasner.
\newblock The cost of poor software quality in the us: A 2020 report.
\newblock \emph{Proc. Consortium Inf. Softw. QualityTM (CISQTM)}, pages 1--46, 2021.

\bibitem[Landauer et~al.(2023)Landauer, Onder, Skopik, and Wurzenberger]{LANDAUER2023100470}
Max Landauer, Sebastian Onder, Florian Skopik, and Markus Wurzenberger.
\newblock Deep learning for anomaly detection in log data: A survey.
\newblock \emph{Machine Learning with Applications}, 12:\penalty0 100470, 2023.
\newblock ISSN 2666-8270.
\newblock \doi{https://doi.org/10.1016/j.mlwa.2023.100470}.
\newblock URL \url{https://www.sciencedirect.com/science/article/pii/S2666827023000233}.

\bibitem[Le and Zhang(2021)]{Neurallog}
V.~Le and H.~Zhang.
\newblock Log-based anomaly detection without log parsing, nov 2021.
\newblock URL \url{https://doi.ieeecomputersociety.org/10.1109/ASE51524.2021.9678773}.

\bibitem[Le and Zhang(2023{\natexlab{a}})]{logparsingchatgpt}
V.~Le and H.~Zhang.
\newblock Log parsing: How far can chatgpt go?
\newblock In \emph{2023 38th IEEE/ACM International Conference on Automated Software Engineering (ASE)}, pages 1699--1704, Los Alamitos, CA, USA, sep 2023{\natexlab{a}}. IEEE Computer Society.
\newblock \doi{10.1109/ASE56229.2023.00206}.
\newblock URL \url{https://doi.ieeecomputersociety.org/10.1109/ASE56229.2023.00206}.

\bibitem[Le and Zhang(2022)]{le2022}
Van-Hoang Le and Hongyu Zhang.
\newblock Log-based anomaly detection with deep learning: how far are we?
\newblock In \emph{Proceedings of the 44th International Conference on Software Engineering}, ICSE '22, page 1356–1367, New York, NY, USA, 2022. Association for Computing Machinery.
\newblock ISBN 9781450392211.
\newblock \doi{10.1145/3510003.3510155}.
\newblock URL \url{https://doi.org/10.1145/3510003.3510155}.

\bibitem[Le and Zhang(2023{\natexlab{b}})]{LogPPT}
Van-Hoang Le and Hongyu Zhang.
\newblock Log parsing with prompt-based few-shot learning.
\newblock In \emph{Proceedings of the 45th International Conference on Software Engineering}, ICSE '23, page 2438–2449. IEEE Press, 2023{\natexlab{b}}.
\newblock ISBN 9781665457019.
\newblock \doi{10.1109/ICSE48619.2023.00204}.
\newblock URL \url{https://doi.org/10.1109/ICSE48619.2023.00204}.

\bibitem[LeCun et~al.(2015)LeCun, Bengio, and Hinton]{lecun2015deep}
Yann LeCun, Yoshua Bengio, and Geoffrey Hinton.
\newblock Deep learning.
\newblock \emph{nature}, 521\penalty0 (7553):\penalty0 436, 2015.

\bibitem[Li et~al.(2020)Li, Chen, Jing, He, and Yu]{swisslog}
Xiaoyun Li, Pengfei Chen, Linxiao Jing, Zilong He, and Guangba Yu.
\newblock {Swisslog: Robust and unified deep learning based log anomaly detection for diverse faults}.
\newblock \emph{Proceedings - International Symposium on Software Reliability Engineering, ISSRE}, 2020-Octob:\penalty0 92--103, 2020.
\newblock ISSN 10719458.
\newblock \doi{10.1109/ISSRE5003.2020.00018}.

\bibitem[Lin et~al.(2016)Lin, Zhang, Lou, Zhang, and Chen]{lin2016log}
Qingwei Lin, Hongyu Zhang, Jian-Guang Lou, Yu~Zhang, and Xuewei Chen.
\newblock {Log Clustering Based Problem Identification for Online Service Systems}.
\newblock In \emph{Proceedings of the 38th International Conference on Software Engineering, ICSE - Companion Volume}, pages 102--111, Austin, TX, USA, 2016. ACM.
\newblock \doi{10.1145/2889160.2889232}.
\newblock URL \url{https://doi.org/10.1145/2889160.2889232}.

\bibitem[Liu et~al.(2008)Liu, Ting, and Zhou]{isolation2008}
Fei~Tony Liu, Kai~Ming Ting, and Zhi-Hua Zhou.
\newblock Isolation forest.
\newblock In \emph{2008 Eighth IEEE International Conference on Data Mining}, pages 413--422, 2008.
\newblock \doi{10.1109/ICDM.2008.17}.

\bibitem[Liu et~al.(2024)Liu, Tao, Meng, Yao, Zhao, and Yang]{logprompt}
Yilun Liu, Shimin Tao, Weibin Meng, Feiyu Yao, Xiaofeng Zhao, and Hao Yang.
\newblock Logprompt: Prompt engineering towards zero-shot and interpretable log analysis.
\newblock In \emph{Proceedings of the 2024 IEEE/ACM 46th International Conference on Software Engineering: Companion Proceedings}, ICSE-Companion '24, page 364–365, New York, NY, USA, 2024. Association for Computing Machinery.
\newblock ISBN 9798400705021.
\newblock \doi{10.1145/3639478.3643108}.
\newblock URL \url{https://doi.org/10.1145/3639478.3643108}.

\bibitem[Locke et~al.(2022)Locke, Li, Chen, Shang, and Liu]{logsummerization-survey}
Steven Locke, Heng Li, Tse-Hsun~Peter Chen, Weiyi Shang, and Wei Liu.
\newblock Logassist: Assisting log analysis through log summarization.
\newblock \emph{IEEE Transactions on Software Engineering}, 48\penalty0 (9):\penalty0 3227--3241, 2022.
\newblock \doi{10.1109/TSE.2021.3083715}.

\bibitem[Lowe et~al.(2023)]{Achiam2023GPT4TR}
Long Ouyang Jeff Wu Xu Jiang Diogo Almeida Carroll L. Wainwright Pamela Mishkin Paul Christiano Jan Leike~Ryan Lowe et~al.
\newblock Gpt-4 technical report.
\newblock 2023.
\newblock URL \url{https://api.semanticscholar.org/CorpusID:257532815}.

\bibitem[Lu et~al.(2018)Lu, Wei, Li, and Wang]{CNN}
Siyang Lu, Xiang Wei, Yandong Li, and Liqiang Wang.
\newblock Detecting anomaly in big data system logs using convolutional neural network.
\newblock \emph{IEEE Access}, 6:\penalty0 21929--21940, 2018.
\newblock \doi{10.1109/ACCESS.2018.2811530}.

\bibitem[Ma et~al.(2024)Ma, Chen, Kim, Chen, and Wang]{LLMparser}
Zeyang Ma, An~Ran Chen, Dong~Jae Kim, Tse-Hsun Chen, and Shaowei Wang.
\newblock Llmparser: An exploratory study on using large language models for log parsing.
\newblock In \emph{Proceedings of the IEEE/ACM 46th International Conference on Software Engineering}, ICSE '24, New York, NY, USA, 2024. Association for Computing Machinery.
\newblock ISBN 9798400702174.
\newblock \doi{10.1145/3597503.3639150}.
\newblock URL \url{https://doi.org/10.1145/3597503.3639150}.

\bibitem[McInnes et~al.(2017)McInnes, Healy, and Astels]{hdbscan}
Leland McInnes, John Healy, and Steve Astels.
\newblock hdbscan: Hierarchical density based clustering.
\newblock \emph{J. Open Source Softw.}, 2\penalty0 (11):\penalty0 205, 2017.

\bibitem[Meng et~al.(2019)Meng, Liu, Zhu, Zhang, Pei, Liu, Chen, Zhang, Tao, Sun, et~al.]{LogAnomaly}
Weibin Meng, Ying Liu, Yichen Zhu, Shenglin Zhang, Dan Pei, Yuqing Liu, Yihao Chen, Ruizhi Zhang, Shimin Tao, Pei Sun, et~al.
\newblock Loganomaly: Unsupervised detection of sequential and quantitative anomalies in unstructured logs.
\newblock In \emph{IJCAI}, volume~19, pages 4739--4745, 2019.

\bibitem[Mishra et~al.(2018)Mishra, Rohaninejad, Chen, and Abbeel]{snail}
Nikhil Mishra, Mostafa Rohaninejad, Xi~Chen, and Pieter Abbeel.
\newblock A simple neural attentive meta-learner, 2018.
\newblock URL \url{https://arxiv.org/abs/1707.03141}.

\bibitem[MistralAI(2024)]{mistralai}
MistralAI.
\newblock Mistral technologies, 2024.
\newblock URL \url{https://mistral.ai/technology/}.
\newblock accessed 2024-09-24.

\bibitem[Mohammed et~al.(2020)Mohammed, Rawashdeh, and Abdullah]{mohammed2020machine}
Roweida Mohammed, Jumanah Rawashdeh, and Malak Abdullah.
\newblock Machine learning with oversampling and undersampling techniques: overview study and experimental results.
\newblock In \emph{2020 11th international conference on information and communication systems (ICICS)}, pages 243--248. IEEE, 2020.

\bibitem[Mosbach et~al.(2023)Mosbach, Pimentel, Ravfogel, Klakow, and Elazar]{mosbach-etal-2023-shot}
Marius Mosbach, Tiago Pimentel, Shauli Ravfogel, Dietrich Klakow, and Yanai Elazar.
\newblock Few-shot fine-tuning vs. in-context learning: A fair comparison and evaluation.
\newblock In Anna Rogers, Jordan Boyd-Graber, and Naoaki Okazaki, editors, \emph{Findings of the Association for Computational Linguistics: ACL 2023}, pages 12284--12314, Toronto, Canada, July 2023. Association for Computational Linguistics.
\newblock \doi{10.18653/v1/2023.findings-acl.779}.
\newblock URL \url{https://aclanthology.org/2023.findings-acl.779}.

\bibitem[Oliner and Stearley(2007)]{bgl-spirit}
Adam Oliner and Jon Stearley.
\newblock What supercomputers say: A study of five system logs, 2007.

\bibitem[OpenAI(2024{\natexlab{a}})]{fine-tuning-format}
OpenAI.
\newblock Fine-tuning: preparing your dataset, 2024{\natexlab{a}}.
\newblock URL \url{https://platform.openai.com/docs/guides/fine-tuning}.
\newblock accessed 2024-01-02.

\bibitem[OpenAI(2024{\natexlab{b}})]{openai-models}
OpenAI.
\newblock Openai models, 2024{\natexlab{b}}.
\newblock URL \url{https://platform.openai.com/docs/models}.
\newblock accessed 2024-05-10.

\bibitem[Ouyang et~al.(2022)Ouyang, Wu, Jiang, Almeida, Wainwright, Mishkin, Zhang, Agarwal, Slama, Ray, Schulman, Hilton, Kelton, Miller, Simens, Askell, Welinder, Christiano, Leike, and Lowe]{Ouyang2022TrainingLM}
Long Ouyang, Jeff Wu, Xu~Jiang, Diogo Almeida, Carroll~L. Wainwright, Pamela Mishkin, Chong Zhang, Sandhini Agarwal, Katarina Slama, Alex Ray, John Schulman, Jacob Hilton, Fraser Kelton, Luke~E. Miller, Maddie Simens, Amanda Askell, Peter Welinder, Paul~Francis Christiano, Jan Leike, and Ryan~J. Lowe.
\newblock Training language models to follow instructions with human feedback.
\newblock \emph{ArXiv}, abs/2203.02155, 2022.
\newblock URL \url{https://api.semanticscholar.org/CorpusID:246426909}.

\bibitem[Pei et~al.(2024)Pei, Z.Liu, Li, Zhang, Zhang, Zhang, Chen, Pei, and Xie]{selfLog}
C.~Pei, Z.Liu, Jianhui Li, E.~Zhang, L.~Zhang, H.~Zhang, W.~Chen, D.~Pei, and G.~Xie.
\newblock Self-evolutionary group-wise log parsing based on large language model.
\newblock In \emph{2024 IEEE 35th International Symposium on Software Reliability Engineering (ISSRE)}, Tsukuba, Japan, 2024. IEEE Computer Society.

\bibitem[Polikar(2012)]{ensemble1}
Robi Polikar.
\newblock Ensemble learning.
\newblock \emph{Ensemble machine learning: Methods and applications}, pages 1--34, 2012.

\bibitem[Schuster and Paliwal(1997)]{BiLSTM}
M.~Schuster and K.K. Paliwal.
\newblock Bidirectional recurrent neural networks.
\newblock \emph{IEEE Transactions on Signal Processing}, 45\penalty0 (11):\penalty0 2673--2681, 1997.
\newblock \doi{10.1109/78.650093}.

\bibitem[Shvachko et~al.(2010)Shvachko, Kuang, Radia, and Chansler]{hadoop}
Konstantin Shvachko, Hairong Kuang, Sanjay Radia, and Robert Chansler.
\newblock The hadoop distributed file system, 2010.

\bibitem[Team(2025)]{qwen2.5-1m}
Qwen Team.
\newblock Qwen2.5-1m: Deploy your own qwen with context length up to 1m tokens, January 2025.
\newblock URL \url{https://qwenlm.github.io/blog/qwen2.5-1m/}.

\bibitem[Touvron et~al.(2023)Touvron, Martin, Stone, and Albert]{touvron2023llama}
Hugo Touvron, Louis Martin, Kevin Stone, and Peter Albert.
\newblock Llama 2: Open foundation and fine-tuned chat models, 2023.

\bibitem[Turnbull(2018)]{turnbull2018monitoring}
James Turnbull.
\newblock \emph{Monitoring with Prometheus}.
\newblock Turnbull Press, 2018.

\bibitem[Unsloth(2024)]{Unsloth}
Unsloth.
\newblock Unsloth fine-tuning package, 2024.
\newblock URL \url{https://github.com/unslothai/unsloth}.
\newblock accessed 2024-11-30.

\bibitem[Vaswani et~al.(2017)Vaswani, Shazeer, Parmar, Uszkoreit, Jones, Gomez, Kaiser, and Polosukhin]{AttentionIA}
Ashish Vaswani, Noam Shazeer, Niki Parmar, Jakob Uszkoreit, Llion Jones, Aidan~N. Gomez, \L{}ukasz Kaiser, and Illia Polosukhin.
\newblock Attention is all you need.
\newblock In \emph{Proceedings of the 31st International Conference on Neural Information Processing Systems}, NIPS'17, page 6000–6010, Red Hook, NY, USA, 2017. Curran Associates Inc.
\newblock ISBN 9781510860964.

\bibitem[Vavilapalli et~al.(2013)Vavilapalli, Murthy, Douglas, Agarwal, Konar, Evans, Graves, Lowe, Shah, Seth, et~al.]{vavilapalli2013apache}
Vinod~Kumar Vavilapalli, Arun~C Murthy, Chris Douglas, Sharad Agarwal, Mahadev Konar, Robert Evans, Thomas Graves, Jason Lowe, Hitesh Shah, Siddharth Seth, et~al.
\newblock Apache hadoop yarn: Yet another resource negotiator.
\newblock In \emph{Proceedings of the 4th annual Symposium on Cloud Computing}, pages 1--16, 2013.

\bibitem[Wang et~al.(2024)Wang, Ma, Feng, Zhang, Yang, Zhang, Chen, Tang, Chen, Lin, et~al.]{llmagent}
Lei Wang, Chen Ma, Xueyang Feng, Zeyu Zhang, Hao Yang, Jingsen Zhang, Zhiyuan Chen, Jiakai Tang, Xu~Chen, Yankai Lin, et~al.
\newblock A survey on large language model based autonomous agents.
\newblock \emph{Frontiers of Computer Science}, 18\penalty0 (6):\penalty0 186345, 2024.

\bibitem[Wang et~al.(2022)Wang, Yang, Li, Ma, He, Xiao, Liu, and Yang]{maddc}
Xiaolei Wang, Lin Yang, Dongyang Li, Linru Ma, Yongzhong He, Junchao Xiao, Jiyuan Liu, and Yuexiang Yang.
\newblock Maddc: Multi-scale anomaly detection, diagnosis and correction for discrete event logs.
\newblock In \emph{Proceedings of the 38th Annual Computer Security Applications Conference}, ACSAC '22, page 769–784, New York, NY, USA, 2022. Association for Computing Machinery.
\newblock ISBN 9781450397599.
\newblock \doi{10.1145/3564625.3567972}.
\newblock URL \url{https://doi.org/10.1145/3564625.3567972}.

\bibitem[Winteringham(2024)]{Winteringham2024AIAssisted}
Mark Winteringham.
\newblock Software testing with generative ai, 2024.

\bibitem[Xiao et~al.(2024)Xiao, Le, and Zhang]{logbatcher}
Yi~Xiao, Van-Hoang Le, and Hongyu Zhang.
\newblock Demonstration-free: Towards more practical log parsing with large language models.
\newblock In \emph{Proceedings of the 39th IEEE/ACM International Conference on Automated Software Engineering}, ASE '24, page 153–165, New York, NY, USA, 2024. Association for Computing Machinery.
\newblock ISBN 9798400712487.
\newblock \doi{10.1145/3691620.3694994}.
\newblock URL \url{https://doi.org/10.1145/3691620.3694994}.

\bibitem[Xu et~al.(2024)Xu, Yang, Huo, Zhang, and He]{divlog}
Junjielong Xu, Ruichun Yang, Yintong Huo, Chengyu Zhang, and Pinjia He.
\newblock Divlog: Log parsing with prompt enhanced in-context learning.
\newblock In \emph{Proceedings of the IEEE/ACM 46th International Conference on Software Engineering}, ICSE '24, New York, NY, USA, 2024. Association for Computing Machinery.
\newblock ISBN 9798400702174.
\newblock \doi{10.1145/3597503.3639155}.
\newblock URL \url{https://doi.org/10.1145/3597503.3639155}.

\bibitem[Xu et~al.(2009{\natexlab{a}})Xu, Huang, Fox, Patterson, and Jordan]{HDFS}
Wei Xu, Ling Huang, Armando Fox, David Patterson, and Michael Jordan.
\newblock Online system problem detection by mining patterns of console logs, 12 2009{\natexlab{a}}.

\bibitem[Xu et~al.(2009{\natexlab{b}})Xu, Huang, Fox, Patterson, and Jordan]{detecting2009}
Wei Xu, Ling Huang, Armando Fox, David Patterson, and Michael~I. Jordan.
\newblock Detecting large-scale system problems by mining console logs.
\newblock In \emph{Proceedings of the ACM SIGOPS 22nd Symposium on Operating Systems Principles}, SOSP '09, page 117–132, New York, NY, USA, 2009{\natexlab{b}}. Association for Computing Machinery.
\newblock ISBN 9781605587523.
\newblock \doi{10.1145/1629575.1629587}.
\newblock URL \url{https://doi.org/10.1145/1629575.1629587}.

\bibitem[Xu et~al.(2010)Xu, Huang, Fox, Patterson, and Jordan]{Xu2010}
Wei Xu, Ling Huang, Armando Fox, David Patterson, and Michael~I. Jordan.
\newblock {Detecting large-scale system problems by mining console logs}, 2010.

\bibitem[Yang et~al.(2025)Yang, Yu, Li, Liu, Huang, Huang, Jiang, Tu, Zhang, Zhou, Lin, Dang, Yang, Yu, Li, Sun, Zhu, Men, He, Xu, Yin, Yu, Qiu, Ren, Yang, Li, Xu, and Zhang]{qwen2.5}
An~Yang, Bowen Yu, Chengyuan Li, Dayiheng Liu, Fei Huang, Haoyan Huang, Jiandong Jiang, Jianhong Tu, Jianwei Zhang, Jingren Zhou, Junyang Lin, Kai Dang, Kexin Yang, Le~Yu, Mei Li, Minmin Sun, Qin Zhu, Rui Men, Tao He, Weijia Xu, Wenbiao Yin, Wenyuan Yu, Xiafei Qiu, Xingzhang Ren, Xinlong Yang, Yong Li, Zhiying Xu, and Zipeng Zhang.
\newblock Qwen2.5-1m technical report.
\newblock \emph{arXiv preprint arXiv:2501.15383}, 2025.

\bibitem[Yang et~al.(2021)Yang, Chen, Wang, Wang, Jiang, Dong, and Zhang]{PLELog}
Lin Yang, Junjie Chen, Zan Wang, Weijing Wang, Jiajun Jiang, Xuyuan Dong, and Wenbin Zhang.
\newblock Semi-supervised log-based anomaly detection via probabilistic label estimation, 2021.

\bibitem[Yang et~al.(2024)Yang, Kassner, Gribovskaya, Riedel, and Geva]{yang2024largelanguagemodelsperform}
Sohee Yang, Nora Kassner, Elena Gribovskaya, Sebastian Riedel, and Mor Geva.
\newblock Do large language models perform latent multi-hop reasoning without exploiting shortcuts?, 2024.
\newblock URL \url{https://arxiv.org/abs/2411.16679}.

\bibitem[Yao et~al.(2024)Yao, Duan, Xu, Cai, Sun, and Zhang]{yao2024survey}
Yifan Yao, Jinhao Duan, Kaidi Xu, Yuanfang Cai, Zhibo Sun, and Yue Zhang.
\newblock A survey on large language model (llm) security and privacy: The good, the bad, and the ugly.
\newblock \emph{High-Confidence Computing}, page 100211, 2024.

\bibitem[Yu et~al.(2024{\natexlab{a}})Yu, Yao, Fu, Zhong, Xie, Wu, Ma, and He]{boxi}
Boxi Yu, Jiayi Yao, Qiuai Fu, Zhiqing Zhong, Haotian Xie, Yaoliang Wu, Yuchi Ma, and Pinjia He.
\newblock Deep learning or classical machine learning? an empirical study on log-based anomaly detection.
\newblock In \emph{Proceedings of the IEEE/ACM 46th International Conference on Software Engineering}, ICSE '24, New York, NY, USA, 2024{\natexlab{a}}. Association for Computing Machinery.
\newblock ISBN 9798400702174.
\newblock \doi{10.1145/3597503.3623308}.
\newblock URL \url{https://doi.org/10.1145/3597503.3623308}.

\bibitem[Yu et~al.(2024{\natexlab{b}})Yu, Yao, Fu, Zhong, Xie, Wu, Ma, and He]{lightad}
Boxi Yu, Jiayi Yao, Qiuai Fu, Zhiqing Zhong, Haotian Xie, Yaoliang Wu, Yuchi Ma, and Pinjia He.
\newblock Deep learning or classical machine learning? an empirical study on log-based anomaly detection.
\newblock In \emph{Proceedings of the IEEE/ACM 46th International Conference on Software Engineering}, ICSE '24, New York, NY, USA, 2024{\natexlab{b}}. Association for Computing Machinery.
\newblock ISBN 9798400702174.
\newblock \doi{10.1145/3597503.3623308}.
\newblock URL \url{https://doi.org/10.1145/3597503.3623308}.

\bibitem[Yu et~al.(2022)Yu, Luo, Zhou, Si, Zhou, Wang, Feng, and Yan]{metaformer}
Weihao Yu, Mi~Luo, Pan Zhou, Chenyang Si, Yichen Zhou, Xinchao Wang, Jiashi Feng, and Shuicheng Yan.
\newblock Metaformer is actually what you need for vision, 2022.
\newblock URL \url{https://arxiv.org/abs/2111.11418}.

\bibitem[Zhang(2021)]{knnimbalance}
Shichao Zhang.
\newblock Challenges in knn classification.
\newblock \emph{IEEE Transactions on Knowledge and Data Engineering}, 34\penalty0 (10):\penalty0 4663--4675, 2021.

\bibitem[Zhang et~al.(2023)Zhang, Qiu, Castellano, Rifai, Chen, and Pianese]{logparsing-survey}
T.~Zhang, H.~Qiu, G.~Castellano, M.~Rifai, C.~Chen, and F.~Pianese.
\newblock System log parsing: A survey.
\newblock \emph{IEEE Transactions on Knowledge \&amp; Data Engineering}, 35\penalty0 (08):\penalty0 8596--8614, aug 2023.
\newblock ISSN 1558-2191.
\newblock \doi{10.1109/TKDE.2022.3222417}.

\bibitem[Zhang et~al.(2024)Zhang, Jia, Duan, Cai, Li, , and Huang]{LogRAG}
W.~Zhang, T.~Jia, C.~Duan, H~Cai, Y.~Li, , and G.~Huang.
\newblock Leveraging rag-enhanced large language model for semi-supervised log anomaly detection.
\newblock In \emph{2024 IEEE 35th International Symposium on Software Reliability Engineering (ISSRE)}, Tsukuba, Japan, 2024. IEEE Computer Society.

\bibitem[Zhang et~al.(2019)Zhang, Xu, Lin, Qiao, Zhang, Dang, Xie, Yang, Cheng, Li, Chen, He, Yao, Lou, Chintalapati, Shen, and Zhang]{LogRobust}
Xu~Zhang, Yong Xu, Qingwei Lin, Bo~Qiao, Hongyu Zhang, Yingnong Dang, Chunyu Xie, Xinsheng Yang, Qian Cheng, Ze~Li, Junjie Chen, Xiaoting He, Randolph Yao, Jian-Guang Lou, Murali Chintalapati, Furao Shen, and Dongmei Zhang.
\newblock Robust log-based anomaly detection on unstable log data.
\newblock In \emph{Proceedings of the 2019 27th ACM Joint Meeting on European Software Engineering Conference and Symposium on the Foundations of Software Engineering}, ESEC/FSE 2019, page 807–817, New York, NY, USA, 2019. Association for Computing Machinery.
\newblock ISBN 9781450355728.
\newblock \doi{10.1145/3338906.3338931}.
\newblock URL \url{https://doi.org/10.1145/3338906.3338931}.

\bibitem[Zhang et~al.(2022)Zhang, Liu, and Shen]{ensemble3}
Yuzhen Zhang, Jingjing Liu, and Wenjuan Shen.
\newblock A review of ensemble learning algorithms used in remote sensing applications.
\newblock \emph{Applied Sciences}, 12\penalty0 (17):\penalty0 8654, 2022.

\bibitem[Zhi et~al.(2024)Zhi, Cheng, Liu, Zhao, Xu, and Deng]{YALP}
Chen Zhi, Liye Cheng, Meilin Liu, Xinkui Zhao, Yueshen Xu, and Shuiguang Deng.
\newblock Llm-powered zero-shot online log parsing.
\newblock In \emph{2024 IEEE International Conference on Web Services (ICWS)}, pages 877--887, 2024.
\newblock \doi{10.1109/ICWS62655.2024.00106}.

\bibitem[Zhong et~al.(2024)Zhong, Mo, Liu, Liu, Lu, Zhou, Wu, Li, and Wen]{logparserllm}
Aoxiao Zhong, Dengyao Mo, Guiyang Liu, Jinbu Liu, Qingda Lu, Qi~Zhou, Jiesheng Wu, Quanzheng Li, and Qingsong Wen.
\newblock Logparser-llm: Advancing efficient log parsing with large language models, 2024.
\newblock URL \url{https://arxiv.org/abs/2408.13727}.

\bibitem[Zhou and Zhou(2021)]{ensemble2}
Zhi-Hua Zhou and Zhi-Hua Zhou.
\newblock \emph{Ensemble learning}.
\newblock Springer, 2021.

\end{thebibliography}

\appendix

\section{Appendix}\label{sec:appendix}

\subsection{Alternative Ensembling Strategies}\label{sec:appendix-ensembling}
\method employs a majority voting strategy, where in the case of a tie between base models, the label is set to \textit{normal}, as anomalies are rare. This ensemble approach uses binary labels for voting rather than anomaly confidence scores, allowing us to adopt base models that directly output the final label, such as our fine-tuned version of Mistral.
Alternative ensembling strategies include the 
alternative majority voting method, which assigns a random label in the case of a tie, and state-of-the-art meta-learning approaches that adaptively learn from base model performance on validation data. Specifically, SNAIL~\cite{snail} is an attentive CNN-based meta-learner, while MetaFormer~\cite{metaformer} introduces a novel attention module called \textit{token mixer}; both have shown significant improvements over traditional meta-learning approaches. We experimented with these methods using their original implementation code.

Table~\ref{tab:appendix-alternative-ensembling} reports the F1 scores of \method using different ensemble strategies, including the original majority voting (\textit{Majority Voting (alternative)}), 
and two meta-learning approaches, \textit{SNAIL} and \textit{MetaFormer}. Experimental results show that \method's majority voting consistently outperforms these alternatives with an average difference of \SI{5}{\pp}, \SI{3}{\pp}, \SI{2}{\pp}, and \SI{3}{\pp}, respectively, for \unstableadfa, \unstablelogevol, \synhdfs, and \unstablelogevol. Mann-Whitney U tests at the dataset level further confirm that the differences in F1 score between \method and the alternative strategies are statistically significant across all four datasets, except for MetaFormer on \unstablelogevol. Overall, the modified majority voting in \method remains the recommended strategy due to its simplicity and effectiveness.

\begin{table*}[t!]
\centering
    \captionsetup[table]{skip=0pt}
\captionof{table}{F1 scores of using alternative Ensembling Strategies in \method. \label{tab:appendix-alternative-ensembling}}
     \begin{threeparttable}[t]
 \resizebox{\linewidth}{!}{
\begin{tabular}{@{\hspace{1.5\tabcolsep}}c@{\hspace{1.5\tabcolsep}}c@{\hspace{1.5\tabcolsep}}c@{\hspace{1.5\tabcolsep}}c@{\hspace{1.5\tabcolsep}}c@{\hspace{1.5\tabcolsep}}c@{\hspace{1.5\tabcolsep}}c@{\hspace{1.5\tabcolsep}}c@{\hspace{1.5\tabcolsep}}c@{\hspace{1.5\tabcolsep}}c@{\hspace{1.5\tabcolsep}}c@{\hspace{1.5\tabcolsep}}c@{\hspace{1.5\tabcolsep}}c@{\hspace{1.5\tabcolsep}}}
\toprule
\multirow{1}{*}{\textbf{\makecell{Ensembling\\Config}}} &\multicolumn{7}{c}{\textbf{\unstableadfa}} & \multicolumn{3}{c}{\textbf{\unstablelogevol}}  & \multicolumn{1}{c}{\textbf{\synhdfs}} & \multicolumn{1}{c}{\textbf{\synlogevol}}\\

\cmidrule(lr){2-8}
\cmidrule(lr){9-11}
\cmidrule(lr){12-12}
\cmidrule(lr){13-13}

& adduser & hydraFTP & hydraSSH & java & meter & web & average & Hadoop & Spark & average & average & average\\
\midrule

 \multirow{1}{*}{\textit{Majority Voting} (\method)}& \textbf{\num{0.718}} & \textbf{\num{0.784}} &  \textbf{\num{0.723}} &  \textbf{\num{0.642}} &  \textbf{\num{0.682}} &  \textbf{\num{0.672}} &\textbf{ \num{0.704}} & \textbf{\num{0.982}} &  \textbf{\num{0.892}} & \textbf{\num{0.937}}& \textbf{\num{0.972}} & \textbf{\num{0.971}}\\
 \midrule
 \midrule
 \multirow{1}{*}{\textit{Majority Voting} (alternative)}& \num{0.641} & \num{0.711} & \num{0.691} & \num{0.615} & \num{0.656} & \num{0.635}& \num{0.650}* & \num{0.964}& \num{0.840} & \num{0.902}* & \num{0.936}* & \num{0.929}* \\
\midrule
 \multirow{1}{*}{\textit{SNAIL}}& \num{0.671} & \num{0.714} & \num{0.683} & \num{0.624} & \num{0.628} & \num{0.651}& \num{0.661}* & \num{0.965}& \num{0.854} & \num{0.909}* & \num{0.958}* & \num{0.949}* \\
\midrule
 \multirow{1}{*}{\textit{MetaFormer}} & \num{0.666} & \num{0.706} & \num{0.691} & \num{0.615} & \num{0.607} & \num{0.663}& \num{0.658}* & \num{0.976}& \num{0.866} & \num{0.921} & \num{0.954}* & \num{0.951}* \\
\bottomrule
\end{tabular}
}
\begin{tablenotes}
         \item [$*$] \footnotesize{\method yields a significant higher F1 score compared to using the alternative ensembling strategy.}
        \end{tablenotes}
\end{threeparttable}
\end{table*} 

\subsection{Impact of De-duplication}\label{sec:appendix-duplication}
Table~\ref{tab:with-duplicate} reports the F1 scores of \method and baseline models evaluated on both de-duplicated and original test data. The column ``dedup'' indicates whether the test set was de-duplicated from the training set. Overall, all models exhibit inflated F1 scores when the test set is not de-duplicated from the training set, highlighting the impact of data leakage. For instance, NeuralLog shows an average inflation of \SI{16}{\pp}, \SI{2}{\pp}, \SI{1}{\pp}, and \SI{24}{\pp} on \unstableadfa, \unstablelogevol, \synhdfs, and \synlogevol, respectively. Mann-Whitney U tests confirm that the differences are statistically significant for Neurallog. In contrast to the most effective baseline (LightAD), \method keeps the inflation in F1 score below \SI{2}{\pp} across all datasets, where the difference between \method's performance on de-duplicated and original test data is statistically insignificant, confirmed by Mann-Whitney U tests. These results indicate that \method delivers more reliable and robust performance. To avoid any risk of inflated results, all reported performances in this paper are based on de-duplicated test data.

\begin{table*}[t]
\centering
    \captionsetup[table]{skip=0pt}
\captionof{table}{Effectiveness of \method and baselines for \task on \unstableadfa, \unstablelogevol, \synhdfs, and \synlogevol with and without de-duplication. \label{tab:with-duplicate} 
}
     \begin{threeparttable}[t]
 \resizebox{\linewidth}{!}{
\begin{tabular}{@{\hspace{1.5\tabcolsep}}c@{\hspace{1.5\tabcolsep}}c@{\hspace{1.5\tabcolsep}}c@{\hspace{1.5\tabcolsep}}c@{\hspace{1.5\tabcolsep}}c@{\hspace{1.5\tabcolsep}}c@{\hspace{1.5\tabcolsep}}c@{\hspace{1.5\tabcolsep}}c@{\hspace{1.5\tabcolsep}}c@{\hspace{1.5\tabcolsep}}c@{\hspace{1.5\tabcolsep}}c@{\hspace{1.5\tabcolsep}}c@{\hspace{1.5\tabcolsep}}c@{\hspace{1.5\tabcolsep}}c@{\hspace{1.5\tabcolsep}}}
\toprule
 \multirow{2}{*}{\textbf{Model}} & \multirow{2}{*}{\textbf{Dedup}} &\multicolumn{7}{c}{\textbf{\unstableadfa}} & \multicolumn{3}{c}{\textbf{\unstablelogevol}}  & \multicolumn{1}{c}{\textbf{\synhdfs}} & \multicolumn{1}{c}{\textbf{\synlogevol}}\\

\cmidrule(lr){3-9}
\cmidrule(lr){10-12}
\cmidrule(lr){13-13}
\cmidrule(lr){14-14}

& & adduser & hydraFTP & hydraSSH & java & meter & web & average & Hadoop & Spark & average & average & average\\
\midrule

 \multirow{2}{*}{\method}& Yes & \num{0.718} & \num{0.784} &  \textbf{\num{0.723}} &  \num{0.642} &  \num{0.682} &  \num{0.672} & \num{0.704} & \num{0.982} &  \num{0.892} & \textbf{\num{0.937}}& \num{0.972} & \textbf{\num{0.971}}\\
 & No & \num{0.723} & \num{0.790} & \num{0.726} & \num{0.652} & \num{0.701} & \num{0.673} & \num{0.711}& \num{0.996}& \num{0.895} & \num{0.945} & \num{0.974} & \num{0.987} \\
 \midrule
 \multirow{2}{*}{\textit{LightAD}} & Yes & \num{0.725} & \num{0.754} & \num{0.666} & \num{0.639} & \num{0.679} & \num{0.602} & \num{0.677} & \num{0.980}&\num{0.829} & \num{0.898}  & \num{0.959}  & \num{0.956} \\
  & No & \num{0.745} & \num{0.778} & \num{0.745} & \num{0.646} & \num{0.695} & \num{0.618} & \num{0.704}* & \num{0.995}& \num{0.876} & \num{0.935}* & \num{0.968} & \num{0.981}* \\
\midrule
 \multirow{2}{*}{\textit{NeuralLog}} & Yes & \num{0.412} & \num{0.388} & \num{0.368} & \num{0.511} & \num{0.461} & \num{0.457} & \num{0.433} & \num{0.948}&\num{0.834} & \num{0.891}  & \num{0.946}  & \num{0.765} \\
  & No & \num{0.606} & \num{0.681} & \num{0.601} & \num{0.570} & \num{0.645} & \num{0.492} & \num{0.599}* & \num{0.961}& \num{0.859} & \num{0.910}* & \num{0.951} & \num{0.905}* \\
\midrule
\multirow{2}{*}{\textit{LogRobust}}& Yes & \num{0.524} & \num{0.408} & \num{0.374} & \num{0.558} & \num{0.651} & \num{0.449} & \num{0.494} & \num{0.927}&\num{0.757} & \num{0.833}  & \num{0.760}  & \num{0.941} \\
 & No & \num{0.636} & \num{0.597} & \num{0.504} & \num{0.662} & \num{0.660} & \num{0.496} & \num{0.592}*& \num{0.981}& \num{0.789} & \num{0.885}* & \num{0.929}* & \num{0.966}* \\
\midrule
 \multirow{2}{*}{\textit{CNN}}& Yes & \num{0.641} & \num{0.750} & \num{0.750} & \num{0.635} & \num{0.711} & \num{0.621} & \num{0.685} & \num{0.980}&\num{0.840} & \num{0.910}  & \num{0.942}  & \num{0.918} \\
  & No & \num{0.703}& \num{0.765} & \num{0.777} & \num{0.644
  } & \num{0.724} & \num{0.634} &\num{0.707}* & \num{0.989}& \num{0.863} & \num{0.925} & \num{0.961}* & \num{0.925} \\
\midrule
 \multirow{2}{*}{\textit{PLELog}}& Yes & \num{0.361} & \num{0.388} & \num{0.473} & \num{0.430} & \num{0.253} & \num{0.233} & \num{0.356} & \num{0.709}&\num{0.165} & \num{0.437}  & \num{0.499}  & \num{0.164} \\
  & No & \num{0.405} & \num{0.428} & \num{0.494} & \num{0.443} & \num{0.311} & \num{0.277} & \num{0.393}* & \num{0.761}& \num{0.223} & \num{0.492}* & \num{0.528}* & \num{0.236}* \\
\midrule
\multirow{2}{*}{\textit{LogAnomaly}}& Yes & \num{0.291} & \num{0.451} & \num{0.480} & \num{0.422} & \num{0.218} & \num{0.343} & \num{0.368} & \num{0.310}&\num{0.216} & \num{0.263}  & \num{0.446}  & \num{0.304} \\
 & No & \num{0.305} & \num{0.464} & \num{0.486}& \num{0.399} & \num{0.237} & \num{0.351} & \num{0.373} & \num{0.619}& \num{0.212} & \num{0.415}* & \num{0.498}* & \num{0.336}* \\
\midrule
\multirow{2}{*}{\textit{DeepLog}}& Yes & \num{0.292} & \num{0.481} & \num{0.458} & \num{0.339} & \num{0.253} & \num{0.353} & \num{0.363} & \num{0.367}&\num{0.122} & \num{0.244}  & \num{0.741}  & \num{0.253} \\
 & No & \num{0.340} & \num{0.499}& \num{0.450} & \num{0.341} & \num{0.249} & \num{0.369}& \num{0.374} & \num{0.685}& \num{0.141} & \num{0.413}* & \num{0.776}* & \num{0.291}* \\
\midrule
\multirow{2}{*}{\textit{LogCluster}}& Yes & \num{0.334} & \num{0.418} & \num{0.461} & \num{0.348} & \num{0.175} & \num{0.304} & \num{0.340} & \num{0.430}&\num{0.485} & \num{0.458}  & \num{0.519}  & \num{0.714} \\
 & No & \num{0.326} & \num{0.431} & \num{0.523} & \num{0.317} & \num{0.211} & \num{0.336} & \num{0.357}* & \num{0.798}& \num{0.614} & \num{0.706}* & \num{0.759}* & \num{0.786}* \\
\midrule
\multirow{2}{*}{\textit{PCA}}& Yes & \num{0.165} & \num{0.144} & \num{0.140} & \num{0.158} & \num{0.241} & \num{0.197} & \num{0.174} & \num{0.360}&\num{0.108} & \num{0.234}  & \num{0.404}  & \num{0.103} \\
 & No & \num{0.155} & \num{0.152} & \num{0.163} & \num{0.277} & \num{0.211} & \num{0.198} & \num{0.192}* & \num{0.454}& \num{0.097}& \num{0.275}* & \num{0.567}* & \num{0.189}* \\
\bottomrule
\end{tabular}
}
\begin{tablenotes}
         \item [$*$] \footnotesize{The same model shows a significant F1 score difference between deduplicated and original test data.}
        \end{tablenotes}
\end{threeparttable}
\end{table*}

\subsection{Baselines with Limited Data}\label{sec:appendix-limited-data}
Table~\ref{tab:appendix-limited-data} presents the effectiveness of \method compared to baselines when all models are trained on the same limited dataset. \method consistently achieves the highest F1 score across all datasets under this setting. For instance, \method outperforms the supervised baselines LightAD, NeuralLog, LogRobust, and CNN by \SI{10}{\pp}, \SI{16}{\pp}, \SI{30}{\pp}, and \SI{14}{\pp}, respectively, on average for \unstableadfa. This advantage is important because supervised models such as NeuralLog and CNN depend on large labeled datasets to achieve high accuracy. Mann-Whitney U tests confirm that the observed performance gaps between \method and each baseline are statistically significant 
across all datasets, demonstrating that none of the baselines match \method’s performance with limited data. Notably, \method trained with limited data even outperforms baselines trained with full datasets in terms of predictive effectiveness. Detailed results are presented and discussed in RQ1 (~\S~\ref{sec:results_rq1}).

\begin{table*}[t]
\centering
    \captionsetup[table]{skip=0pt}
\captionof{table}{Effectiveness of \method and Baselines Trained with Limited Data on \unstableadfa, \unstablelogevol, \synhdfs, and \synlogevol.\label{tab:appendix-limited-data} 
}
     \begin{threeparttable}[t]
 \resizebox{\linewidth}{!}{
\begin{tabular}
{@{\hspace{1.5\tabcolsep}}c@{\hspace{1.5\tabcolsep}}c@{\hspace{1.5\tabcolsep}}c@{\hspace{1.5\tabcolsep}}c@{\hspace{1.5\tabcolsep}}c@{\hspace{1.5\tabcolsep}}c@{\hspace{1.5\tabcolsep}}c@{\hspace{1.5\tabcolsep}}c@{\hspace{1.5\tabcolsep}}c@{\hspace{1.5\tabcolsep}}c@{\hspace{1.5\tabcolsep}}c@{\hspace{1.5\tabcolsep}}c@{\hspace{1.5\tabcolsep}}c@{\hspace{1.5\tabcolsep}}}
\toprule
 \multirow{2}{*}{\textbf{Model}} &\multicolumn{7}{c}{\textbf{\unstableadfa}} & \multicolumn{3}{c}{\textbf{\unstablelogevol}}  & \multicolumn{1}{c}{\textbf{\synhdfs}} & \multicolumn{1}{c}{\textbf{\synlogevol}}\\

\cmidrule(lr){2-8}
\cmidrule(lr){9-11}
\cmidrule(lr){12-12}
\cmidrule(lr){13-13}

& adduser & hydraFTP & hydraSSH & java & meter & web & average & Hadoop & Spark & average & average & average\\
\midrule

 \multirow{1}{*}{\method}& \textbf{\num{0.718}} & \textbf{\num{0.784}} &  \textbf{\num{0.723}} &  \textbf{\num{0.642}} &  \textbf{\num{0.682}} &  \textbf{\num{0.672}} & \textbf{\num{0.704}} & \textbf{\num{0.982}} &  \textbf{\num{0.892}} & \textbf{\num{0.937}}& \textbf{\num{0.972}} & \textbf{\num{0.971}}\\
 \midrule
 \multirow{1}{*}{\textit{LightAD}}& \num{0.587} & \num{0.720} & \num{0.633} & \num{0.564} & \num{0.520} & \num{0.557} & \num{0.596}* & \num{0.973}& \num{0.821} & \num{0.897}* & \num{0.925}* & \num{0.915}* \\
\midrule
 \multirow{1}{*}{\textit{NeuralLog}}& \num{0.607} & \num{0.645} & \num{0.630} & \num{0.577} & \num{0.438} & \num{0.328} & \num{0.537}* & \num{0.975}& \num{0.232} & \num{0.603}* & \num{0.914}* & \num{0.270}* \\
\midrule
 \multirow{1}{*}{\textit{LogRobust}}& \num{0.393} & \num{0.551} & \num{0.424} & \num{0.392} & \num{0.299} & \num{0.374} & \num{0.405}* & \num{0.851}& \num{0.254} & \num{0.552}* & \num{0.930}* & \num{0.903}* \\
\midrule
 \multirow{1}{*}{\textit{CNN}}& \num{0.534} & \num{0.703} & \num{0.603} & \num{0.545} & \num{0.422} & \num{0.549} & \num{0.559}* & \num{0.819}& \num{0.217} & \num{0.518}* & \num{0.925}* & \num{0.863}* \\
 \midrule
 \multirow{1}{*}{\textit{PLELog}}& \num{\fpeval{0.718-0.223}} & \num{\fpeval{0.784-0.327}} & \num{\fpeval{0.723-0.560}} & \num{\fpeval{0.642-0.498}} & \num{\fpeval{0.682-0.580}} & \num{\fpeval{0.672-0.350}} & \num{0.280}* & \num{0.309}& \num{0.313} & \num{0.311}* & \num{0.378}* & \num{0.160}* \\
\midrule
 \multirow{1}{*}{\textit{LogAnomaly}}& \num{0.310} & \num{\fpeval{0.784-0.425}} & \num{\fpeval{0.723-0.377}} & \num{\fpeval{0.642-0.297}} & \num{\fpeval{0.682-0.504}} & \num{\fpeval{0.672-0.398}} & \num{0.302}* & \num{0.298}& \num{0.105} & \num{0.201}* & \num{0.152}* & \num{0.251}* \\
\midrule
 \multirow{1}{*}{\textit{DeepLog}}& \num{\fpeval{0.718-0.457}} & \num{\fpeval{0.784-0.426}} & \num{\fpeval{0.723-0.314}} & \num{\fpeval{0.642-0.355}} & \num{\fpeval{0.682-0.528}} & \num{\fpeval{0.672-0.400}} & \num{0.290}* & \num{0.355}& \num{0.061} & \num{0.208}* & \num{0.188}* & \num{0.219}* \\
\midrule
 \multirow{1}{*}{\textit{LogCluster}}& \num{\fpeval{0.718-0.492}} & \num{0.340} & \num{\fpeval{0.723-0.360}} & \num{\fpeval{0.642-0.361}} & \num{\fpeval{0.682-0.497}} & \num{\fpeval{0.672-0.418}} & \num{0.274}* & \num{0.460}& \num{0.321} & \num{0.390}* & \num{0.547}* & \num{0.445}* \\
 \midrule
 \multirow{1}{*}{\textit{PCA}}& \num{\fpeval{0.718-0.641}} & \num{\fpeval{0.784-0.648}} & \num{\fpeval{0.723-0.536}} & \num{\fpeval{0.642-0.478}} & \num{\fpeval{0.682-0.645}} & \num{\fpeval{0.672-0.606}} & \num{0.111}* & \num{0.134}& \num{0.051} & \num{0.092}* & \num{0.103}* & \num{0.120}* \\
\bottomrule
\end{tabular}
}
\begin{tablenotes}
        \item [$*$] \footnotesize{\method yields a significant higher F1 score than compared baselines.}
       \end{tablenotes}
\end{threeparttable}
\end{table*}

\subsection{Effectiveness on \synlogevol with Sequence Level Changes}~\label{sec:appendix-synevol}
Table~\ref{tab:injectionrateresults_synevol}  presents the effectiveness of \method compared to baselines on \synlogevol under varying log sequence instability levels. \method consistently achieves the highest F1 score across all injection ratios, surpassing LightAD by \SI{1.3}{\pp} on average ($98.2\%-96.9\%$). A Mann-Whitney U test confirms this difference is statistically insignificant, indicating comparable effectiveness between \method and LightAD. However, \method achieves this performance while being trained on only \SI{22.96}{\percent} of unique log sequences, reducing usage of labeled data by \SI{77.04}{\pp}. Unlike in \synhdfs, where only LightAD and \method remain robust across instability levels, all supervised methods---including \method, LightAD, NeuralLog, LogRobust, and CNN---do not show a strong correlation between performance and increasing log instability in \synlogevol.  In contrast, semi-supervised and unsupervised methods exhibit a clear decline in precision, recall, and F1 score as instability increases. One possible reason is that changes at the log sequence level in \synlogevol often involve minor modifications, such as adding or removing a single template, which may have a limited impact on \task. For instance, at a \SI{30}{\percent} injection ratio, \SI{55}{\percent} of changes involve just one template, making the overall sequence structure relatively stable despite modifications. 

\begin{table*}[htbp]
\centering
\captionsetup[table]{skip=0pt}
\captionof{table}{Effectiveness of \method and baselines under different sequence-level injection ratios on \synlogevol. 
\label{tab:injectionrateresults_synevol}}
\footnotesize{%
\begin{threeparttable}[htbp]
 \resizebox{\linewidth}{!}{
\begin{tabular}{@{\hspace{0.5\tabcolsep}}c@{\hspace{0.5\tabcolsep}}c@{\hspace{0.8\tabcolsep}}c@{\hspace{0.5\tabcolsep}}c@{\hspace{0.5\tabcolsep}}c@{\hspace{0.5\tabcolsep}}c@{\hspace{0.5\tabcolsep}}c@{\hspace{0.5\tabcolsep}}c@{\hspace{0.5\tabcolsep}}c@{\hspace{0.5\tabcolsep}}c@{\hspace{0.5\tabcolsep}}c@{\hspace{0.5\tabcolsep}}c@{\hspace{0.5\tabcolsep}}c@{\hspace{0.5\tabcolsep}}}
\toprule
\multirow{3}{*}{\textbf{Data}}& \multirow{3}{*}{\textbf{Unstable}} & \multirow{3}{*}{\textbf{M}} & \multicolumn{1}{c}{\textbf{limited data}} & \multicolumn{9}{c}{\textbf{full training set}}\\
\cmidrule(lr){4-4} 
\cmidrule(lr){5-13}
&&& \multicolumn{5}{c}{\textbf{supervised}} & \textbf{Semi-S} & \multicolumn{4}{c}{\textbf{Unsupervised}}\\
\cmidrule(lr){4-8} 
\cmidrule(lr){9-9}
\cmidrule(lr){10-13}
& & & \method & LightAD & NeuralLog& LogRobust& CNN & PLELog & LogAnomaly& DeepLog& LogCluster& PCA\\
\midrule
  \multirow{3}{*}{0\%}& \multirow{3}{*}{No}& P &  \num{0.999}   &\num{0.999} & \num{0.999} & \num{0.941} & \num{0.999}& \num{0.172} & \num{0.501} & \num{0.512}& \num{0.771} & \num{0.072}\\ 

& &R & \num{0.969}  & \num{0.939} & \num{0.636} & \num{0.969} & \num{0.878}& \num{0.129} & \num{0.393} & \num{0.443}& \num{0.818} & \num{0.471} \\ 

 && F1 & \textbf{\num{0.984}} & \num{0.968} & \num{0.777} & \num{0.952} & \num{0.935}& \num{0.243} & \num{0.441} & \num{0.475}& \num{0.794} & \num{0.125}\\ 
\midrule
\midrule
 \multirow{3}{*}{5\%} & \multirow{3}{*}{Yes}& P & \num{0.999} & \num{0.999} & \num{0.999} & \num{0.969} & \num{0.999} & \num{0.179} & \num{0.388} & \num{0.384} & \num{0.783} & \num{0.057}  \\ 

& & R & \num{0.971} & \num{0.942} & \num{0.628} & \num{0.914} & \num{0.885} & \num{0.141} & \num{0.440} & \num{0.428} & \num{0.828} & \num{0.457}  \\ 

 & & F1 & \textbf{\num{0.985}} & \num{0.970} & \num{0.771} & \num{0.941} & \num{0.939} & \num{0.158} & \num{0.394} & \num{0.405} & \num{0.805} & \num{0.102} \\ 
\midrule

 \multirow{3}{*}{10\%}& \multirow{3}{*}{Yes}&P & \num{0.999} & \num{0.999} & \num{0.999} & \num{0.971} & \num{0.999} & \num{0.177} & \num{0.326} & \num{0.271} & \num{0.769} & \num{0.072}  \\ 

& &R & \num{0.973} & \num{0.945} & \num{0.648} & \num{0.918} & \num{0.891} & \num{0.145} & \num{0.459} & \num{0.432} & \num{0.810} & \num{0.471} \\ 

 && F1 & \textbf{\num{0.986}} & \num{0.972} & \num{0.786} & \num{0.944} & \num{0.942} & \num{0.160} & \num{0.382} & \num{0.333} & \num{0.789} & \num{0.125}  \\ 
\midrule

 \multirow{3}{*}{15\%}& \multirow{3}{*}{Yes} & P & \num{0.999} & \num{0.999} & \num{0.999} & \num{0.971} & \num{0.999} & \num{0.163} & \num{0.224} & \num{0.229} & \num{0.769} & \num{0.063} \\ 

& &R & \num{0.973} & \num{0.945} & \num{0.621} & \num{0.918} & \num{0.891} & \num{0.141} & \num{0.351} & \num{0.378} & \num{0.810} & \num{0.475} \\ 

 && F1 & \textbf{\num{0.986}} & \num{0.972} & \num{0.766} & \num{0.944} & \num{0.942} & \num{0.151} & \num{0.273} & \num{0.285} & \num{0.789} & \num{0.112}  \\ 
\midrule
 \multirow{3}{*}{20\%}& \multirow{3}{*}{Yes}& P  & \num{0.999} & \num{0.999} & \num{0.999} & \num{0.973} & \num{0.999} & \num{0.189} & \num{0.205} & \num{0.200} & \num{0.804} & \num{0.053} \\ 

& &R & \num{0.975} & \num{0.951} & \num{0.583} & \num{0.902} & \num{0.878} & \num{0.154} & \num{0.365} & \num{0.439} & \num{0.804} & \num{0.470} \\ 

 & &F1  & \textbf{\num{0.987}} & \num{0.975} & \num{0.738} & \num{0.936} & \num{0.935} & \num{0.170} & \num{0.263} & \num{0.274} & \num{0.804} & \num{0.095}  \\ 
\midrule

 \multirow{3}{*}{25\%}& \multirow{3}{*}{Yes}&P & \num{0.999} & \num{0.999} & \num{0.999} & \num{0.975} & \num{0.999} & \num{0.167} & \num{0.180} & \num{0.165} & \num{0.800} & \num{0.071}  \\ 

& &R& \num{0.953} & \num{0.930} & \num{0.651} & \num{0.930} & \num{0.906} & \num{0.141} & \num{0.441} & \num{0.418} & \num{0.837} & \num{0.511}  \\ 

 & &F1 & \textbf{\num{0.976}} & \num{0.963} & \num{0.788} & \num{0.952} & \num{0.951} & \num{0.153} & \num{0.256} & \num{0.236} & \num{0.818} & \num{0.125} \\ 
\midrule

 \multirow{3}{*}{30\%}& \multirow{3}{*}{Yes}& P  & \num{0.999} & \num{0.999} & \num{0.999} & \num{0.972} & \num{0.972} & \num{0.174} & \num{0.162} & \num{0.125} & \num{0.659} & \num{0.058}  \\ 

& &R & \num{0.950} & \num{0.925} & \num{0.600} & \num{0.900} & \num{0.875} & \num{0.145} & \num{0.475} & \num{0.425} & \num{0.775} & \num{0.434} \\ 

 & &F1& \textbf{\num{0.974}} & \num{0.961} & \num{0.750} & \num{0.935} & \num{0.921} & \num{0.158} & \num{0.242} & \num{0.193} & \num{0.712} & \num{0.102}  \\ 

\midrule

 \multirow{3}{*}{Average}& \multirow{3}{*}{Yes}& P  & \num{0.999}& \num{0.999}& \num{0.999}& \num{0.972}& \num{0.994}& \num{0.175}& \num{0.247}& \num{0.229}& \num{0.764}& \num{0.062}\\ 

& &R & \num{0.966}& \num{0.94}& \num{0.622}& \num{0.914}& \num{0.888}& \num{0.144}& \num{0.422}& \num{0.42}& \num{0.811}& \num{0.47} \\ 

 & &F1&\textbf{\num{0.982}}& \num{0.969}& \num{0.767}*& \num{0.942}*& \num{0.938}*& \num{0.158}*& \num{0.302}*& \num{0.288}*& \num{0.786}*& \num{0.11} *\\ 
\bottomrule
\end{tabular}
}
\begin{tablenotes}
         \item [$*$] \footnotesize{\method yields a significant higher F1-score than compared baseline.}
        \end{tablenotes}
\end{threeparttable}

}
\end{table*}

\end{document}